\documentclass[aps,prd,twocolumn,nofootinbib,superscriptaddress,preprintnumbers,balancelastpage,longbibliography]{revtex4-2}
\pdfoutput=1
\usepackage[english]{babel}

\usepackage{amsfonts}
\usepackage{amsmath,mathtools,physics,xfrac}
\usepackage{mathrsfs}
\usepackage[normalem]{ulem}
\usepackage{graphicx}
\usepackage{afterpage}
\usepackage{float}
\usepackage{rotating}
\usepackage{multirow}
\usepackage{tabularx}
\usepackage{booktabs}
\usepackage{fancyhdr}
\usepackage[utf8]{inputenc}
\usepackage{theorem}
\usepackage{moreverb}
\usepackage{euscript}
\usepackage{psfrag}
\usepackage{slashed}
\usepackage{makecell}
\usepackage{adjustbox}
\usepackage{dcolumn}
\usepackage{bm}
\usepackage[dvipsnames]{xcolor}
\usepackage{hyperref}
\usepackage{siunitx}
\usepackage{orcidlink}
\DeclareSIUnit\CCD{CCD}
\DeclareSIUnit\year{yr}

\definecolor{darkviolet}{rgb}{0.58, 0.0, 0.83}
\usepackage{dsfont}

\hypersetup{
     colorlinks   = true,
     citecolor    = darkviolet,
     urlcolor     = darkviolet,
     linkcolor    = darkviolet
}

\definecolor{darkblue}{rgb}{0.0,0.0,0.75}
\definecolor{darkred}{rgb}{0.6,0.0,0}
\definecolor{darkgreen}{rgb}{0.0,0.6,0.}

\begin{document}

\title{Towards Quantum-Dot Detectors as Barcodes for Dark Matter Interactions}

\author{Marek Matas~\orcidlink{0000-0002-8250-8187}}
\thanks{\href{mailto:marek.matas@fjfi.cvut.cz}{marek.matas@fjfi.cvut.cz}}
\affiliation{Faculty of Nuclear Sciences and Physical Engineering, Czech Technical University in Prague, Czech Republic}

\author{Andrea Gallo Rosso~\orcidlink{0000-0002-4664-5504}}
\thanks{\href{mailto:andrea.gallo.rosso@fysik.su.se}{andrea.gallo.rosso@fysik.su.se}}
\affiliation{Stockholm University and The Oskar Klein Centre for Cosmoparticle Physics, Alba Nova, Stockholm, Sweden}

\author{Antonio Cammarata~\orcidlink{0000-0002-5691-0682}} 
\affiliation{Department of Control Engineering, Faculty of Electrical Engineering, Technicka 2, 16627 Prague 6, Czech Technical University in Prague, Prague, Czech Republic}

\author{Nora Hoch~\orcidlink{0000-0001-7227-2556}}
\affiliation{Laboratory of Nuclear Science, Massachusetts Institute of Technology, Cambridge, MA, USA}

\author{Carlos~Blanco~\orcidlink{0000-0001-8971-834X}}
\thanks{\href{mailto:carlosblanco@psu.edu}{carlosblanco@psu.edu}}
\affiliation{Institute for Gravitation and the Cosmos, The Pennsylvania State University, University Park, PA, USA}
\affiliation{Department of Physics, Princeton University, Princeton, NJ, USA}
\affiliation{Stockholm University and The Oskar Klein Centre for Cosmoparticle Physics, Alba Nova, Stockholm, Sweden}

\author{Jan~Conrad \orcidlink{0000-0001-9984-4411}} 
\affiliation{Stockholm University and The Oskar Klein Centre for Cosmoparticle Physics, Alba Nova, Stockholm, Sweden}

\author{Rouven Essig~\orcidlink{0000-0002-3066-0486}}
\affiliation{C.N.~Yang Institute for Theoretical Physics, Stony Brook University, NY, USA}

\author{Tim Linden~\orcidlink{0000-0001-9888-0971}}
\affiliation{Stockholm University and The Oskar Klein Centre for Cosmoparticle Physics, Alba Nova, Stockholm, Sweden}
\affiliation{Erlangen Centre for Astroparticle Physics (ECAP), Friedrich-Alexander-Universität Erlangen-Nürnberg, Erlangen, Germany}

\author{Lindley Winslow~\orcidlink{0000-0002-9970-108X}}
\affiliation{Laboratory of Nuclear Science, Massachusetts Institute of Technology, Cambridge, MA, USA}

\begin{abstract}
Quantum dots are tunable semiconductor nanocrystals that can be produced at industrial scales. We present the first \emph{ab initio} calculation of the scattering of dark matter  on electrons bound in quantum dots. The momentum-dependence of a quantum dot's electronic response depends on its morphology and on the dark matter mass, interaction operator, mediator coupling, and mediator mass. Therefore, the relative rates across an array of distinct quantum dot targets form a ``barcode'' that carries information about the nature of the dark matter interaction. We project the sensitivity of a detector concept in which a collection of independent target subunits, each loaded with silicon quantum dots of a particular morphology, are read out by Skipper CCDs. Given a future signal, this barcode could discriminate between interaction operators and mediator types. We quantify the discrimination power for a benchmark pair of models as a function of readout noise and exposure.
\end{abstract}

\maketitle

\section{Introduction}

The identity of dark matter (DM) remains one of the central mysteries of modern physics. Direct searches for dark matter at electroweak scales and at masses heavier than a proton have successfully probed large regions of parameter space~\cite{XENON:2025vwd,LZ:2024zvo,PandaX:2024qfu}. However, many theoretically motivated dark matter candidates are expected to have masses below a GeV, where nuclear-recoil searches become kinematically limited~\cite{Battaglieri:2017aum,Essig:2022dfa}. Searching for MeV-GeV dark matter requires methods sensitive to inelastic processes, which can detect energy deposits as small as an electronvolt. 

Sub-GeV dark matter searches largely focus on probing the coupling between dark matter and the electron~\cite{Essig:2011nj}, since electrons are lighter and can be found in bound states which provide a channel for inelastic scattering. Some of the most sensitive direct detection experiments use bulk silicon to look for the small energy depositions following a dark matter scattering event with a target electron~\cite{Essig:2015cda,DAMIC-M:2025luv,SENSEI:2024yyt,SENSEI:2023zdf,Oscura:2022vmi}. Other technologies, including organic and inorganic scintillators as well as bolometers, have also made significant headway in probing the existing parameter space~\cite{Derenzo:2016fse,Blanco:2019lrf,EDELWEISS:2020fxc,SuperCDMS:2025dha,XENON:2026qow,XENON:2025vwd,PhysRevLett.130.261001,PandaX:2025rrz}. Luminescent chromophores, such as organic scintillators, are interesting, since they could be scaled up to ton-year exposures~\cite{Blanco:2021hlm,Blanco:2022pkt}. In such experiments, a luminescent signal emerges from radiative de-excitation following a bound-to-bound electronic transition. It was previously pointed out that colloidal nanocrystals, in particular PbS quantum dots, provide similar scaling advantages, along with lower threshold energies, and the potential for time-coincident 2$\gamma$ signals~\cite{Blanco:2022cel}.  

\begin{figure}[t!]
\centering
  \includegraphics[width=0.48\textwidth]{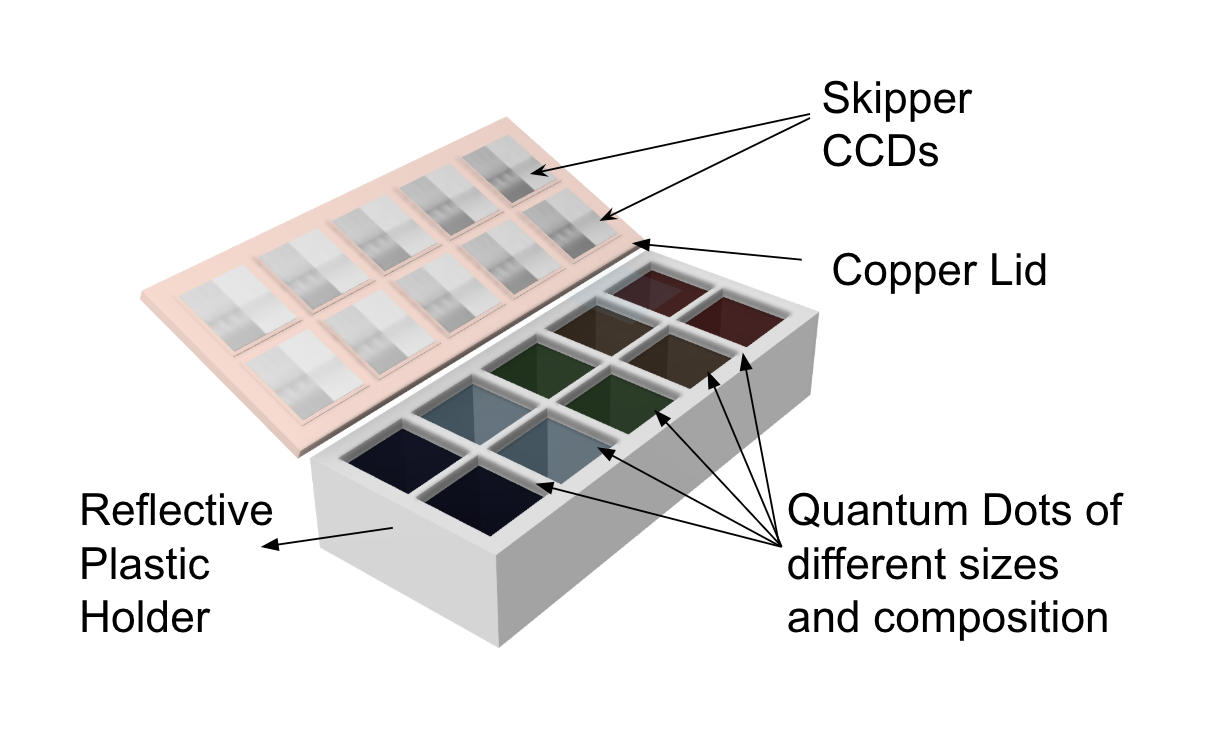}
\caption{A CAD rendering of a possible experimental detector design, based on quantum dot (QD) technology and using Skipper CCDs~\cite{Tiffenberg:2017aac} to detect photon signals from the QDs. Here, each section of the detector consists of a \SI{10}{\centi\meter\squared} area Skipper CCD and \SI{1}{\kilo\gram} of QD target material. Each QD-based scintillator volume, in our benchmark setup, is composed of 50\% QDs by weight, loaded in a solid matrix. There are five pairs of volumes, one for each QD combination of diameter and surface termination. We consider independent pairs to separate backgrounds originating in the active volume from those in the photon readout.}
\label{fig:3ex-mockup}
\end{figure}

Quantum dots (QDs) are nano-scale crystals, herein of semiconducting material, whose low-energy electronic states are strongly influenced by the characteristic size of the crystal~\cite{Brus1984,Alivisatos1996,EfrosBrus2021}. These QDs are clusters of hundreds to thousands of atoms, and their electronic states have characteristics of both molecules and bulk crystals, converging on the bulk properties when the QD becomes sufficiently large. Selecting the QD size tunes the band-edge properties, notably the electronic gap~\cite{Brus1984,Alivisatos1996,Murray1993,EfrosBrus2021,wolkin1999electronic,kanemitsu1997visible,hannah2012origin,takai2017size,gert2017tight,dohnalova2013surface}. QDs can be manufactured at industrial scales and are commercially available in large (g--kg) quantities, usually as colloidal suspensions in organic solvents.

Here, we consider colloidal Si QDs as the primary target of DM scattering, and compare the specific rate to that of bulk Si. We show that Si QDs can be a key complement to bulk-crystal experiments, and show that QDs provide a way path towards multi-kg Si-based searches with ability to discriminate between DM models due to the inherent tunability of the QDs. We aim to determine how the tunable QD morphology affects their ability to become electronically excited following a dark-matter electronic recoil. We consider two characteristic variables, QD diameter and the type of surface ligand. In colloidal suspensions, the size can be tuned during the synthesis of these materials, and this process has been industrially refined due to the photonic and electronic applications of quantum dots. To stabilize the colloidal suspension, and since the atoms on the exterior of the nanocrystals are very reactive due to dangling bonds, QDs are coated with covalently bonded surface ligands. Here, we consider two kinds of surface passivation: hydrogen (H-terminated) and hydrocarbon chains (alkyl-terminated). We select H-terminated Si QDs, since they are expected to have a band structure most like that of bulk Si~\cite{reboredo2005theory,dohnalova2013surface}, making them good benchmarks for the effects of electronic gap tunability--- although they are known to be dim chromophores prone to oxidation~\cite{wolkin1999electronic}.  Alkyl-terminated QDs, on the other hand, are known to be bright, fast emitters in the optical spectrum~\cite{dohnalova2013surface}, with near-unity internal quantum efficiency~\cite{Sangghaleh_2015}. We find that carbon-based ligands significantly affect the electronic states of the QDs and therefore their expected excitation rates, on average improving the QD sensitivity over their H-terminated counterparts.

We show that semiconducting quantum dots used as the primary target in dark matter detectors provide information complementary to bulk-crystal semiconductors. The per-mass interaction rate in QDs is suppressed by a factor of a few compared to the bulk, as shown in Figs.~\ref{fig:rates-QDs-O1short}--\ref{fig:rates-QDs-O3short}, and the signal consists predominantly of single-photon events, so that little information about the recoil spectrum remains when observing with a single QD morphology. However, scintillation experiments need only be instrumented at the surface for photon readout. Therefore, the readout, along with its dark-count background, scales with detector area rather than with target mass. In principle, QDs are readily deployed at kg-scales by loading them into a solid matrix, where the total mass of the detector volume is composed of active QDs at $\mathcal{O}(50\%)$ by weight.

Furthermore, the size-tunable electronic gap and the choice of surface ligands provide a way to probe the nature of the dark matter interaction in the event of a signal. We compute the expected rates for a complete set of couplings between dark matter and the electrons in the QDs, and consider an array of independently instrumented detector volumes, each with a QD-based target scintillator, as shown in~\autoref{fig:3ex-mockup}. In our benchmark setup, there are five pairs of scintillator targets, each pai made of 1~kg of QD-based scintillator hosting QDs of a particular diameter and surface ligand, for five distinct size--ligand combinations in total. Both the QD size and the surface ligands affect the expected rates across the target volumes; the relative rates between these targets are what we refer to as a ``barcode.'' While the barcode can be thought of as a proxy for the energy spectrum, it constitutes a new set of experimentally tunable parameters: varying the nanocrystal size and surface composition within a single experimental platform gives access to many targets with distinct energy--momentum electron distributions, which can be exploited to identify the nature of the DM interaction.

The paper is structured as follows. In \autoref{sec:formalism}, we summarize the effective-field-theory framework for DM--electron scattering in finite QD targets, while \autoref{sec:QDmodeling} describes the \emph{ab initio} construction of the Si QD models, their excitation thresholds, and the treatment of screening. In \autoref{sec:materialresponse}, \autoref{sec:gapandoverlaps}, and \autoref{sec:optimalsetup}, we present the resulting material responses, disentangle the roles of the QD band gap and electronic form factors, and discuss the identification of favorable QD morphologies; \autoref{sec:barcore} develops the QD barcode strategy for interaction identification, and \autoref{sec:detectorresponse} presents detector-level sensitivity projections including instrumental backgrounds. We summarize our conclusions in \autoref{sec:conclusions}, and provide supporting figures and data in the appendices.

\section{Interaction Formalism}\label{sec:formalism}
This work builds upon the formalism presented in Ref.~\cite{Catena:2021qsr}, constructed for DM scattering in crystalline materials (see also Refs.~\cite{Catena:2019gfa, Catena:2022fnk, PhysRevD.98.123003, Catena:2019hzw} for further review). In contrast to the case of bulk crystals, here we model QDs as finite objects fully contained within the simulation volume and dispersed randomly within a solvent. This setup prevents the formation of a crystal momentum that would arise from phase differences between periodically repeating sites. We therefore do not include the entire Brillouin zone in the scattering and set the crystal momenta of the initial and final states to zero, $|\mathbf{k}| = |\mathbf{k'}| = 0$. Therefore, all information about the initial- and final-state momenta of the electronic states is encoded in the plane-wave expansion of the unit-cell wavefunction. This is expanded in terms of the Bloch-state plane wave basis in the limiting case of no crystal momentum, as

\begin{equation}
\psi_{i}(\mathbf{x})=\frac{1}{\sqrt{V}}\sum_\mathbf{G}{u_i(\mathbf{G})e^{i\mathbf{G}\cdot \mathbf{x}}}\, ,
\label{eq:psi}
\end{equation}
where $\psi_{i}(\mathbf{x})$ is the initial-state wavefunction corresponding to the electronic band $i$ with energy $E_i$. The basis set is given in terms of the reciprocal lattice vectors $\mathbf{G}$ that represent the electronic momentum, $V$ is the volume of the crystal, and $u_i(\mathbf{G})$ are the Bloch coefficients. An analogous expression holds for the final state $\psi_{i'}(\mathbf{x})$. 

We are working in the non-relativistic limit in which one can rewrite the free scattering amplitude in terms of the momentum transfer $\mathbf{q}$ and the transverse relative velocity $\mathbf{v}_{\rm el}^\perp$~\cite{Catena:2021qsr} as

\begin{equation}
 \label{eq:Mnr}
\mathcal{M}(\mathbf{q},\mathbf{v}_{\rm el}^\perp) = \sum_k \left(c_k^s +c^\ell_k \frac{q_{\rm ref}^2}{|\mathbf{q}|^2} \right) \,\langle \mathcal{O}_k \rangle  \,.
\end{equation}
Here, $\mathcal{O}_k$ are all the allowed effective operators for a spin-1/2 DM particle listed in~\autoref{tab:operators}. In this paper, we do not discuss the origin of these operators in UV-complete theories, which may, however, provide additional constraints on them. In this approach, the interaction can be mediated by a heavy (light) mediator, represented here by the couplings $c_k^s$ ($c^\ell_k$), where the superscripts denote short- and long-range interactions. The reference momentum associated with the light mediator is given by $q_{\rm ref}= \alpha m_e$, where $\alpha$ is the fine structure constant and $m_e$ is the mass of an electron.

We can then write the momentum-space transition element between states $i$ and $i'$ as

\begin{align}
    \overline{\left| \mathcal{M}_{i\rightarrow i'}\right|^2}\equiv \overline{\left|\int  \frac{{\rm d}^3 \ell}{(2 \pi)^3} \, \widetilde{\psi}_i'^*(\boldsymbol{\ell}+\mathbf{q})  
\mathcal{M}(\mathbf{q},\mathbf{v}_{\rm el}^\perp)
\widetilde{\psi}_i(\boldsymbol{\ell}) \right|^2}\,,
\label{eq:transition_amplitude}
\end{align}
where $\boldsymbol{\ell}$ is the electron momentum, $\widetilde{\psi}_i$ and $\widetilde{\psi}_i'$ are the initial- and final-state wavefunctions in the momentum space, and the transverse relative velocity $\mathbf{v}_{\rm el}^\perp$ (perpendicular to $\mathbf{q}$ for elastic scatterings) is given by 

\begin{equation}
    \mathbf{v}_{\rm el}^\perp=\mathbf{v}-\frac{\boldsymbol{\ell}}{m_e}-\frac{\mathbf{q}}{(2 \mu_{\chi e})}\,.
\end{equation}
Here $\mathbf{v}$ is the dark matter velocity, and $\mu_{\chi e}$ is the DM-electron reduced mass.

After the expansion of the scattering amplitude in terms of the ratio of the electron momentum and mass, one can identify the scalar and vector electronic form factors, 

\begin{align}
f_{i \rightarrow i' }(\mathbf{q})&=\int \mathrm{d}^3x \, \psi_{i' }^*(\mathbf{x})\,e^{i\mathbf{x}\cdot\mathbf{q}} \,\psi_{i}(\mathbf{x}) \label{eq:f}\\
\mathbf{f}_{i\rightarrow i' }(\mathbf{q})&=\int \mathrm{d}^3x \, \psi_{i' }^*(\mathbf{x}) \,e^{i\mathbf{x}\cdot\mathbf{q}}\, \frac{i\nabla_\mathbf{x}}{m_e} \psi_{i}(\mathbf{x})\, .
\label{eq:fvec}
\end{align}
These contribute to the final cross section in the $|\mathbf{k}| = |\mathbf{k'}| = 0$ limit, since the individual QDs are not embedded in a periodic structure.

The momentum transfer $\mathbf{q}$ can equivalently be expressed as the difference between the initial and final state electron momenta $\mathbf{q}= \mathbf{G}^\prime-\mathbf{G}$ and, substituting in the Bloch form of the wavefunctions, one obtains

\begin{align}
 f_{i\rightarrow i^\prime}^\prime &= \sum_{\mathbf{G}}u_{i^\prime}^*\left(\mathbf{G}+\mathbf{q}\right) u_i\left(\mathbf{G}\right)\label{eq:f_prime} \\
    \mathbf{f}_{i\rightarrow i^\prime}^\prime &= -\frac{1}{m_e}\sum_{\mathbf{G}}u_{i^\prime}^*\left(\mathbf{G}+\mathbf{q}\right)u_i\left(\mathbf{G}\right)\cdot\mathbf{G} \, .
\label{eq:fvec_prime}
\end{align}

The form factors $f_{i\rightarrow i^\prime}^\prime$ and $\mathbf{f}_{i\rightarrow i^\prime}^\prime$ can be combined into five independent contributions to the material response functions

\begin{align}
B_1 =& \left| f_{i\rightarrow i^\prime }^\prime \right|^2 \\
B_2=&-\frac{\mathbf{q}}{m_e} \cdot(f_{i\rightarrow i^\prime }^\prime) (\mathbf{f}_{i\rightarrow i^\prime }^\prime)^* \\
B_3=&\left| \mathbf{f}_{i\rightarrow i^\prime }^\prime \right|^2  \\
B_4=& \left|\frac{\mathbf{q}}{m_e} \cdot \mathbf{f}_{i\rightarrow i^\prime }^\prime \right|^2 \\
B_5=& i\frac{\mathbf{q}}{m_e} \cdot \left[\mathbf{f}_{i\rightarrow i^\prime }^\prime \times \left(\mathbf{f}_{i\rightarrow i^\prime }^\prime\right)^*\right],
\end{align}
where the velocity integral suppresses the contribution of the functions $\mathbf{B_6}$ and $\mathbf{B_7}$, which are given in~\cite{Catena:2021qsr}. Here we follow the findings presented in Ref.~\cite{Catena:2024rym} and correct for the missing negative sign of the term $B_2$. The five material responses that encode the energy--momentum dependence of the interaction, based on the properties of the target, are expressed as 

\begin{align}
    \overline{W}_l(q,\Delta E)&=(4\pi)^2V_\text{cell}\frac{\Delta E}{q^2}\sum_{\mathbf{G'}-\mathbf{G}} \sum_{ii^\prime}\,B_l \, \nonumber\\
        &\times\delta(\left|\mathbf{G'} -\mathbf{G} \right|-q) \nonumber\\
        &\times\delta(\Delta E -E_{i'}+E_{i})
    \label{eq:W_scalar_2D_2}\,,
\end{align}
where $\Delta E = E_{i'}-E_{i}$ is the energy deposited in the QD by the DM-electron interaction, and $V_\text{cell}$ is the volume of the unit cell (fiducial volume $V = N_\text{cell} V_\text{cell}$). 

The final expression for the event rate is given by

\begin{align}
\mathscr{R}&=\frac{n_\chi N_\text{cell} }{128\pi m_\chi^2 m_e^2}\int \mathrm{d} (\ln\Delta E)\int \mathrm{d} q \, q \,\widehat{\eta}\left(q, \Delta E
\right)
\nonumber\\
&\times \sum_{l=1}^{5} \Re\left(R_l^*(q,v) \overline{W}_l(q,\Delta E)\right)\,,
\label{eq:R_crystal_2D}
\end{align}
where $n_\chi=0.4\text{ GeV/cm}^{3}/m_\chi$ is the local DM number density~\cite{Catena:2009mf}, $N_\text{cell}$ is the number of unit cells, $m_\chi$ is the DM-particle mass, and $\widehat{\eta}\left(q, \Delta E \right)$ denotes the velocity integral of $f_\chi(\mathbf{v})/v$ applied to the velocity-dependent parts of $R_l$, following Ref.~\cite{Catena:2021qsr}; for velocity-independent operators it reduces to the standard inverse mean speed $\widehat{\eta}(v_{\rm min})$ defined below~\cite{Essig:2015cda}. The dark matter responses $R_l(q,v)$ are functions of the couplings in~\autoref{eq:Mnr} and encode the model dependence of the scattering. These are identical to those identified in Ref.~\cite{Catena:2021qsr}.

Throughout this work, we adopt a standard halo model, i.e. a truncated Maxwell-Boltzmann distribution boosted to the detector frame as

\begin{align}\label{eq:maxwell-boltzmann}
    f_\chi(\mathbf{v})&= \frac{1}{N_{\rm esc}\pi^{3/2}v_0^3}\exp\left[-\frac{(\mathbf{v}+\mathbf{v}_\oplus)^2}{v_0^2} \right]
    \nonumber\\
    &\times \Theta\left(v_{\rm esc}-|\mathbf{v}+\mathbf{v}_\oplus|\right)\,,
\end{align}
where $v_0=220~\text{km/s}$~\cite{Kerr:1986hz}, the galactic escape velocity is $v_{\rm esc} = 544~\text{km/s}$~\cite{Smith:2006ym}, and the velocity of the Earth in the galactic rest frame is~$|\mathbf{v_\oplus}|= 244~\text{km/s}$~\cite{Catena:2009mf}. The normalization factor is chosen to unit-normalize the distribution as
\begin{align}
N_{\rm esc}\equiv \erf(v_{\rm esc}/v_0)-2 (v_{\rm esc}/v_0)\exp(-v_{\rm esc}^2/v_0^2)/\sqrt{\pi}.
\end{align}
The inverse mean speed is given by the following~\cite{Essig:2015cda}, 
\begin{align}
\widehat{\eta}(v_{\text{min}} ) &= \int{\rm d}^3v \,\frac{f_\chi(\mathbf{v})}{v} \Theta(v-v_{\text{min}}) \,
\end{align}
where the minimal velocity a DM particle must have to deposit energy $\Delta E$ and momentum $q$ is given by 
\begin{align}
v_{\rm min}(q, \Delta E)=\frac{q}{2 m_\chi } + \frac{\Delta E}{q } \,.
\label{eq:vmin}
\end{align}

\begin{table}[t]
    \centering
    \begin{tabular*}{\columnwidth}{@{\extracolsep{\fill}}ll@{}}
    \toprule
      $\mathcal{O}_1 = \mathds{1}_{\chi e}$ & $\mathcal{O}_9 = i\mathbf{S}_\chi\cdot\left(\mathbf{S}_e\times\frac{ \mathbf{q}}{m_e}\right)$  \\
        $\mathcal{O}_3 = i\mathbf{S}_e\cdot\left(\frac{ \mathbf{q}}{m_e}\times \mathbf{v}^{\perp}_{\rm el}\right)$ &   $\mathcal{O}_{10} = i\mathbf{S}_e\cdot\frac{ \mathbf{q}}{m_e}$   \\
        $\mathcal{O}_4 = \mathbf{S}_{\chi}\cdot \mathbf{S}_e$ &   $\mathcal{O}_{11} = i\mathbf{S}_\chi\cdot\frac{ \mathbf{q}}{m_e}$   \\                                                                             
        $\mathcal{O}_5 = i\mathbf{S}_\chi\cdot\left(\frac{ \mathbf{q}}{m_e}\times \mathbf{v}^{\perp}_{\rm el}\right)$ &  $\mathcal{O}_{12} = \mathbf{S}_{\chi}\cdot \left(\mathbf{S}_e \times \mathbf{v}^{\perp}_{\rm el} \right)$ \\                                                                                                                 
        $\mathcal{O}_6 = \left(\mathbf{S}_\chi\cdot\frac{ \mathbf{q}}{m_e}\right) \left(\mathbf{S}_e\cdot\frac{ \mathbf{q}}{m_e}\right)$ &  $\mathcal{O}_{13} =i \left(\mathbf{S}_{\chi}\cdot  \mathbf{v}^{\perp}_{\rm el}\right)\left(\mathbf{S}_e\cdot \frac{ \mathbf{q}}{m_e}\right)$ \\   
        $\mathcal{O}_7 = \mathbf{S}_e\cdot  \mathbf{v}^{\perp}_{\rm el}$ &  $\mathcal{O}_{14} = i\left(\mathbf{S}_{\chi}\cdot \frac{ \mathbf{q}}{m_e}\right)\left(\mathbf{S}_e\cdot  \mathbf{v}^{\perp}_{\rm el}\right)$  \\
        $\mathcal{O}_8 = \mathbf{S}_{\chi}\cdot  \mathbf{v}^{\perp}_{\rm el}$  & $\mathcal{O}_{15} = i\mathcal{O}_{11}\left[ \left(\mathbf{S}_e\times  \mathbf{v}^{\perp}_{\rm el} \right) \cdot \frac{ \mathbf{q}}{m_e}\right] $ \\       
    \bottomrule
    \end{tabular*}
    \caption{Effective interaction operators for spin-1/2 DM--electron interactions in the non-relativistic limit~\cite{Fan:2010gt,Fitzpatrick:2012ix,Catena:2019gfa}. $\mathbf{S}_{\chi}$ ($\mathbf{S}_{e}$) are the DM (electron) spin operators, and $\mathds{1}_{\chi e}$ denotes the identity operator. The operator $\mathcal{O}_2 = \left(\mathbf{v}^{\perp}_{\rm el}\right)^2$ is conventionally omitted, as it does not arise at leading order from relativistic UV models.}
\label{tab:operators}
\end{table}

\section{Modeling Quantum Dots}\label{sec:QDmodeling}
In this section, we describe how we obtain the wavefunctions of bound electrons in the Bloch-state plane wave basis (\autoref{eq:psi}). In this study, we use the open-source density-functional-theory code Quantum ESPRESSO~\cite{Giannozzi_2009,Giannozzi_2017,10.1063/5.0005082}. This software is interfaced with QEdark-EFT~\cite{urdshals_2023_7836577}, an extension of QEdark~\cite{Essig:2015cda} that calculates the material responses entering the DM rate.

\subsection{Ab Initio QD Simulation}
\label{sec:qdsetup}

\begin{figure*}
\centering
  \centering
  \includegraphics[width=0.8\textwidth]{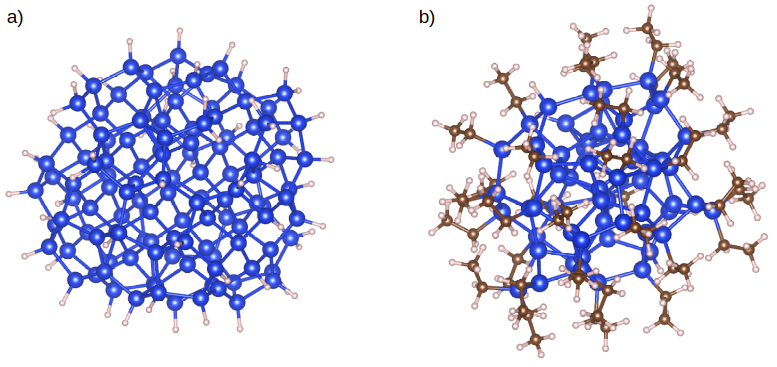}
\caption{Model geometry example of a) hydrogen-terminated and b) alkyl-terminated quantum dots with 121 and 63 silicon atoms, respectively (121-Si-H and 63-Si-alk systems). Blue, brown, and white spheres represent the position of the silicon, carbon, and hydrogen atoms, respectively.}
\label{fig:3D-QD-example}
\end{figure*}

To model the geometries of the quantum dots used in this study, we begin with the experimental structure of bulk silicon reported in the literature \cite{tobbens2001}. We truncate the periodic structure to obtain silicon clusters of different sizes, which we name $n$-Si, where $n$ is the number of Si atoms. In each $n$-Si cluster, the dangling bonds at the surface are saturated by adding H or alkyl groups (C$_2$H$_5$) to complete the 4-fold tetrahedral coordination as in the parent bulk geometry. In this way, we create the $n$-Si-H and $n$-Si-alk quantum dot models, where the species used to passivate the surface is indicated by the respective names. An example of the morphology of two quantum dots is shown in \autoref{fig:3D-QD-example}, while the remaining ones and the bulk are shown in Appendix~\ref{app:morphology}, \autoref{fig:3D-QDs}.

The ground-state geometries, charge densities, wavefunctions, and derived quantities of all the quantum dot models have been obtained by solving the time-independent Schr\"odinger equation within the density functional theory (DFT) framework \cite{PhysRev.136.B864,1965PhRv..140.1133K} as implemented in the Quantum ESPRESSO software \cite{Giannozzi_2009,Giannozzi_2017,10.1063/5.0005082}. We select the energy functional by benchmarking all compatible combinations of exchange and correlation options implemented in the software. The benchmark is done on the \emph{bulk structure} as follows. 
\begin{itemize}
    \item For each energy functional, we relax the atomic positions and lattice parameters, then select the functional that best reproduces the experimental geometry
    \item We use the relaxed geometry to calculate the band gap, varying the energy functional, then select the one that provides the best agreement with the experimental value.
\end{itemize}
This ensures that the geometry of the clusters is properly relaxed and the corresponding ground state wavefunction accurately calculated. After benchmarking, we select the Perdew-Zunger \cite{PhysRevB.23.5048} formulation to obtain the relaxed geometric configurations, and the Slater exchange functional without correlation to calculate the ground state wavefunction, band gap, and related properties.

Since we aim to simulate isolated (dilute) quantum dots, we insert the model geometries in a cubic unit cell with a minimum lattice parameter of 25 \AA{}, to prevent interactions between adjacent cell replicas. We model each QD as an isolated cluster in vacuum; the dielectric environment of the host matrix, which can shift the electronic levels and alter the surface states, is left for future work. We sample only the $\Gamma$ point in the reciprocal space. Atomic species are specified via scalar-relativistic norm-conserving pseudopotentials \cite{PhysRevB.88.085117} to preserve the unit-normalization of the Bloch-state basis set. The plane-wave basis energy cutoff ($E_\text{cut}$) is set to 29 Ry, while the convergence criteria are chosen to be 10$^{-8}$ Ry and 10$^{-6}$ Ry/\AA{} for the self-consistent-field and geometry-optimization cycles, respectively.

 We test the convergence of all numerical parameters with respect to the expected DM excitation rates, which enables us to use a smaller energy cutoff and box size than is usually necessary for solid-state calculations. Since DFT is known to underestimate the band gap, we use a scissor correction to fix its value in the postprocessing code QEdark-EFT, as discussed in the next section.

\subsection{Transition Threshold Energy}

\begin{figure}[t]
    \centering
    \includegraphics[width=\linewidth]{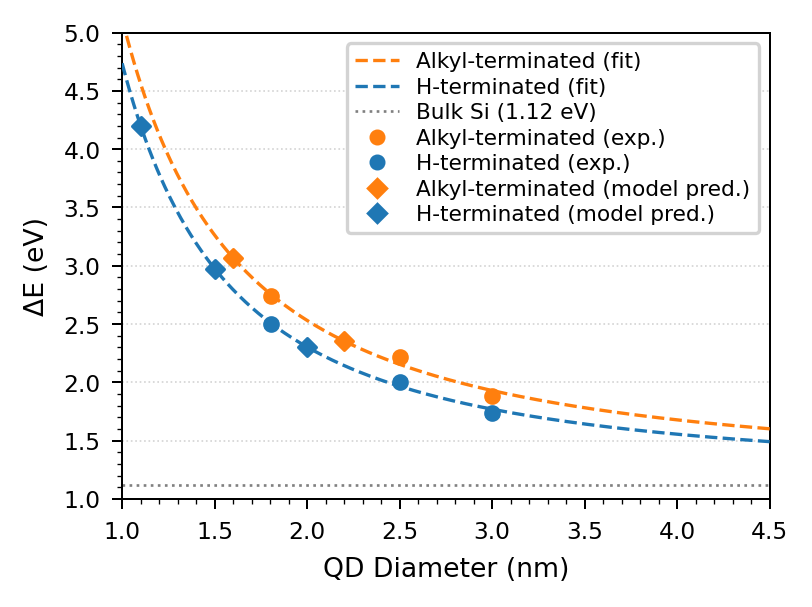}
    \caption{We show the vertical transition energy (optical gap) $\Delta E$ of H-terminated (blue) and Alkyl-terminated (orange) quantum dots. The dashed lines show our best-fit model based on the experimental values in Ref.~\cite{dohnalova2013surface} (circles), which is used to predict the transition-energy threshold for our quantum dots (diamonds). }
    \label{fig:opticalgaps}
\end{figure}

To drive a transition from the ground state to an excited state, the dark matter must deposit a minimum threshold energy. In the literature, this is sometimes called the optical gap, i.e., the minimum energy $\Delta E$ needed to drive a vertical (optical) transition. Generically, the energy levels in quantum dots are shifted with respect to those of bulk materials. This \textit{quantum confinement} can be heuristically understood as arising from confining the electrons (or excitons) in potentials whose size is on the order of their wavelength. This effect is what leads to the size-tunable absorption and emission of QDs.

While the optical gap of strongly-confining QDs can be estimated directly~\cite{Blanco:2022cel}, here we adopt a data-tuned model. We fit a two-parameter model to the experimental optical gaps of Si-H and Si-alk QDs reported in Ref.~\cite{dohnalova2013surface}, given by the following,
\begin{equation}
    \Delta E(d) = \alpha \left(\frac{\text{nm}}{d}\right)^2 + \beta \, \frac{\text{nm}}{d} + \Delta E_\text{gap}^\text{bulk} \ ,
\end{equation}
where $\Delta E$ is the optical gap, $\Delta E_\text{gap}^\text{bulk} = 1.12\,\text{eV}$ is the bulk gap of silicon, $d$ is the diameter of the quantum dots, and the coefficients $\alpha$ and $\beta$ (both in eV) are fit numerically. We show these best-fit models in~\autoref{fig:opticalgaps} as dashed lines. The experimental values from Ref.~\cite{dohnalova2013surface} are shown as filled circles, and the values adopted for our QDs are shown as filled diamonds. We present the best-fit parameters for our optical-gap models in~\autoref{tab:best-fit}. These models provide an excellent fit to the experimentally-measured rising edge of the photon-absorption cross section as shown in the supplemental material of Ref.~\cite{dohnalova2013surface}, for both kinds of surface termination (Si-H and Si-alk). The optical gaps we adopt in this study are shown in~\autoref{tab:gap-values}. Hereafter, we refer to this fitted optical gap simply as the QD gap (or band gap), noting that for QDs it includes the exciton binding energy.

\begin{table}[t!]
    \centering
    \begin{tabular}{ccc}
        QD & $\alpha$ [eV] & $\beta$ [eV]\\ \hline \hline
        $n$-Si-alk  & 2.35 & 1.65 \\
        $n$-Si-H  & 2.50 & 1.12 \\
    \end{tabular}
    \caption{Best-fit parameters for our optical-gap models. We adopt separate models for $n$-Si-alk and $n$-Si-H quantum dots.}
    \label{tab:best-fit}
\end{table}

\begin{table}
    \centering
    \begin{tabular}{ccc}
        QD & $d$ (nm) & $\Delta E$ (eV)\\ \hline \hline
        29-Si-H & 1.1 & 4.2\\
        64-Si-H & 1.5 & 3.0\\
        121-Si-H & 2.0 & 2.3\\
        29-Si-alk & 1.6 & 3.0\\
        63-Si-alk & 2.2 & 2.4\\
    \end{tabular}
    \caption{The diameters $d$ and optical gaps $\Delta E$ of the studied quantum dots. The optical gaps, used as excitation energy thresholds, are computed from the best-fit models in~\autoref{fig:opticalgaps} and~\autoref{tab:best-fit}. }
    \label{tab:gap-values}
\end{table}

\subsection{Emission Energy and Quantum Yield}\label{sec:emissionQY}
The energy necessary for excitation is different, and in general larger, than the energy emitted in the form of a photoluminescence photon. This so-called Stokes shift is due to non-radiative relaxation of the exciton down to the band edge. The Stokes shift in commercially available blue-green Si QDs is characteristically about 50~nm, which corresponds to about 0.3~eV. Such commercially available QDs have measured photoluminescence quantum yields (QY) of 10--60\%, with ensemble quantum yields of 30--70\% reported for ligand-passivated Si nanocrystals in Ref.~\cite{Sangghaleh_2015}). We adopt a benchmark value of $\text{QY}=50\%$.\footnote{Spec-sheet values for commercially available dots are taken, for example, from QDs supplied by CD Bioparticles or Applied Quantum Materials.} 

A signal photon emitted by a QD must exceed the $\sim$1.1~eV absorption threshold of the silicon CCDs used for readout (see \autoref{sec:detectorresponse}). Accounting for the $\sim$0.3~eV Stokes shift, this requires Si QDs with optical gaps above about 1.4~eV. For simplicity, we take $\text{QY}=50\%$ to be independent of the deposited energy, i.e.~a single photon is emitted with 50\% probability regardless of the energy the dark matter deposits. Here, we ignore the (much smaller) two-photon signal, which was discussed in~\cite{Blanco:2022cel}.

\subsection{Reciprocal Lattice Vector Basis Set Scaling}
For large unit cells, necessary to contain QDs with many atoms, the size of the reciprocal lattice vector basis set proves to be the limiting factor. The large-frequency limit of this basis set is given by the choice of the electron wavefunction energy cutoff $E_{\mathrm{cut}}$ as
\begin{equation}
    \frac{\left|\mathbf{G}\right|^2}{2m_e}\leq E_\mathrm{cut}\,,
\end{equation}
within our limiting case of $|\mathbf{k}|=0$. This parameter regulates the large plane-wave frequencies that otherwise have no physically motivated constraint. Since this cutoff is a numerical convergence parameter, we check the convergence of the calculated observables of interest and set it to 29\,Ry (corresponding to $q_\text{max}\sim20$\,keV).
On the other hand, the low-frequency limit of the basis set is given by the size of the simulation box, as the longest-wavelength plane waves must be commensurate with the cell (for our simulation volume of a 30\,\AA{} side, this corresponds to $q_\text{min}\sim0.4$\,keV). 

Since the value of $E_{\mathrm{cut}}$ is fixed by the large-momentum part of the electronic wavefunction still contributing to the scattering, the number of basis vectors per spatial direction grows linearly with the real-space simulation box size $a$, and the full basis set grows as $a^3$. To calculate the form factors, one loops over the basis set twice: once over the momentum transfers $\Delta\mathbf{G}$ and once over the vectors $\mathbf{G}$ entering the sum in~\eqref{eq:f_prime}. At fixed initial and final states, the calculation time $t_{\text{calc}}$ therefore scales as
\begin{equation}
    t_{\text{calc}} \sim a^6\ .
\end{equation}
The need for an increase in simulation-box size is driven by increasing the number of atoms within the studied QD. A larger number of atoms linearly increases the number of electronic states in the form factor. This number grows with the volume as $\sim a^3$ for both the initial and final state, leading to a combined scaling of \begin{equation}
    t_{\text{calc}} \sim a^{12}\ .
\end{equation}

This scaling is inherent to the task, and there is no straightforward way to bypass it. In addition, evaluating the coefficient $u_{i'}\left(\mathbf{G} + \Delta\mathbf{G}\right)$ in~\eqref{eq:f_prime} requires locating the vector $\mathbf{G} + \Delta\mathbf{G}$ (and its corresponding Bloch coefficient) in the basis set. Implemented as a search, this adds another loop over the entire basis set, bringing the total CPU-time scaling to
\begin{equation}
    t_{\text{calc}} \sim a^{15}\ .
\end{equation}
Therefore, a small increase in the simulation box size can make the simulation intractable. Since this extra loop only performs a lookup, it can be eliminated with a precalculated hash map, which reduces the lookup to constant time and restores the inherent scaling of $t_{\text{calc}} \sim a^{12}$. This speedup is negligible for small unit cells, but proves to be essential for larger systems, such as the QDs containing $\mathcal{O}(100)$ atoms with $a=30\,\si{\angstrom}$. Our modification to the QEdark-EFT code~\cite{urdshals_2023_7836577, Essig:2015cda} has been made publicly available~\cite{data}.

\subsection{Screening}\label{sec:screening}
Since the single-electron excitations described above lack collective screening effects, an a posteriori correction is used. Following the work reported in Refs.~\cite{Trickle:2019nya, Dreyer:2023ovn, PhysRevB.47.9892}, we add a correction to the material responses as $W_{\text{screened}}(q, \Delta E) = W_{}(q, \Delta E)/|\epsilon|^2$ with $\epsilon(q, \Delta E)$ given by

\begin{align}\label{eq:screening}
    \epsilon(q, \Delta E) & =  1 + \\ \nonumber &\bigg[ \frac{1}{\epsilon_0 - 1} + \tau \bigg(\frac{q}{q_{TF}}\bigg)^2 + \frac{q^4}{4m_e^2 \omega_p^2} - \bigg( \frac{\Delta E}{\omega_p} \bigg)^2 \bigg]^{-1}.
\end{align}
The values for the parameters of this dielectric function are listed in~\autoref{tab:screening}. As reported in Ref.~\cite{Catena:2024rym}, this prescription applies to the $\mathcal{O}_1$ operator, which couples to the electron number density, but not necessarily to the other operators listed in~\autoref{tab:operators}: their in-medium corrections involve other generalized susceptibilities and do not reduce to a simple $1/|\epsilon|^2$ suppression. A non-trivial expression involving the dielectric function is nevertheless expected for each operator.

\begin{table}[t!]
    \caption{Parameters used in the dielectric function calculation in Eq.~\eqref{eq:screening} for bulk silicon as modeled in~\cite{Dreyer:2023ovn, PhysRevB.47.9892}.}
    \centering
    \begin{tabular}{c@{\hspace{0.4cm}}c@{\hspace{0.4cm}}c@{\hspace{0.4cm}}c}\hline\hline
        $\epsilon_0$&$\tau$&$\omega_p$ [eV]&$q_\mathrm{TF}$[keV]\\\hline
        11.3 & 1.563 & 16.6 & 4.13 \\
        \hline
    \end{tabular}
    \label{tab:screening}
\end{table}

\begin{figure*}
\centering
  \centering
  \includegraphics[width=0.48\textwidth]{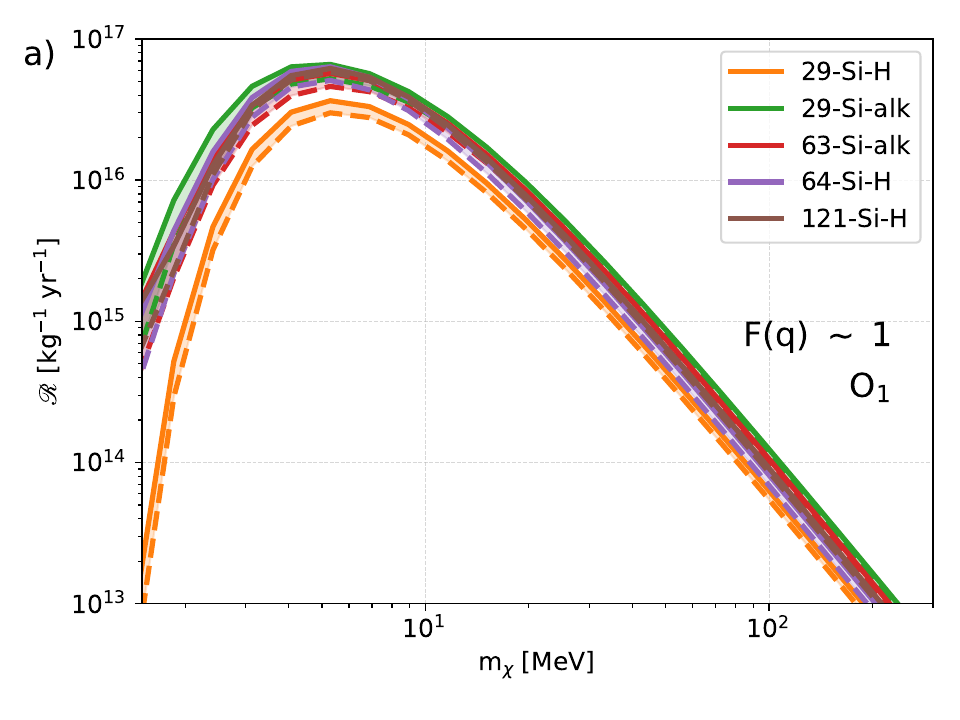}
  \includegraphics[width=0.48\textwidth]{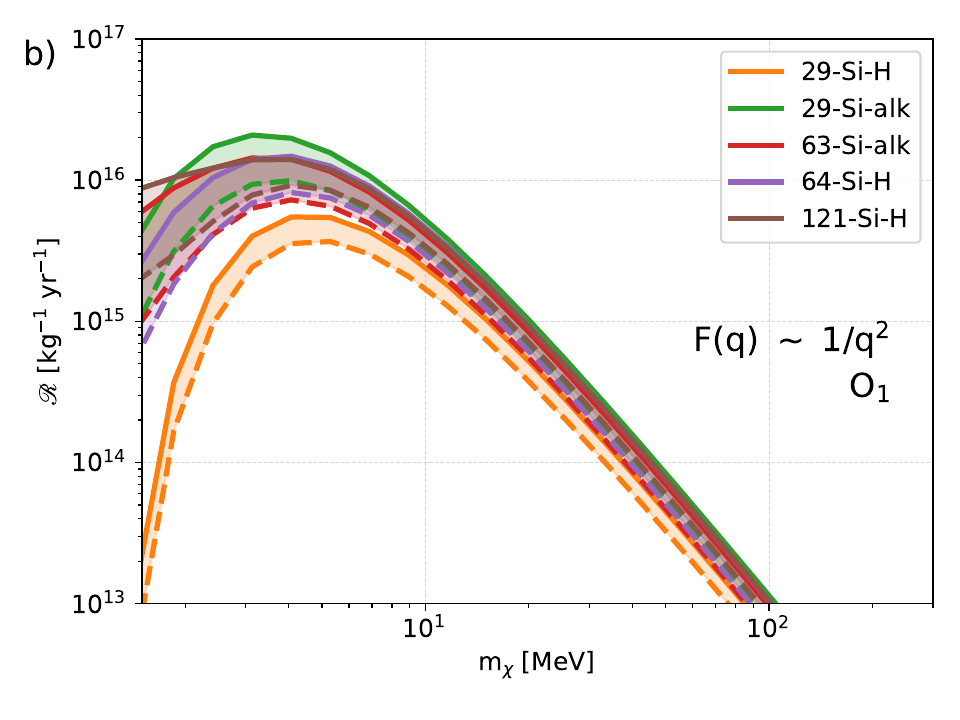}
\caption{The effect of screening on the interaction rate for the $\mathcal{O}_1$ operator in the heavy (panel \textbf{a}) and light (panel \textbf{b}) mediator limits, with $c_1^s=1$ ($c_1^\ell=1$), for the five considered quantum dots. Solid lines show the unscreened rates; dashed lines show the rates with maximal screening, corresponding to that of bulk silicon. Because a finite crystal cannot support collective screening on length scales beyond its own size, the screening in QDs is weaker than in the bulk, and the actual rates are expected to lie within the shaded bands. In-medium effects become negligible for DM masses above $m_\chi\sim10$\,MeV. For heavy mediators, screening plays a sub-dominant role as seen in the vanishing widths of the ranges for $F(q)\sim 1$.}
\label{fig:screening}
\end{figure*}

Furthermore, the dielectric-function screening we adopt was derived in the context of bulk crystals. We expect that for QDs, its effects will be suppressed due to their small size. To estimate the impact of the screening effects on the interactions mediated by the $\mathcal{O}_1$ operator, we present a range bounded by the limiting cases in~\autoref{fig:screening}. The solid lines show the unscreened scattering rate, the dashed lines show the rate with full bulk-silicon screening, and the shaded region denotes the range where the actual rate is expected to lie. 

While it is beyond the scope of this work to express the screening of the rate for all effective operators, one can see that its effect decreases and becomes insignificant for DM masses above $\sim$10\,MeV. For these heavier masses, the energy and momentum transfer are large enough for the collective effects to become subdominant, which covers most of our region of interest. Furthermore, we are focusing on energy depositions below 20\,eV, where the all-electron reconstruction is not required to capture the probed momentum-dependence of the wavefunction~\cite{Dreyer:2023ovn}. Therefore, the predictions presented herein are shown without the inclusion of in-medium effects.

If one wanted to include the screening effects in full and to focus on the region below $m_\chi \sim 10$\,MeV, a correction for the other effective operators would be necessary, as well as a rigorous calculation of the dielectric function from first principles, using codes such as DarkELF~\cite{Knapen:2021bwg} or QCDark~\cite{Dreyer:2023ovn,Dreyer:2026bmz}.

\section{Material Response}\label{sec:materialresponse}

Here we present the material responses calculated for the five considered quantum dots (listed in \autoref{tab:gap-values} and visualized in \autoref{fig:3D-QD-example} and \autoref{fig:3D-QDs}), along with bulk silicon for comparison. The material response scales with the number of atoms within the considered unit cell (\autoref{eq:W_scalar_2D_2}), and the normalization to the total mass of the detector is done later for the final rate (\autoref{eq:R_crystal_2D}). We therefore scale the presented responses by the number of silicon atoms contained within the QD to compare their performance directly. 

\autoref{fig:responses-QDs-W3} shows the responses for $W_3$, while responses $W_1$ and $W_4$ are shown in Appendix~\ref{app:responses}. The dashed black line bounds the kinematically accessible region, which satisfies
\begin{align}
v_{\rm min}< v_{\rm esc}\ ;
\label{eq:vmin_limit}
\end{align}
the shaded region below it is inaccessible.
Here, for illustrative purposes, one can neglect the first term in \autoref{eq:vmin} since the typical momentum transfers are much lower than the considered DM masses. If all-electron effects were included in this formalism, the difference between the large- and low-momentum support of the responses would be less striking~\cite{Griffin:2021znd,Dreyer:2023ovn,Dreyer:2026bmz}. However, since we focus on energy depositions below 20\,eV, the impact on the expected rates is not significant, as the low-momentum region contributes substantially to the response.

We can see that the material responses for the various QDs differ in a number of aspects. First, the change in the band gap produces a rigid shift of the responses toward larger energy transfers, with the hydrogen-terminated QD (with 29 silicon atoms) having the largest gap. This degrades the sensitivity, as less of the electron phase space is in the kinematically allowed region. 

Second, we can see that the composition of the surface passivation has a significant effect on the energy--momentum distribution of the response. While the hydrogen-terminated QDs are more similar to that of the bulk, the alkyl-terminated ones show a ridge at about 15\,eV of deposited energy for all electronic momenta. This is because carbon atoms contribute fully occupied $2s$ and $2p$ bands. These hybridize with the silicon surface into delocalized Si--C surface states that have a large momentum dispersion and energy below the silicon valence band. This opens up new possibilities for transitions at the order of $\Delta E \sim$10--20\,eV. On the other hand, localized $1s$ states of hydrogen saturate the dangling bonds of silicon while keeping the energy--momentum electronic structure largely intact. 

Furthermore, at fixed surface passivation, changing the number of silicon atoms in the core not only alters the band gap but also slightly modifies the response distribution. This is not easily visible from the logarithmic plot of the responses, but it manifests in the interaction rates presented below.

\begin{figure*}
\centering
  \centering
  \includegraphics[width=0.48\textwidth]{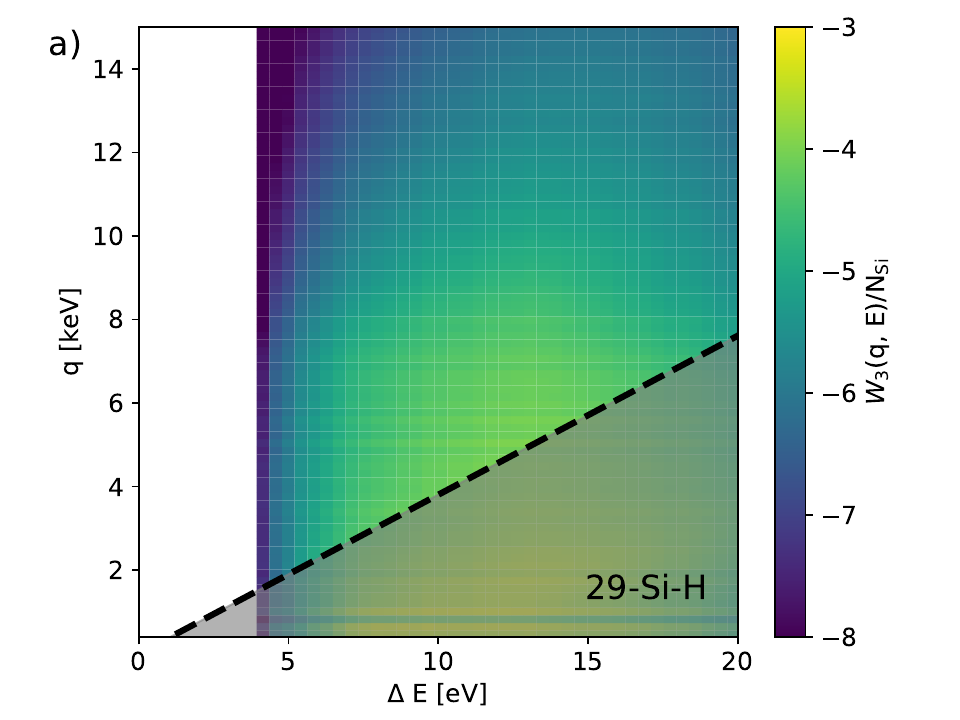}
  \includegraphics[width=0.48\textwidth]{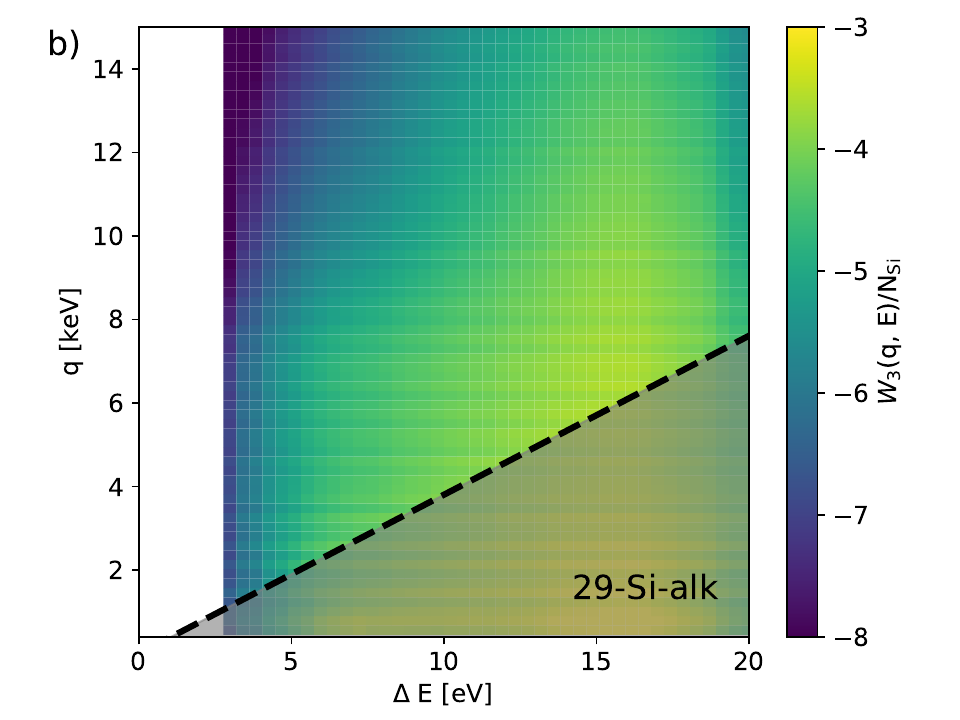}
  \includegraphics[width=0.48\textwidth]{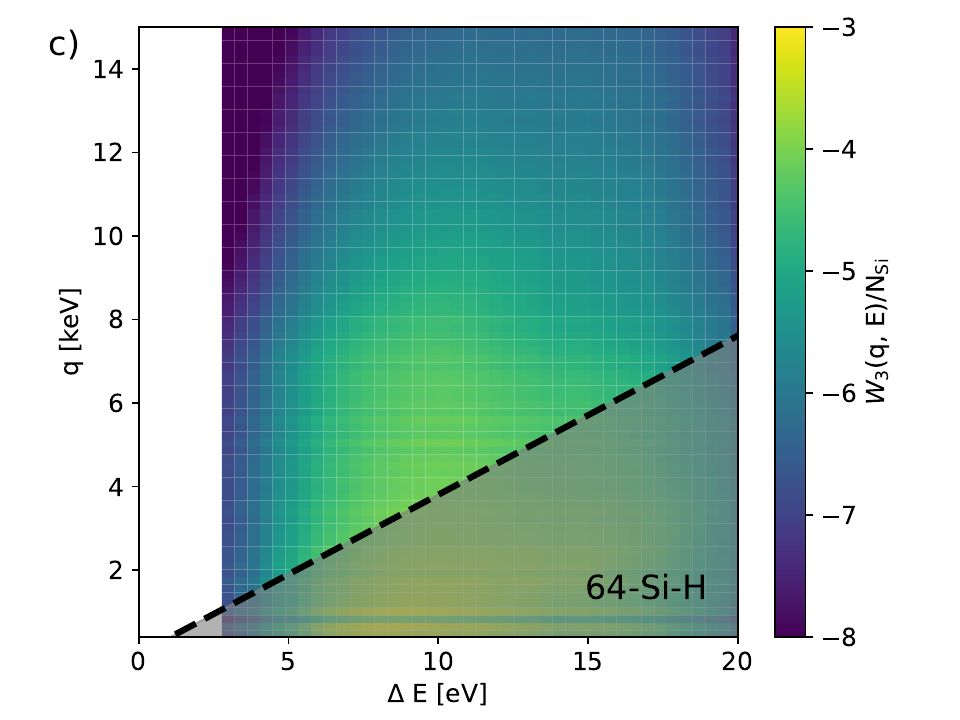}
  \includegraphics[width=0.48\textwidth]{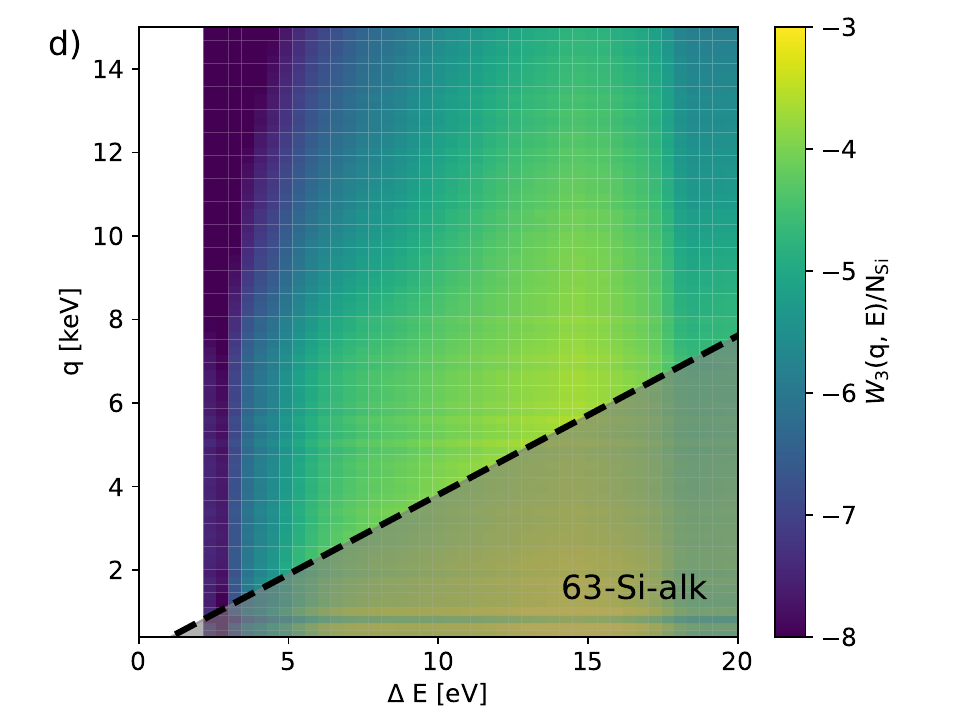}
  \includegraphics[width=0.48\textwidth]{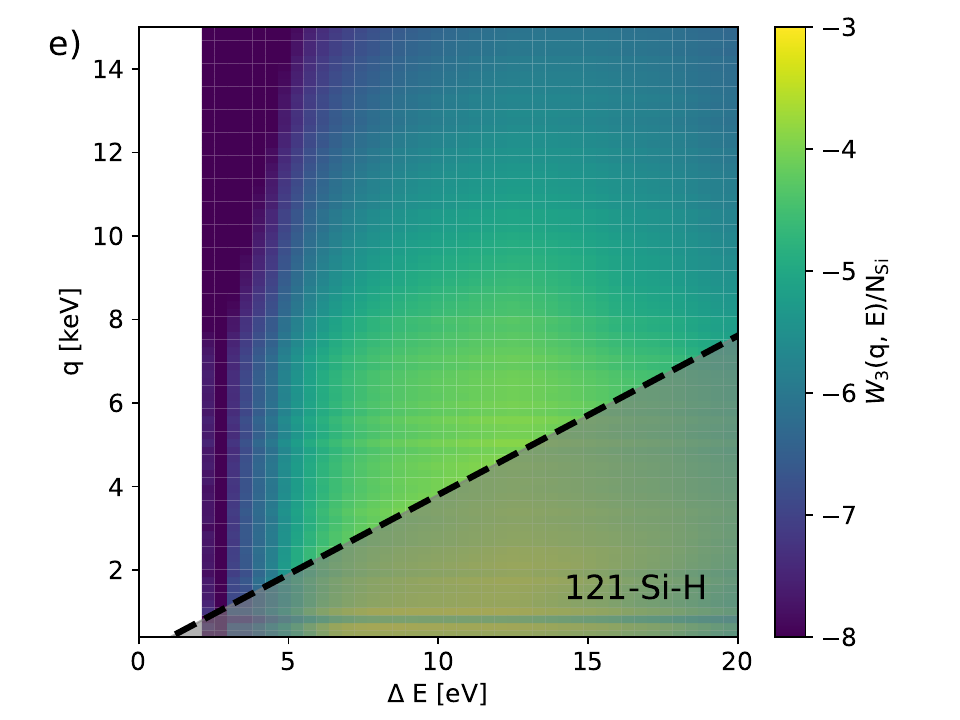}
  \includegraphics[width=0.48\textwidth]{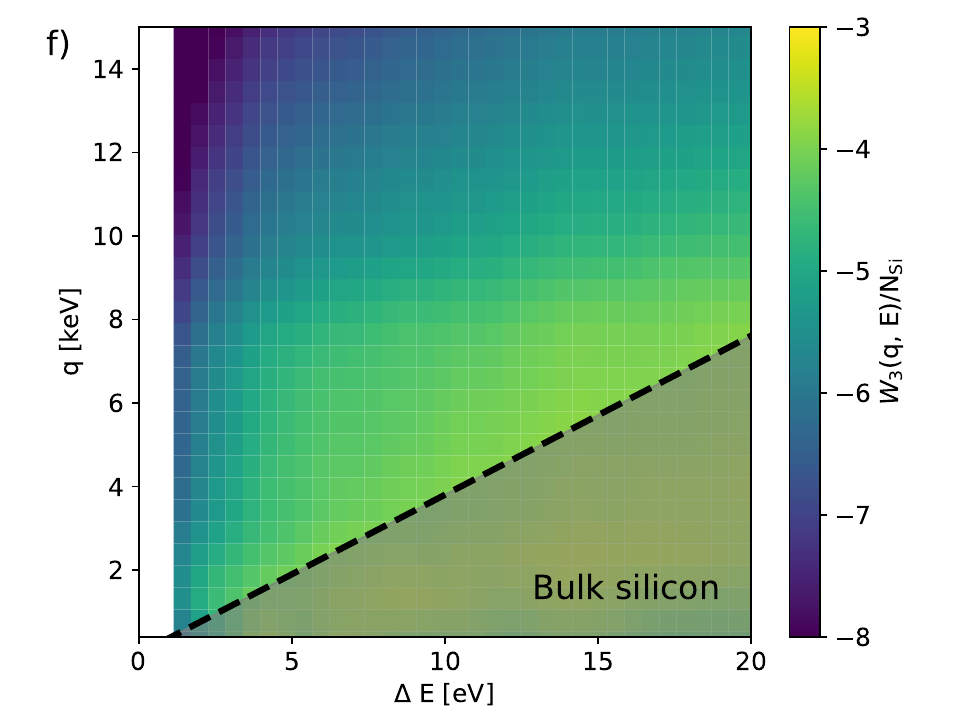}
\caption{Material response function $W_3$ normalized by the number of silicon atoms in a unit cell for all considered QDs, as well as for the bulk Si crystal. The shaded region below the dashed black line shows the kinematically inaccessible region. In the shaded region, the minimal DM velocity required to induce an excitation exceeds the galactic escape velocity in the large-mass limit.}
\label{fig:responses-QDs-W3}
\end{figure*}

\section{Separating the Band Gap and Form Factor Effects}\label{sec:gapandoverlaps}

As mentioned in the previous section, the effect of the QD composition on the material response is twofold: \textit{i}) the size of the QD changes its band gap due to quantum confinement effects; and \textit{ii}) the altered wavefunction of the QD and its surface changes the form factors calculated in \autoref{eq:f} and \autoref{eq:fvec}.
In this section, we split the two contributions and examine how the form factor and QD band gap contribute to the final event rate and sensitivity.

A larger band gap of the QD induces a smaller material response for the DM interaction. The kinematically allowed phase space shrinks as we move the response in $\Delta E$ by shifting the valence states to higher energies. This increases the energy of all valence-conduction transitions as
\begin{equation}\label{eq:shift}
\Delta E = \left(E_{i'} - E_i\right) + \Delta E_{\text{gap}}\,,
\end{equation}
where $\Delta E_{\text{gap}}$ is the increase of the gap relative to its bulk value.
The shift of the material response to larger energy transfers and its effect on the allowed phase space can be equivalently expressed by shifting the minimal velocity necessary for the interaction (see \autoref{eq:vmin}).

\begin{figure}[t!]
\centering
  \includegraphics[width=0.48\textwidth]{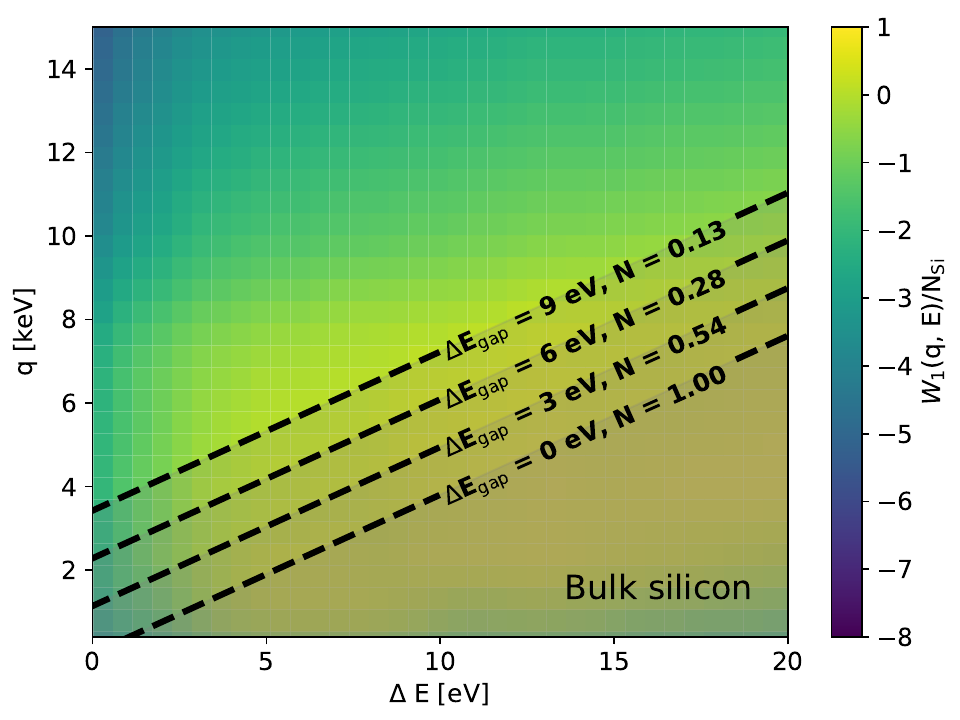}
\caption{The effect of a constant band gap shift on the kinematically-allowed phase space of the $W_1$ response of bulk silicon. The various black dashed lines denote different shifts of the band gap of the bulk silicon crystal. Below these lines, the minimal DM velocity required to induce an excitation exceeds the galactic escape velocity. The integral of the response over the allowed phase space, normalized to that for the original band gap, is denoted $N$.}
\label{fig:attenuation}
\end{figure}

This effect is shown in \autoref{fig:attenuation}, where we plot the $W_1$ response of bulk silicon with multiple dashed black lines denoting the inaccessible part of the energy--momentum phase space. We can see that the increase in band gap removes the part of the material response that is the dominant contribution to its overall integral (note the logarithmic scale on the color bar).  

In \autoref{fig:attenuation}, we also show the integrated material response over all allowed energies and momenta for four values of band gap shift $\Delta E_{\text{gap}}$ normalized to that of the original value for bulk silicon. We can see that an increase in band gap of 3\,eV results in a 50\% suppression, while increasing the band gap by 9\,eV suppresses the integral roughly by a factor of 8. Furthermore, the total rate is strongly suppressed through the decrease in the number of DM particles available to cause an excitation. If the band gap is too large, DM particles have to have a large velocity to overcome it. The population of fast-moving DM particles described by the Maxwell-Boltzmann distribution (\autoref{eq:maxwell-boltzmann}) drops exponentially, significantly suppressing the final rate even for small band gap increases.

\begin{figure}
\centering
  \centering
  \includegraphics[width=0.48\textwidth]{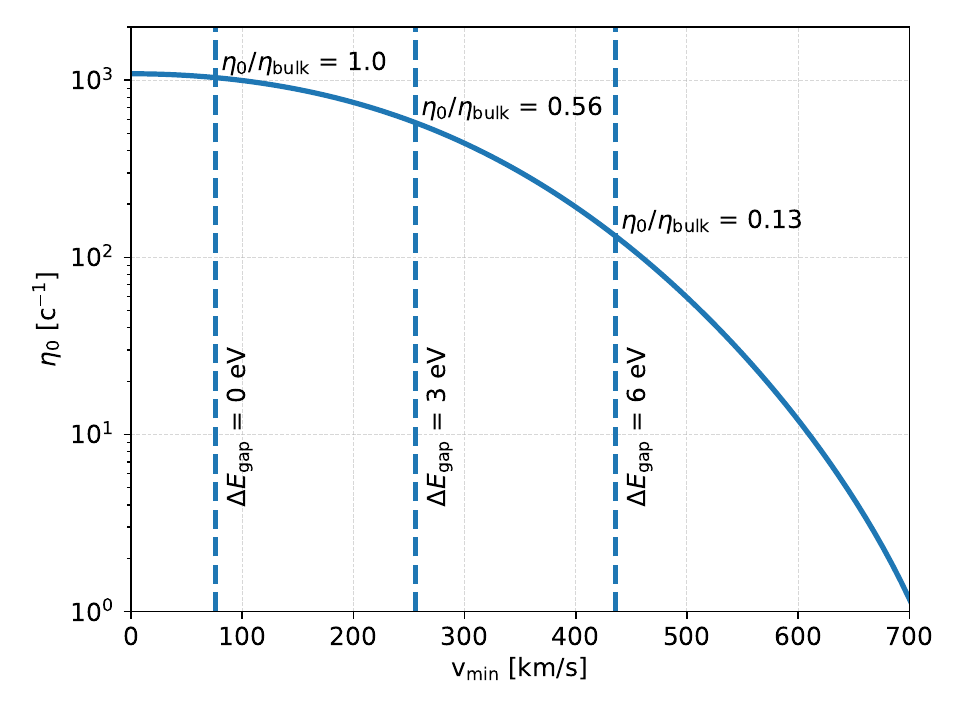}
\caption{The inverse mean speed of DM particles plotted for several values of the minimal required DM velocity $v_\text{min}$. An increase in the band gap of the material $\Delta E_\text{gap}$ results in a strong suppression with respect to that of the bulk band gap. Several values of $\Delta E_\text{gap}$ are tested and the relative suppression $\eta_0 / \eta_\text{bulk}$ is shown for a typical momentum transfer of $q=5$\,keV and an asymptotically large DM particle mass.}
\label{fig:eta}
\end{figure}

In \autoref{fig:eta}, we plot the inverse mean speed as a function of the band gap of the system while assuming a typical momentum transfer of $q=5$\,keV (where the bulk silicon material response is the strongest). We focus on the asymptotic large masses, where we can neglect the $\frac{q}{2m_\chi}$ term in \autoref{eq:vmin}. We see that an increase in the band gap of 3\,eV reduces the rate for this dominant momentum transfer by an additional factor of 2, and an increase of 6\,eV results in a suppression of a factor of 8.

\begin{figure*}
\centering
  \centering
  \includegraphics[width=0.48\textwidth]{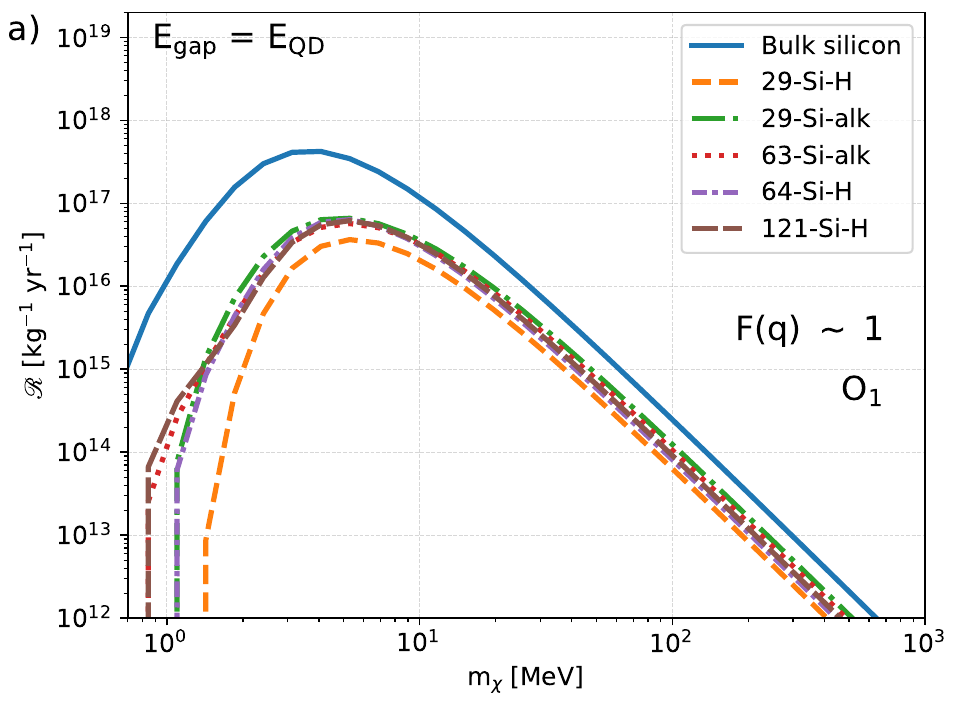}
  \includegraphics[width=0.48\textwidth]{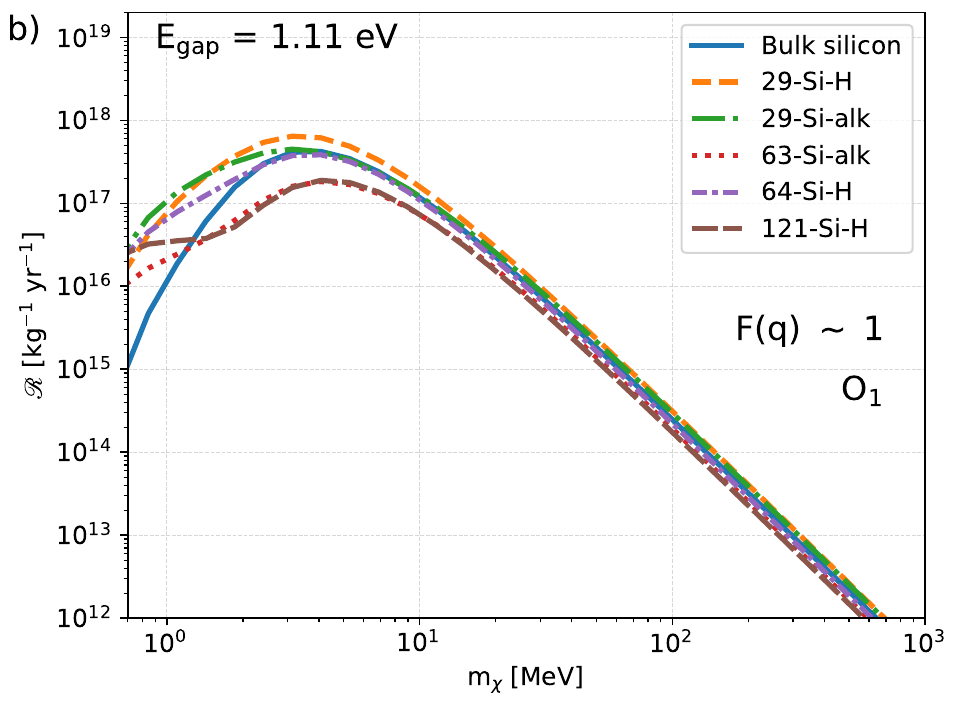}
\caption{Scattering rates for the $\mathcal{O}_1$ operator in the heavy mediator limit with the QD band gaps obtained from our fit (panel \textbf{a}) and for all QD band gaps set artificially (for illustrative purposes) to that of bulk silicon (panel \textbf{b}) with the coupling constant $c_1^s=1$.}
\label{fig:rates-QDs-O1short}
\end{figure*}

To estimate the effects on detector sensitivity, one has to calculate the final rate. The rate convolves the material response with the DM velocity distribution and the DM response functions, rather than sampling them at fixed points. In \autoref{fig:rates-QDs-O1short}, we show the calculated interaction rates for the $\mathcal{O}_1$ operator in the heavy mediator limit, where we have set the coupling constant to $c_1 = 1$. The panel \textbf{a} shows the rate calculated with the value of the band gap set to the one obtained from our fit (\autoref{tab:gap-values}). In panel \textbf{b}, we show a rate calculated with a band gap set to that of the bulk for all QDs. 

As expected, we see that the decrease of the band gap in panel \textbf{b} increases the rates and makes them comparable to the case of solid silicon for all QDs. There is a slight variation in QD performance in the mid-mass region, originating from the altered energy--momentum phase space of the electron form factor. This results in the 29-Si-H QDs outperforming the bulk, while the larger QDs tend to underperform with respect to bulk Si. The larger natural QD band gap degrades sensitivity in the low-$m_\chi$ limit, where light DM particles do not have enough energy to exceed the excitation threshold.

\begin{figure*}
\centering
  \centering
  \includegraphics[width=0.48\textwidth]{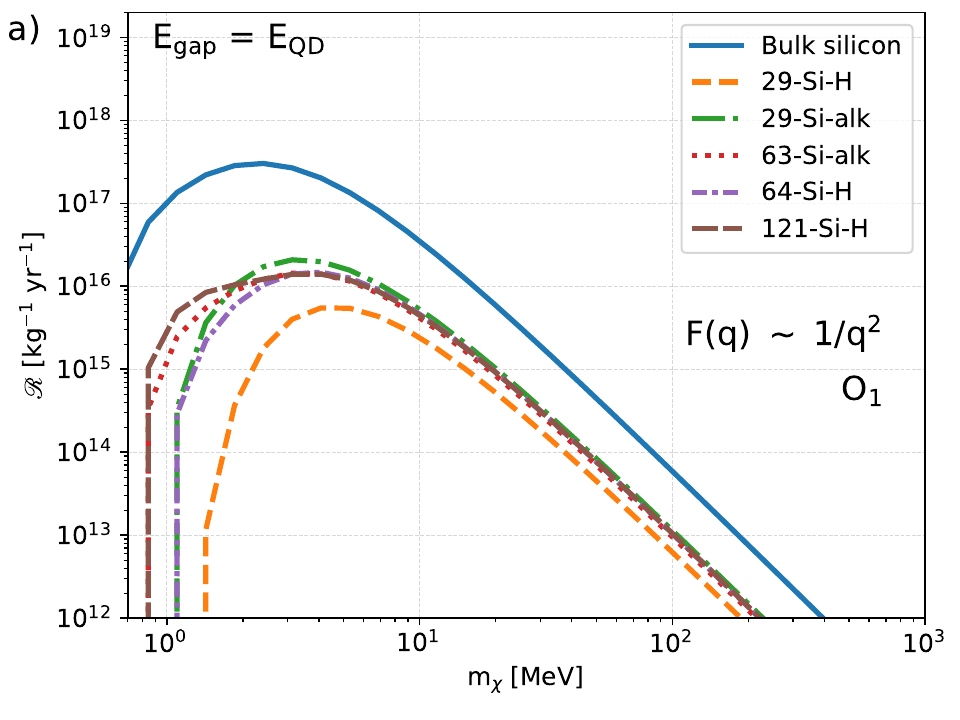}
  \includegraphics[width=0.48\textwidth]{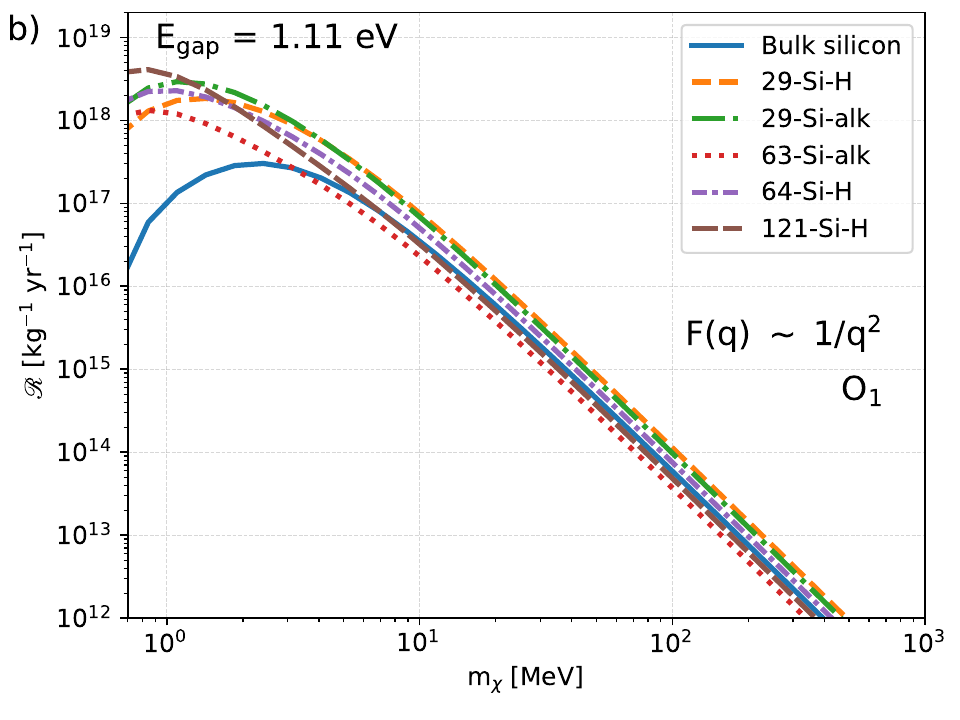}
\caption{Scattering rates for the $\mathcal{O}_1$ operator in the light mediator limit with the QD band gaps obtained from our fit (panel \textbf{a}) and for all QD band gaps set artificially (for illustrative purposes) to that of bulk silicon (panel \textbf{b}) with the coupling constant $c_1^\ell=1$.}
\label{fig:rates-QDs-O1long}
\end{figure*}

The situation changes significantly for the case of a light mediator (see \autoref{fig:rates-QDs-O1long}), which suppresses the rate for large momentum transfers as $\sim 1/q^2$. The rate is strongly affected by both the form factor and the value of the band gap. QDs with band gaps artificially set to the bulk band gap outperform bulk silicon at low masses and cluster around the bulk rate at large masses. QDs with properly modeled band gaps show a different behavior and present a non-trivial rate ordering even in the large-$m_\chi$ limit. 

\begin{figure*}
\centering
  \centering
  \includegraphics[width=0.48\textwidth]{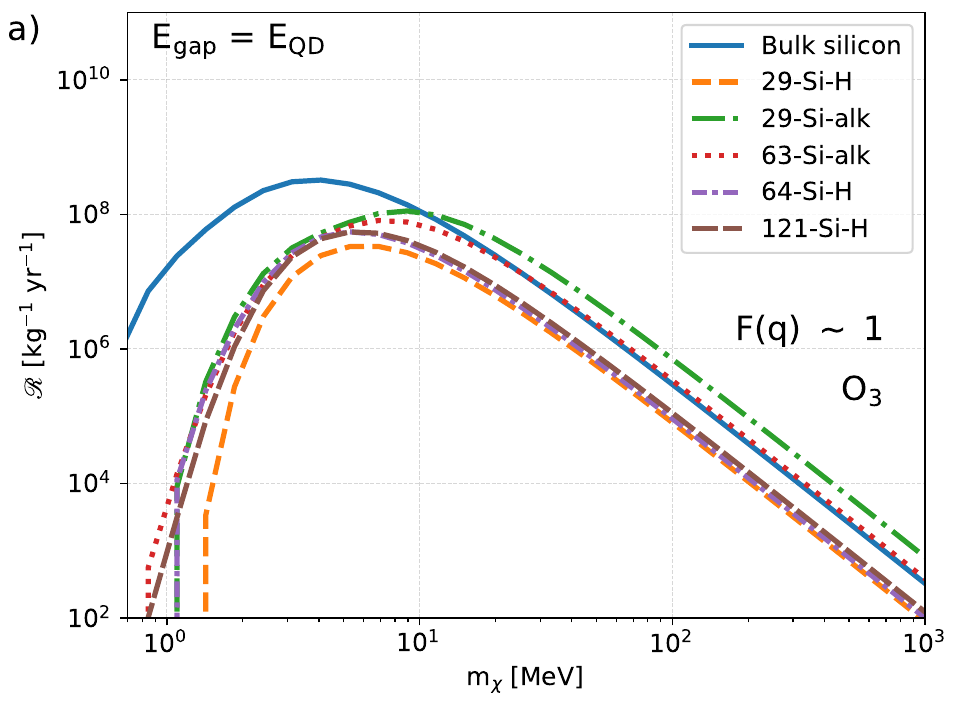}
  \includegraphics[width=0.48\textwidth]{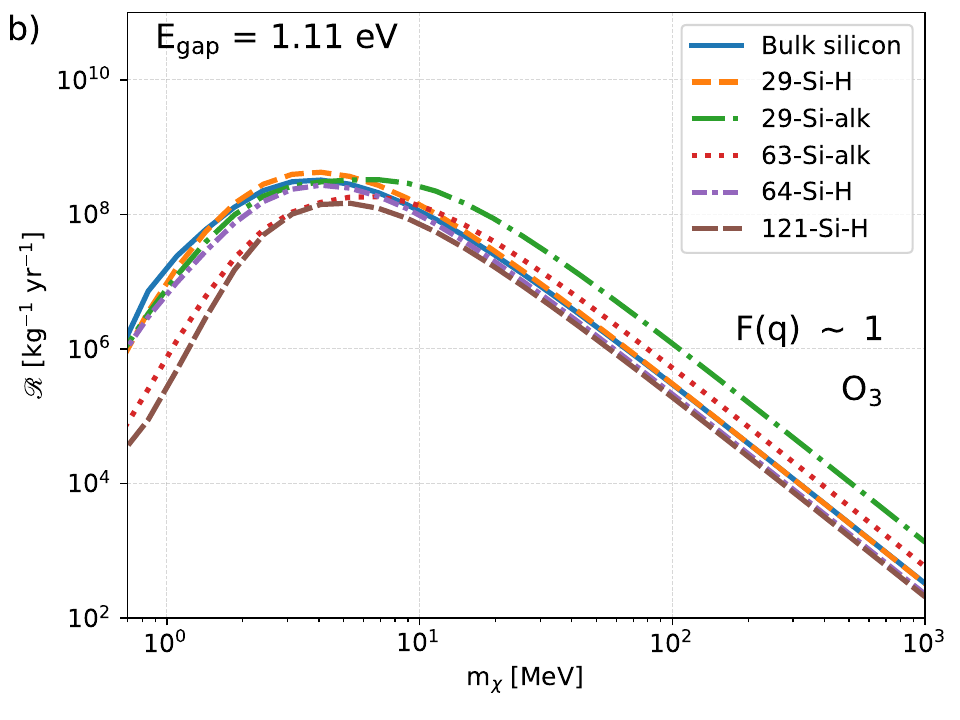}
\caption{Scattering rates for the $\mathcal{O}_3$ operator in the heavy mediator limit with the QD band gaps obtained from our fit (panel \textbf{a}) and for all QD band gaps set artificially (for illustrative purposes) to that of bulk silicon (panel \textbf{b}) with the coupling constant $c_3^s=1$.}
\label{fig:rates-QDs-O3short}
\end{figure*}

For the operator $\mathcal{O}_3$, which contains the cross product between the momentum transfer and the transverse DM velocity ($\mathcal{O}_3 \sim q \times \mathbf{v}^{\perp}_{\rm el}$), the rate ordering both in the low-mass and large-mass regions shows a different pattern (see \autoref{fig:rates-QDs-O3short}). The energy--momentum dependence of the material response for the alkyl-terminated QDs favors the interaction so much that, even with the larger band gap, their performance exceeds that of the bulk at large DM masses, while falling short for lighter DM masses. 

In this study, we observe that both the band gap and the form factor contribute significantly to the final scattering rate and that their combination is imprinted as a unique variation of rates for a given DM model. This variation, dependent on the interaction operator and DM mass scale, motivates us to search for the optimal QD, or combinations thereof, for DM direct detection searches, which we discuss in the next section.

\section{The Salient Effects of QD Surface Passivation and Size}\label{sec:optimalsetup}

To identify the optimal QD setup for general DM detection, we choose to integrate the expected rates over all mass scales of interest---from the lowest energy deposits accessible up to large dark matter masses, where the interaction phase space is already strongly constrained ($m_\chi \sim 1\,\mathrm{GeV}$)~\cite{XENON10:2011prx, XENON:2019gfn, XENON:2023cxc}. This then serves as a proxy for the QD performance and for the probability that an observable signal will be detected with this setup. 

\begin{figure*}
\centering
  \centering
  \includegraphics[width=0.48\textwidth]{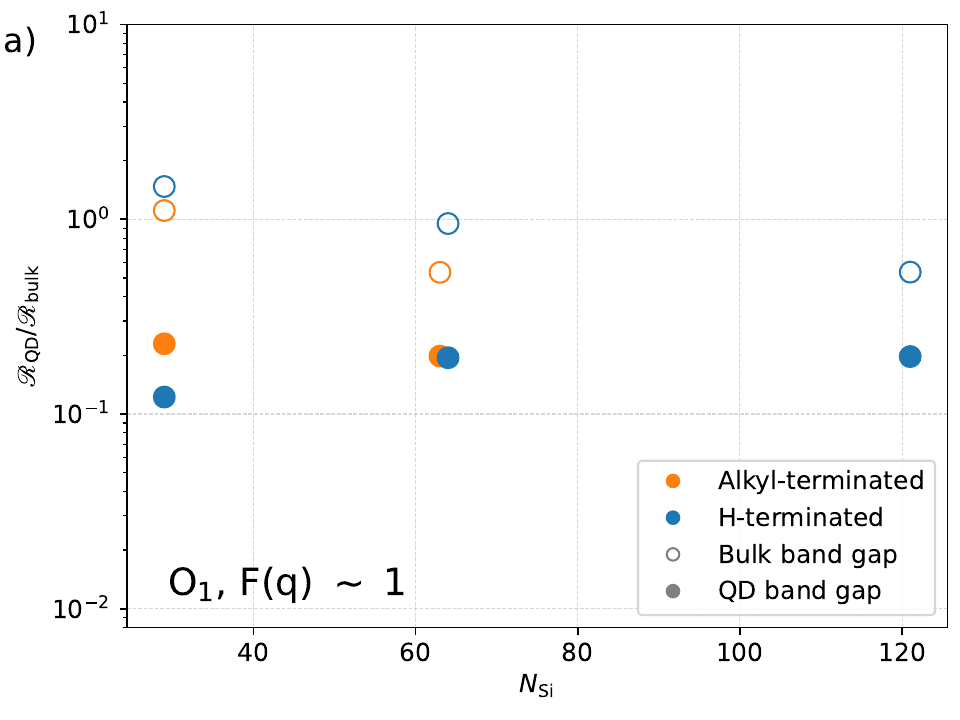}
  \includegraphics[width=0.48\textwidth]{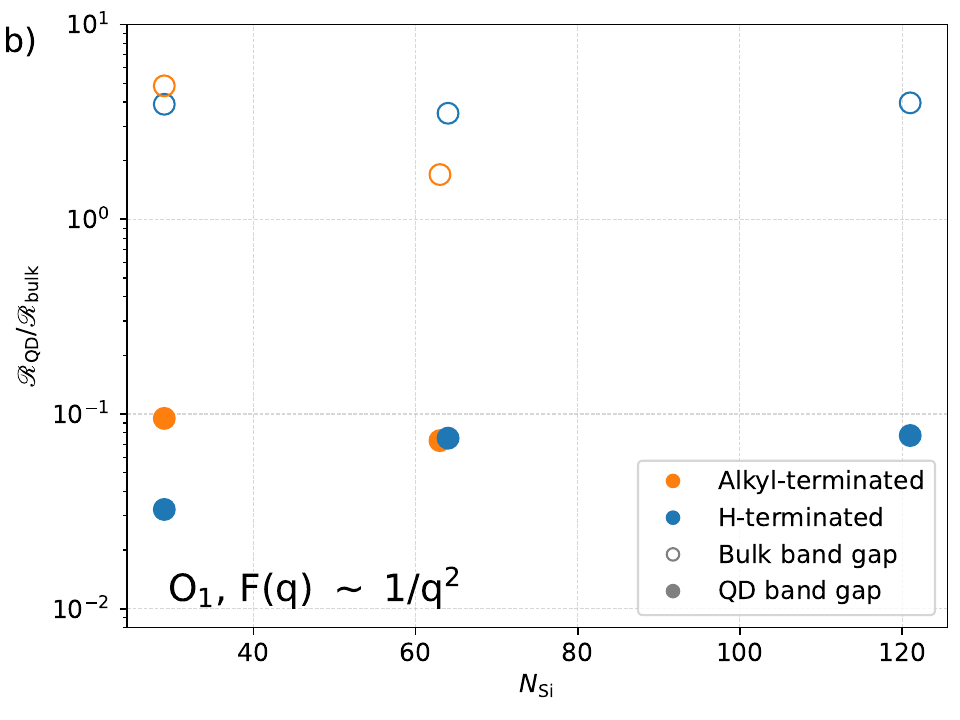}
  \includegraphics[width=0.48\textwidth]{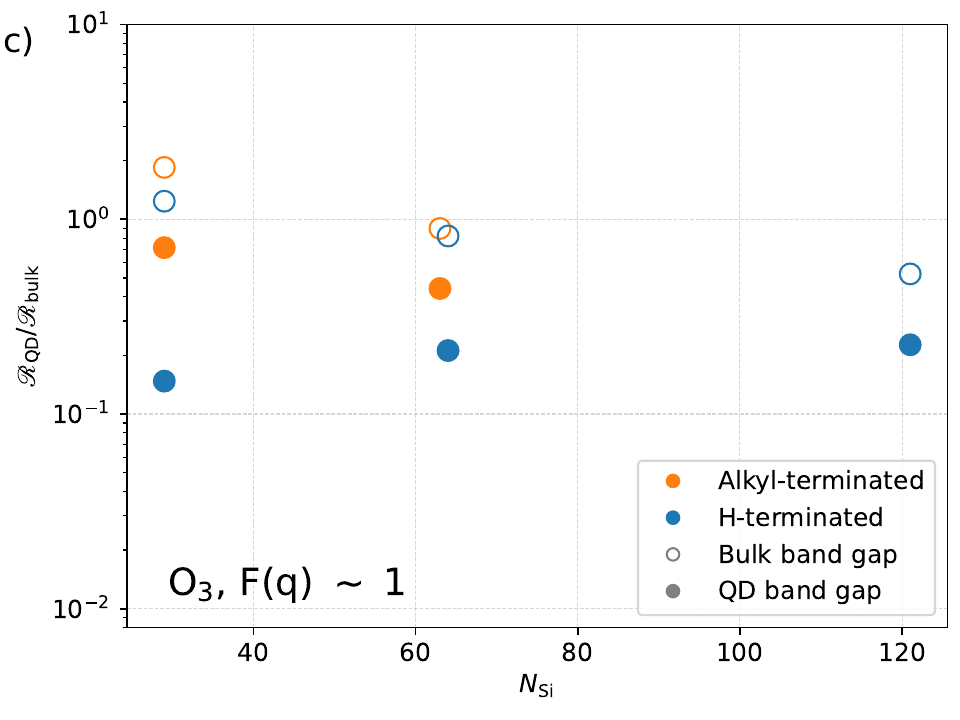}
  \includegraphics[width=0.48\textwidth]{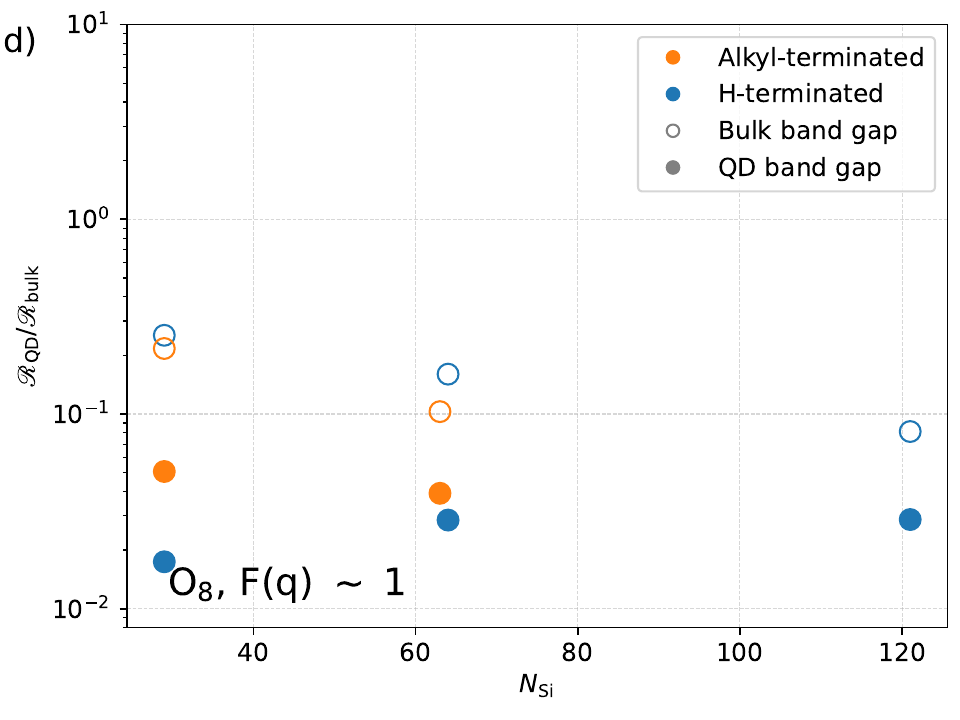}
\caption{An integral of the interaction rates for various QDs and interaction operators over the allowed DM phase space up to the scale where the interaction phase space is already strongly constrained ($m_\chi \sim 1$\,GeV)~\cite{XENON10:2011prx, XENON:2019gfn, XENON:2023cxc}. The interaction rates are divided by the rate obtained for bulk silicon, and their dependence on the number of silicon atoms in the core ($N_{\text{Si}}$) is shown. To disentangle the effects of the band gap variation and form factor, we show interaction rates for QDs with their band gap artificially (for illustration purposes) set to that of the bulk (open circles) as well as to the actual band gap (filled circles).}
\label{fig:size-dependence-QDs}
\end{figure*}

We have normalized the expected event rate to that of bulk silicon and show its dependence both on the size of the QD and on the surface passivation in \autoref{fig:size-dependence-QDs}. To be able to split up the contribution to the rate ordering coming from the band gap alteration and from the energy--momentum dependence of the electronic states, we keep the QDs with an artificially-set bulk band gap in the figure for comparison.

The general trend is that the alkyl-terminated QDs tend to have larger rates when compared to their hydrogen-terminated counterparts. However, the difference is not as significant for the 63/64-silicon QDs and for the $\mathcal{O}_1$ operator in the heavy mediator limit. Due to its small band gap, bulk silicon generally shows the most sensitivity, except for the case of the $\mathcal{O}_3$ operator. 

The size-dependent trends in the rate appear to be opposite in $n$-Si-H when compared to $n$-Si-alk. If the band gap is set to be the same for all QDs, the general trend is that the form factor for larger QDs suppresses the rate. This effect is partially balanced by the decrease in the band gap. This opposing effect can overcome the contribution of the form factor and, for example, the rates in $n$-Si-H QDs increase with increasing size. The opposite holds for $n$-Si-alk QDs. While their band gap decreases with an increase in size, it is not enough to counter the suppression coming from a different momentum-space distribution. This results in a decrease in the overall rate compared to bulk silicon for larger QDs. From the direct comparison of the overall sensitivities, one can see that the best QD candidates tend to be the small, alkyl-passivated ones. However, as shown in \autoref{fig:size-dependence-QDs}, the various kinds of QDs display a specific rate pattern that depends on the interaction type and mediator mass. We use this fact to develop a strategy for dark matter interaction identification in the next section.

\section{QD Detectors as a Barcode for DM Interaction}\label{sec:barcore}

The size and surface passivation of the QDs provide two tunable variables that alter the expected rate differently for different interaction types. This effect comes from changes in the electronic wavefunctions as well as the optical gap. The momentum-space structure of the electronic states is affected by the surface chemistry as well as the size of the QD. Meanwhile, the optical gap is affected, primarily through quantum confinement, by the size of the QD, which alters the accessible electronic states. 

This dependence on size and composition manifests itself differently for the various interaction operators and mediator types, which are sensitive to the exact momentum dependence of the material response. We can use this fact to build an independent set of QD detector units that, in case a DM signal is observed, will produce a pattern of excitation rates. The rate pattern could be used to help identify the type of DM interaction, its mediator, and potentially also the mass of the DM particle.

This contrasts with other solid-state detectors, where one would typically rely on the energy spectrum of the interaction to distinguish the DM signal from background and to extract information on its type. This is not possible in the case of QDs: excited states relax non-radiatively to the band edge, so the emitted photon energy retains little information about the deposited energy~\cite{Blanco:2022cel}.

\begin{figure*}[t!]
\centering
  \includegraphics[width=0.48\textwidth]{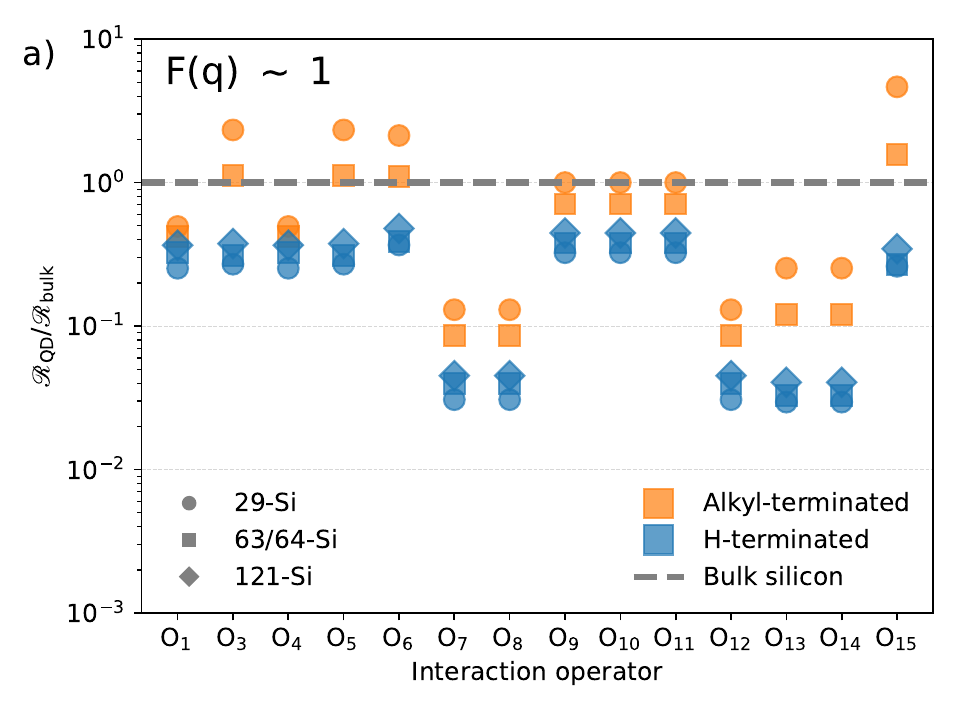}
  \includegraphics[width=0.48\textwidth]{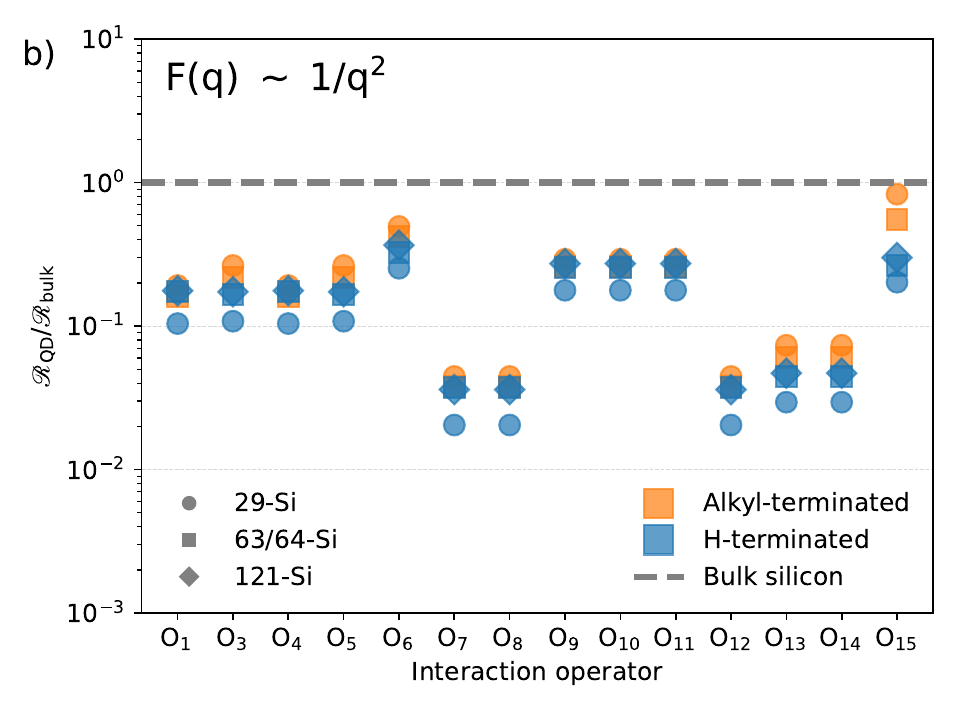}
\caption{Relative interaction rates in the large mass limit for all interaction operators in the heavy (panel \textbf{a}) and light (panel \textbf{b}) mediator limit. The rates expected for various QD setups are normalized to those of bulk silicon at the same exposure to obtain a rate ordering that can be used for identifying interaction operators and mediator types. This approach takes advantage of the tunability of the energy--momentum properties of QDs due to the admixture of the surface states as well as due to the quantum confinement effects. Several QD detectors operating in parallel could, therefore, serve as a dark matter identification platform.}
\label{fig:barcodes-QDs}
\end{figure*}

We find that a combination of five different silicon-QD detector units of two termination types allows us to differentiate several of the interaction operators (\autoref{fig:barcodes-QDs}). This differentiation can be further strengthened by considering additional core materials and terminations, which would contribute new electronic states at different energies and momenta, broadening the statistical power of the interaction barcode.

While the DM-SM interaction remains unknown, the operators presented in \autoref{tab:operators} can be thought of as a basis set in which any model can be expanded. Although the presented operators might not individually be responsible for the DM interaction, their combination can describe any non-relativistic DM model~\cite{Catena:2022fnk}. Once a DM-compatible signal is observed in a QD detector, one can, in principle, expand the target basis by selecting several different core materials and a number of surface passivations. Preparing each combination in three different sizes would lead to $\mathcal{O}(100)$ potential targets, each with a unique energy--momentum electron dependence and DM response. This would enable mapping the observed rate hierarchy onto the operator basis, yielding information on the nature of the DM interaction. To our knowledge, this property is unique to QDs: no other material class can be \emph{tuned} across so many variations while maintaining a similar energy threshold and an identical readout.  We note that while the separation of the interaction operators is theoretically possible, a significant challenge will be to read out the signal and control backgrounds, especially for the large exposures needed to identify a potential dark matter signal below current bounds. We discuss this further in \autoref{sec:detectorresponse}.

\section{Sensitivity Including Detector Response}\label{sec:detectorresponse}

In this section, we present the projected sensitivities
for a QD setup run as a counting experiment.
To enable a quantitative study, we assume a device composed of
\SI{10}{\kilo\gram} of active mass, exposed for 1 year.
The QD concentration is taken to be \SI{500}
{\gram\per\kilo\gram} and the photoluminescence quantum yield (QY)
has a benchmark value of 0.5 (see \autoref{sec:emissionQY}). Such high loadings are motivated by demonstrated QD-loaded scintillators with up to 40\% QDs by weight~\cite{yu2022liquid}. Furthermore, the Stokes shift between absorption and emission suppresses the self-absorption of scintillation light, allowing the volume to remain transparent to its own emission even at high concentrations~\cite{dohnalova2013surface}; a quantitative optical-transport study at this loading is left for future work. As shown in \autoref{fig:3ex-mockup}, the active volume is subdivided into ten \SI{1}{\kilo\gram} units, each filled with QDs of a different size and surface passivation, loaded in a solid matrix. In this mock-up, each of the five QD morphologies of \autoref{tab:gap-values} (three H-terminated, two alkyl-terminated) fills a pair of independent units. The benchmark thus corresponds to a total QD exposure of \SI{5}{\kilo\gram\year}, or \SI{1}{\kilo\gram\year} per QD morphology.

To read out the photon signal, we equip each of the ten units with a Skipper-CCD~\cite{Tiffenberg:2017aac} that detects the photons with a geometric efficiency of 1 (all other surfaces are taken to be perfectly reflective). The quantum efficiency,
on the other hand, is assumed to be 0.85~\cite{Villalpando:2023ate}, while the
1-electron and 2-electron dark rates are assumed to be
\SI{1.7e-2}{\per\CCD\per\second} and \SI{3e-5}{\per\CCD\per\second}, 
respectively. These values correspond to the rates measured by DAMIC-M~\cite{DAMIC-M:2025luv}; SENSEI has demonstrated lower rates~\cite{SENSEI:2024yyt}, e.g.~a 1-electron rate of \SI{1.16e-3}{\per\CCD\per\second}, but we use the larger number from DAMIC-M for concreteness. We assume that these rates, which are obtained with individual Skipper-CCDs that are packaged in copper modules, will be the same when the Skipper-CCDs are placed near the QD scintillator volumes. Other photodetectors may be available in the future with lower backgrounds, such as TESs, SNSPDs~\cite{Romani:2024rfh,DeLucia:2024sxp,10.1063/1.1388868,mazin2012superconducting}. 

\begin{figure*}
\centering
  \centering
  \includegraphics[width=0.48\textwidth]{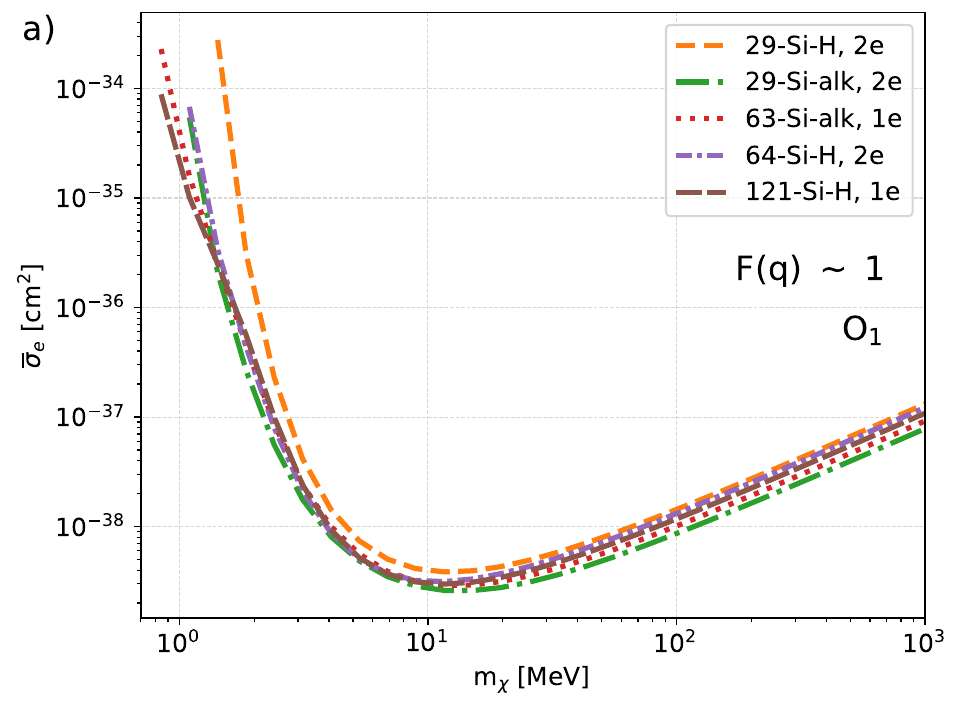}
  \includegraphics[width=0.48\textwidth]{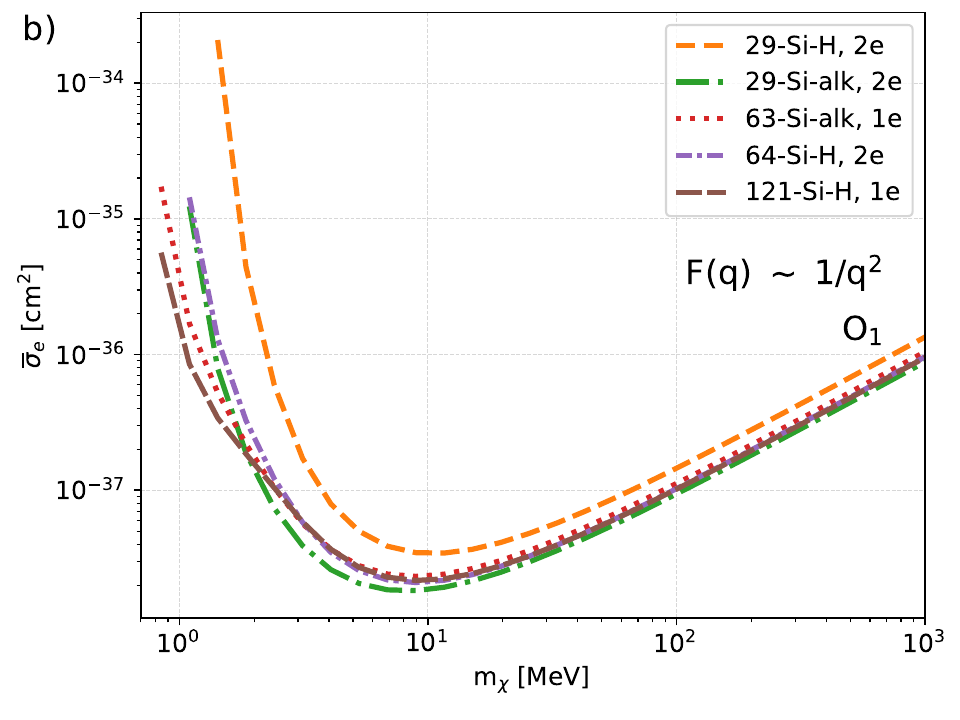}
  \includegraphics[width=0.48\textwidth]{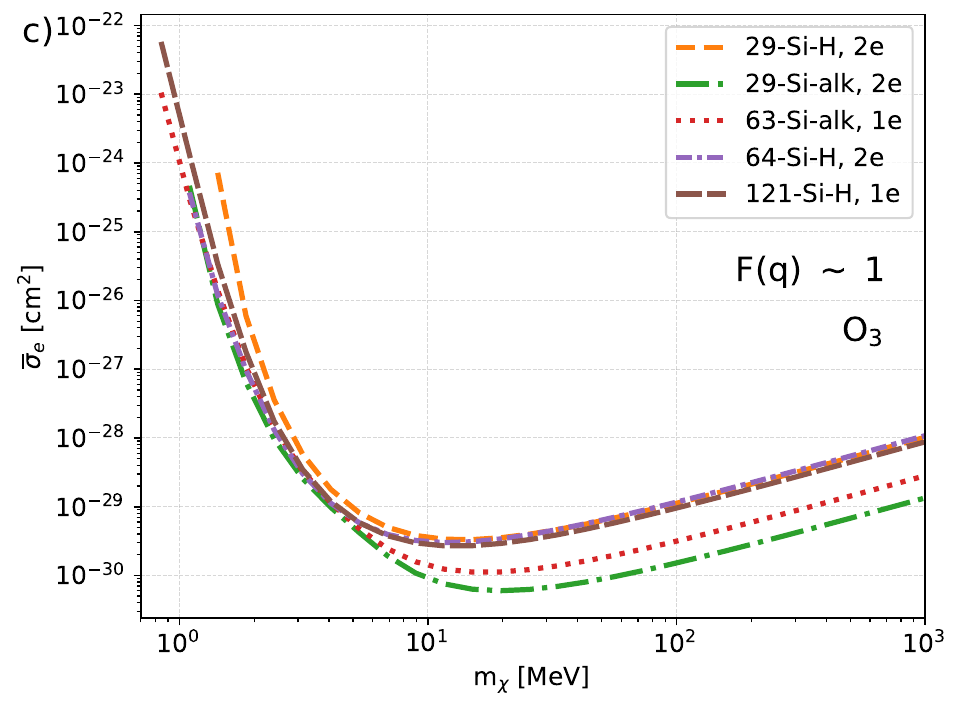}
  \includegraphics[width=0.48\textwidth]{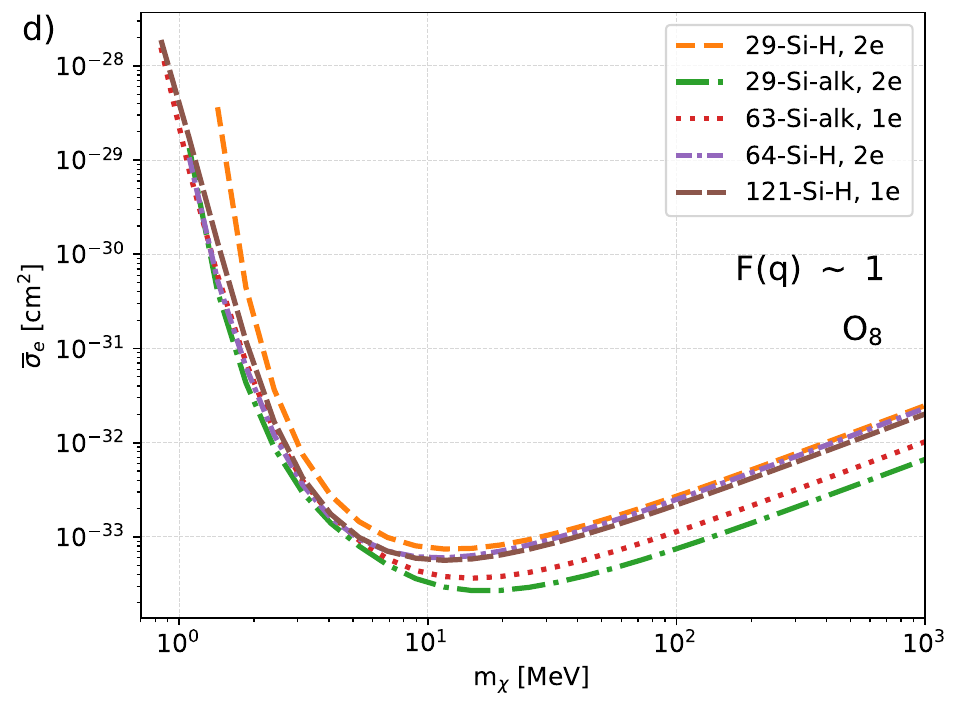}
\caption{90\% C.L.~median upper limits on the reference cross section $\bar{\sigma}_e$ for each QD morphology, for the $\mathcal{O}_1$ operator with a heavy (panel \textbf{a}) and light (panel \textbf{b}) mediator, and for the $\mathcal{O}_3$ (panel \textbf{c}) and $\mathcal{O}_8$ (panel \textbf{d}) operators with a heavy mediator. The exposure is \SI{1}{\kilo\gram\year} per morphology (\SI{5}{\kilo\gram\year} in total). We assume a photoluminescence quantum yield of 0.5, unit photon-collection efficiency, a CCD quantum efficiency of 0.85, and 1-electron (2-electron) dark rates of \SI{1.7e-2}{\per\CCD\per\second} (\SI{3e-5}{\per\CCD\per\second})~\cite{DAMIC-M:2025luv}. The legend indicates the readout channel (1e or 2e) used for each morphology. These curves are per-morphology benchmark sensitivities, intended to compare the effects of QD size and surface termination. They do not represent the combined reach of the full array. For this reason, we do not overlay existing constraints. See text for a comparison with current limits.}
\label{fig:limit-projections}
\end{figure*}

The expected number of events is obtained by evaluating \autoref{eq:R_crystal_2D} for a given coupling strength $c_i$.
The probability that the photon emitted from the QD produces either 1-electron or 2-electron events in the Skipper CCD is 
given by the pair-creation probability distributions reported in~\cite{Ramanathan:2020fwm}, which we compute at the optical gap energy
(\autoref{tab:gap-values}). In reality, the emitted photon is redshifted with respect to the gap by the Stokes shift, which depends on the surface passivation and the surrounding solvent or matrix~\cite{dohnalova2013surface}, and which we therefore do not model explicitly. This choice does not qualitatively affect our projections: since the optical gap rises steeply with decreasing QD diameter (\autoref{fig:opticalgaps}), a Stokes shift of $\mathcal{O}(0.1)$\,eV can be compensated by selecting a slightly smaller dot that emits at the same energy, at a marginal cost in interaction rate from the correspondingly larger excitation threshold. The generation of multiple charge-carrier pairs requires at least about 3~eV. Therefore, some of the larger QDs are not able to generate a 2-electron signal, since their band gap is too small. However, since the 2-electron noise is significantly reduced compared to the 1-electron noise, the QDs that are able to contribute to the 2-electron signal have significantly improved sensitivities, when compared to their 1-electron expected sensitivity. In other words, while the increase in the optical gap decreases the overall rate in the QDs (as described in the previous two sections), sufficiently high-energy photon emission comes with a significant boost in sensitivity due to smaller readout backgrounds. 

\autoref{fig:limit-projections} shows the 90\% C.L.~median upper limits computed with
a combined likelihood on the 1-electron and 2-electron events.
The median limit is computed from 200 Poisson realizations of the expected background.

\begin{figure*}[t!]
\centering
  \includegraphics[width=0.48\textwidth]{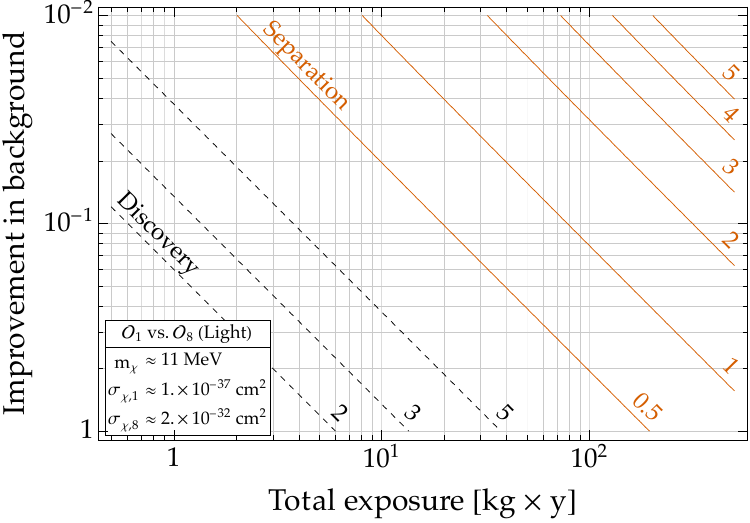}
  \includegraphics[width=0.48\textwidth]{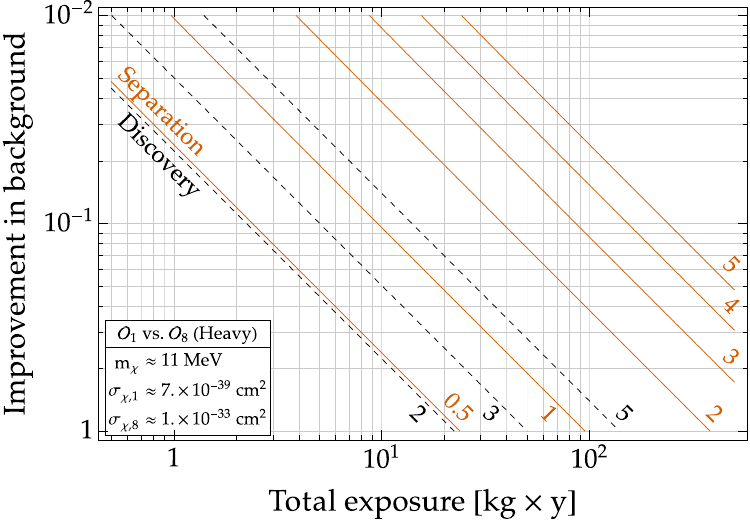}
  \includegraphics[width=0.48\textwidth]{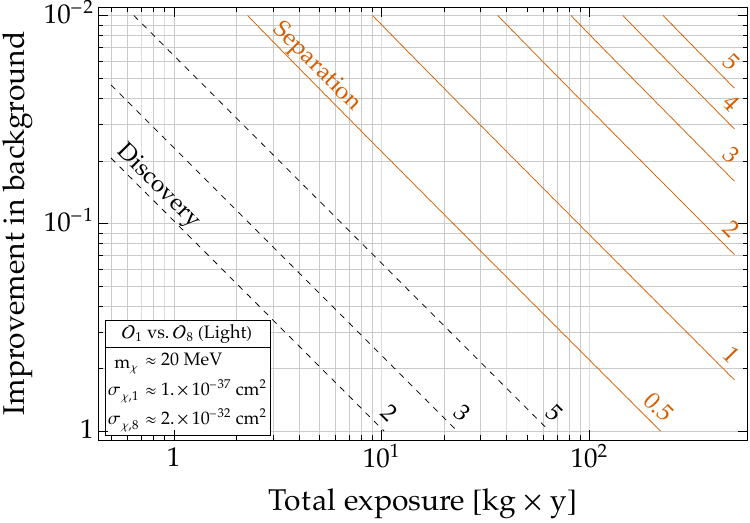}
  \includegraphics[width=0.48\textwidth]{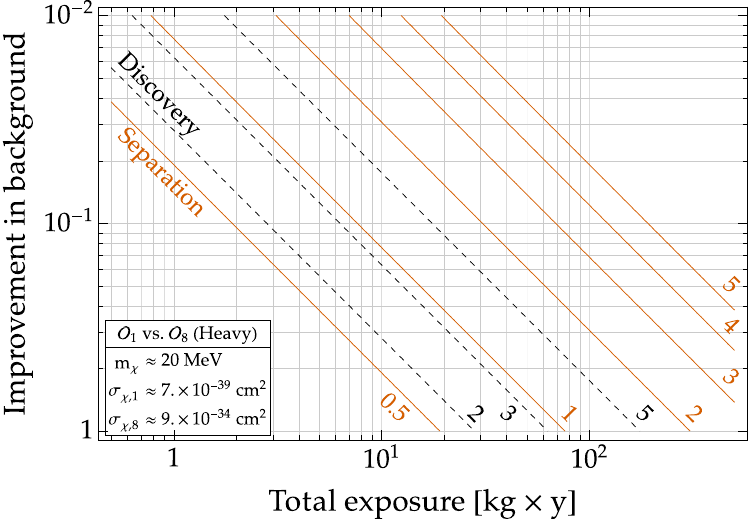}
  \includegraphics[width=0.48\textwidth]{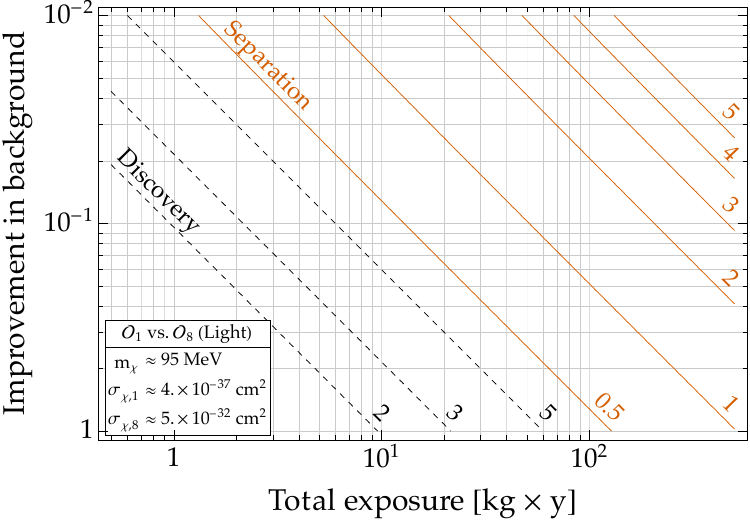}
  \phantom{\includegraphics[width=0.48\textwidth]{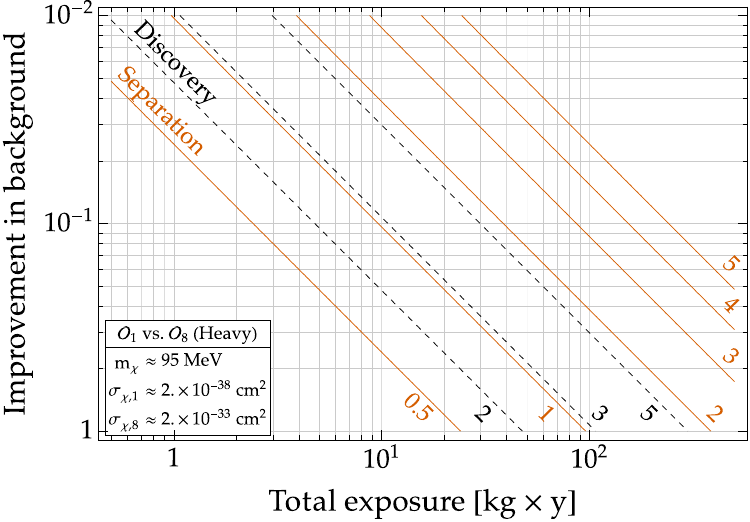}}%
\caption{Gaussian-equivalent significance $Z_{\mathrm{sym}}$ \eqref{eq:Zsym} between the $\mathcal{O}_1$ vs.\ $\mathcal{O}_8$ hypotheses given a light (left) and heavy (right) mediator, assuming a dark matter mass of $m_{\chi}=\SI{11}{\mega\electronvolt}$ (top), $m_{\chi}=\SI{20}{\mega\electronvolt}$ (center), $m_{\chi}=\SI{95}{\mega\electronvolt}$ (bottom). The case of a heavy mediator for a DM mass of $m_{\chi}=\SI{95}{\mega\electronvolt}$ has been omitted because of stringent constraints provided by liquid-xenon experiments~\cite{PhysRevLett.130.261001}. The orange solid curves mark benchmark values of $Z_{\mathrm{sym}}$ in the 2-dimensional space, expressed in number of $\sigma$ equivalent and defined by possible experimental improvements, i.e. reduction in background and increase in exposure; relative to the benchmark of 5~kg~yr  exposure with a background of $1.7\times10^{-2}$/(CCD~s)~\cite{DAMIC-M:2025luv}.  The black dashed lines mark the discovery significance, also expressed in number of $\sigma$ equivalent. The vertical axis gives the background scale factor relative to the benchmark assumptions (e.g., 0.1 corresponds to a tenfold background reduction). }
\label{fig:ModelSep}
\end{figure*}

Despite the modest benchmark exposure, the projected limits are competitive with the leading constraints near $m_\chi \sim 10$\,MeV. For light mediators, the projection reaches roughly a factor of five below the current DAMIC-M bound~\cite{DAMIC-M:2025luv} near the sensitivity floor, probing unexplored parameter space, while for heavy mediators it is comparable to the DAMIC-M floor. Noble-liquid experiments remain more sensitive above $m_\chi \sim 50$\,MeV~\cite{PhysRevLett.130.261001,XENON:2019gfn}, and the eV-scale thresholds of direct-charge CCD readout retain the advantage at the lowest masses. Importantly, this setup can move beyond the state of the art with realistic improvements alone: extending the exposure (e.g., three years of data instead of one) and lowering the CCD dark rates directly deepen the reach, as illustrated by the exposure and background scalings in \autoref{fig:ModelSep}.

We also assess the detector's ability to break the degeneracy between the coupling strength and the operator type. To study the possibility of distinguishing different operators, we focus on the separability of the $\mathcal{O}_1$ and $\mathcal{O}_8$ operators, which show distinct rate patterns. We do so in three benchmark cases: dark matter masses of $m_{\chi}=\SI{11}{\mega\electronvolt}$ and $m_{\chi}=\SI{20}{\mega\electronvolt}$, for the case of both light and heavy mediators, and additionally $m_{\chi}=\SI{95}{\mega\electronvolt}$, for the case of a light mediator.

The experimental setup is the same as above: the ten active units form five pairs, each filled with a distinct QD morphology (see~\autoref{tab:gap-values}). For each group, the Skipper-CCD records 1-electron and 2-electron event counts, yielding a total of $N_\text{bins} = 8$ bins\footnote{The 63-Si-alk and 121-Si-H morphologies have 0 expected 2-electron events.} whose shape carries information on the underlying interaction operator.

\begin{table}
    \centering
    \begin{tabular}{ccccc}
    \hline
        \multirow{2}{*}{Mass}
        & \multicolumn{2}{c}{Light mediator}
        & \multicolumn{2}{c}{Heavy mediator}\\
        & $\sigma_{\chi,1}$ & $\sigma_{\chi,8}$
        & $\sigma_{\chi,1}$ & $\sigma_{\chi,8}$\\
        {[MeV]} & {[\si{\centi\meter\squared}]}& {[\si{\centi\meter\squared}]}& {[\si{\centi\meter\squared}]}& {[\si{\centi\meter\squared}]}\\
        \hline \hline 
        11 & \num{1.0e-37} & \num{1.6e-32}& \num{7.1e-39} & \num{1.1e-33}\\
        20 & \num{1.2e-37} & \num{1.7e-32}& \num{6.7e-39} & \num{9.3e-34}\\
        95 & \num{3.6e-37} & \num{5.4e-32} & --- & ---\\
        \hline
    \end{tabular}
    \caption{Values for $\sigma_{\chi,1}$ and $\sigma_{\chi,8}$ for the benchmark values of $m_{\chi}$ assumed in \autoref{fig:ModelSep}.}
    \label{tab:cSecVal}
\end{table}

We adopt $\mathcal{O}_1$ as the null hypothesis $H_0$, in the two possible cases of light and heavy mediators. Here, $\sigma_{\chi,k}$ denotes the reference DM--electron cross section associated with the coupling $c_k$ of the operator $\mathcal{O}_k$, following the convention of Refs.~\cite{Catena:2019gfa,Catena:2021qsr}. For each case and for a given mass, the value of $\sigma_{\chi,1}$ is listed in~\autoref{tab:cSecVal} and corresponds to the most favorable case not yet excluded~\cite{DAMIC-M:2025luv}. The heavy mediator case for a DM mass of $m_{\chi}=\SI{95}{\mega\electronvolt}$ has been omitted from this table because the constraints provided by the PandaX-4T experiment would lead to requiring unrealistic exposures~\cite{PhysRevLett.130.261001}.

The alternative hypothesis $H_1$, on the other hand, corresponds to the operator $\mathcal{O}_8$ and a cross section $\sigma_{\chi,8}$ that produces the same total number of events as $\mathcal{O}_1$ under $H_0$. That is, 
\begin{equation}\label{eq:Constr1}
    \sum_{i=1}^{N_\text{bins}}\mu_i^{(0)}=\sum_{i=1}^{N_\text{bins}}\mu_i^{(1)}\ ,
\end{equation}
where $\mu_i^{(k)}$ is the expected number of events in the $i$-th bin under $H_k$. The values of $\sigma_{\chi,8}$ that satisfy the constraint \eqref{eq:Constr1} are also listed in \autoref{tab:cSecVal}. Imposing the same number of events eliminates the trivial separation driven purely by differences in the expected overall rate, isolating the discriminating power of the bin-shape information.

The separation between the two hypotheses is quantified by a likelihood ratio between the Poissonian counts $\vec{n}$ of the $N_\text{bins}$ bins:
\begin{equation}\label{eq:LogLP}
    \log\Lambda(\vec{n}) = \log\frac{L(\vec{n}|H_1)}
    {L(\vec{n}|H_0)}
    = \sum_{i=1}^{N_\text{bins}} n_i\log\frac{\mu_i^{(1)}}{\mu_i^{(0)}}\ ,
\end{equation}
where the difference $\mu_i^{(1)} - \mu_i^{(0)}$ cancels out because of
\eqref{eq:Constr1}. We use the Asimov dataset, that is $n_i = \mu_i^{(k)}$ under $H_k$. Asymptotically,
\begin{equation}
    -2\log\Lambda\sim\chi^2_1\ ,
\end{equation}
and since $\chi^2_1$ is related to the square of a standard normal, the Gaussian-equivalent significance $Z$ is
\begin{equation}
    Z_{0\to1} = \sqrt{2 \sum_{i=1}^{N_\text{bins}} 
    \mu_i^{(0)}\log\frac{\mu_i^{(0)}}{\mu_i^{(1)}}}
\end{equation}
for the separation between $H_0$ and $H_1$. In the same way, we can define $Z_{1\to 0}$ for the separation between $H_1$ and $H_0$ as well as the average symmetrical quantity:
\begin{equation}\label{eq:Zsym}
    Z_{\mathrm{sym}} = \sqrt{\frac{Z_{0\to1}^2+Z_{1\to0}^2}{2}}\ .
\end{equation}

The values of $Z_{\mathrm{sym}}$ are reported in \autoref{fig:ModelSep}. They are expressed in Gaussian-equivalent significance ($\sigma$) in the 2-dimensional parameter space defined by the total exposure and improvement in the background. The behavior of $Z_{\mathrm{sym}}$ is dependent on the assumptions both for $m_\chi$ and the nature of the mediator. The Gaussian-equivalent benchmarks for discovery are also reported. Clearly, a separation between the $\mathcal{O}_1$ and $\mathcal{O}_8$ hypotheses at a given confidence level can only follow a discovery of the signal over the background-only hypothesis at the same significance.

We see in \autoref{fig:ModelSep} that large exposures are needed for discovery and even larger ones for separating the operators. To be concrete, we quote the requirements at a reference exposure of \SI{100}{\kilo\gram\year}. For light mediators, such an exposure yields a $5\sigma$ discovery of the benchmark cross sections at the benchmark background rates. For heavy mediators, a $5\sigma$ discovery requires background rates roughly 30\% lower. A $2\sigma$ separation of the operators requires background rates about three times lower than the benchmark for heavy mediators, and twenty to thirty times lower for light mediators. Note that a tenfold background reduction lowers the required exposure by roughly the same factor. It will be challenging to deploy multiple low-background QD detectors and obtain such large exposures, but such an effort will be well worth it in case of a future signal, as it will help identify the particle properties of the DM.

\section{Conclusions}\label{sec:conclusions}

We have presented the first \emph{ab initio} calculation of DM--electron scattering in quantum dots, treating them as finite, non-periodic clusters within density functional theory and computing rates for the complete set of non-relativistic effective operators. Our modification of the QEdark-EFT code, together with the data underlying the results presented in this work, is publicly available on Zenodo~\cite{data}.

We find that the QD response separates into two tunable effects: the size-dependent optical gap, which controls the kinematically accessible phase space, and the surface termination, which shapes the energy--momentum support of the response. Alkyl termination introduces carbon-derived surface states that, on average, enhance the rates relative to H-terminated dots. Per unit mass, the QD rates remain within a factor of a few of those of bulk silicon, and a benchmark \SI{5}{\kilo\gram\year} QD setup, with Skipper-CCDs used to read out the photons, is competitive with leading constraints near $m_\chi \sim 10$\,MeV, with clear scaling paths beyond.

Beyond raw reach, our results show that quantum dots can serve as a set of complementary targets whose relative rates encode information about the underlying interaction. The morphology-dependent response produces a characteristic ordering of rates across core sizes and surface terminations, which depends on the effective operator, the mediator type, and in some cases the DM mass. If a DM-compatible excess is observed, and provided that the required exposures and background control can be achieved, a multi-QD detector could therefore function as an interaction barcode: the pattern of signals across morphologies carries information that is not available from a single target alone. Deploying multiple low-background detector volumes at these exposures will be challenging, but in the event of a signal such a program could shed light on the particle properties of the dark matter.

This strategy relies on a feature that is unique to QDs among solid-state detector materials. Because the optical gap is controlled primarily by quantum confinement, QDs with different sizes, surface terminations, and different core compositions can be engineered to have comparable excitation thresholds and distinct energy--momentum electronic structure. QDs thus provide a route to generating many detector targets within a single experimental platform, in the same energy range, and with the same optical readout concept. To our knowledge, no other material class offers an analogous route to producing $\mathcal{O}(100)$ distinct detector responses while keeping the excitation scale and detector architecture fixed.

The two surface terminations examined in this study, hydrogen and alkyl passivation, were selected to bracket two physically distinct regimes of the QD electronic structure. H-terminated Si QDs, whose band structure closely resembles that of bulk silicon~\cite{reboredo2005theory,dohnalova2013surface}, serve as a natural benchmark for isolating quantum confinement effects. Alkyl-terminated QDs, by contrast, introduce covalently bonded hydrocarbon ligands that hybridize with the surface silicon atoms, generating delocalized Si--C surface states that substantially modify the form factors~\cite{dohnalova2013surface} and enable bright emission with near-unity internal quantum efficiency~\cite{Sangghaleh_2015}. These choices serve as a proof of principle. In practice, the optimal surface passivation for a deployed detector would be governed by experimental requirements---solvent compatibility, colloidal shelf life, and resistance to photo-oxidation, a known limitation of H-terminated Si QDs~\cite{wolkin1999electronic}. A wider palette of surface chemistries, including thiol, amine, or phosphonate ligands, would introduce further variation in the surface electronic structure and enrich the discriminating power of the interaction barcode. Extending the target basis to other semiconductor cores---such as PbS, CdSe, or Ge nanocrystals~\cite{Blanco:2022cel}---could provide additional distinct energy--momentum electron distributions, broadening the statistical basis for interaction identification.

The predictions presented here focus on the large-mass tail of sub-GeV DM and on energy depositions below $20\,\mathrm{eV}$. In this momentum-transfer regime, the collective in-medium effects can be neglected, and the presented results are obtained without applying a screening correction or including all-electron effects. A full treatment of screening in QDs, including its operator dependence and its modification relative to bulk crystals, remains an important direction for future work, particularly for lower DM masses.

The sensitivity projections in~\autoref{sec:detectorresponse} are based on a detector geometry in which each \SI{1}{\kilo\gram} QD unit is instrumented with Skipper-CCDs to read out the photons. 
This configuration leverages the sub-electron noise performance of Skipper-CCDs~\cite{Tiffenberg:2017aac} that have been used by DAMIC-M and SENSEI~\cite{DAMIC-M:2025luv,SENSEI:2024yyt} for DM searches, and exploits the distinction between 1-electron and 2-electron readout channels. The dominant background is the CCD dark current, and we assume 1-electron and 2-electron rates corresponding to the DAMIC-M measurement. For QDs whose optical gap exceeds the two-charge-carrier threshold of approximately 3~eV~\cite{Ramanathan:2020fwm}, the 2-electron signal channel carries substantially reduced background and improved sensitivity, partially offsetting the rate suppression from the larger band gap. Future reductions in the single-electron dark rate, in line with the generational improvements already demonstrated by DAMIC-M and SENSEI~\cite{DAMIC-M:2025luv,SENSEI:2024yyt}, would directly extend the sensitivity reach of a QD-based search, particularly at lower DM masses and for operators with suppressed low-momentum transfer cross sections. Photodetectors with even lower dark counts, such as the TESs and SNSPDs discussed in \autoref{sec:detectorresponse}, could extend the reach further.

\begin{acknowledgments}
We are grateful to the organizers of the 2026 MIAPbP workshop ``Fill the Gap'' for their hospitality, and to Megan Hott for useful discussions.
Furthermore, A.C. and M.M. acknowledge support of the Ministry of Education, Youth and Sports of the Czech Republic through e-INFRA CZ (ID:90254). 
A.C. acknowledges the support of the project ``Robotics and advanced industrial production'' (reg.no. CZ.02.01.01/00/22\_008/0004590) co-funded by the European Union.
M.M. acknowledges the support by the CTU Mobility Project MSCA-F-CZ-III (reg.no. CZ.02.01.01/00/22\_010/0008601).
NH acknowledges support under the GIRA from the Coordinating Panel for Advanced Detectors. NH and LW also acknowledge support from the QuantISED 2.0 program under DOE DE-SC0026237.
RE acknowledges support from the DOE
Grant DE-SC0025309 and the Simons Investigator in Physics award MPS-SIP-00010469. 
TL is supported by the Swedish Research Council under contract
2022-04283 and the Swedish National Space Agency under contract 2023-00242. TL also acknowledges sabbatical support from the Wenner-Gren foundation under contract SSh2024-0037. JC and AGR acknowledge the support from the Swedish Research  Council, Knut and Alice Wallenberg Foundation and Olle Enqvists Foundation.
\end{acknowledgments}

\clearpage

\appendix

\onecolumngrid

\section{Morphology of Studied QDs}\label{app:morphology}
\begin{figure}[!ht]
\centering
  \centering
  \includegraphics[width=0.8\textwidth]{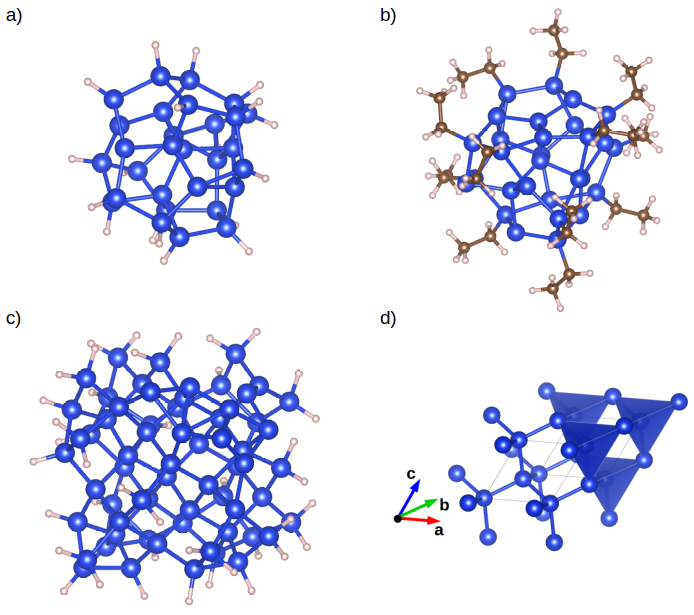}
\caption{Model geometry of a) 29-Si-H, b) 29-Si-alk, and c) 64-Si-H quantum dots. d) Bulk Si model where the tetrahedral coordination is highlighted by the coordination polyhedra; thin black lines indicate the shape of the primitive unit cell, while the colored arrows indicate the relative orientation of the crystallographic axes. The atom color code is the same as in \autoref{fig:3D-QD-example}.}
\label{fig:3D-QDs}
\end{figure}

In this appendix, we show the model geometries not displayed in the main text, complementing the representative QDs shown in \autoref{fig:3D-QD-example}. All clusters are carved from the experimental bulk silicon structure, with every surface dangling bond saturated by the passivating species, and are subsequently relaxed (see \autoref{sec:qdsetup}). Panels a and b of \autoref{fig:3D-QDs} are built from the same 29-atom silicon fragment and differ in their termination, hydrogen in panel a and ethyl groups in panel b. This pair isolates the effect of surface chemistry at fixed core size, which is the comparison used throughout Secs.~\ref{sec:materialresponse}--\ref{sec:optimalsetup}. Since the relaxation responds to the termination, the two relaxed cores are, however, not identical. The ethyl shell also visibly enlarges the passivated cluster, which is reflected in the effective diameters adopted in \autoref{tab:gap-values} (1.1\,nm for 29-Si-H versus 1.6\,nm for 29-Si-alk). Panel c shows the intermediate 64-Si-H cluster. While essentially every atom in the 29-Si clusters is a surface-passivated atom, a smaller proportion needs passivation in larger clusters. The surface termination therefore has the most pronounced effect on the electronic states of the smallest QDs.

Panel d shows the bulk silicon reference. The shaded polyhedra highlight the four-fold tetrahedral coordination of the diamond lattice, the thin lines trace the primitive unit cell, and the arrows give the crystallographic axes. The passivating species complete the same four-fold coordination at the cluster surfaces. The clusters nevertheless differ from the bulk in three ways. They are finite, they are terminated by a surface species, and their relaxed geometries deviate from the ideal bulk arrangement, since the relaxation pushes the core away from the crystalline positions. All three effects diminish as the core grows, and the models converge to the bulk crystal.

\clearpage

\section{Material Response Functions}\label{app:responses}

In this appendix, we provide the material-response functions that enter the rate calculation but are not shown in the main text. The response $W_1$ is built from the scalar form factor $f'_{i\rightarrow i'}$ of \eqref{eq:f_prime}, through the combination $B_1$ defined in \autoref{sec:formalism}, and is the response probed by density-type couplings such as $\mathcal{O}_1$. The response $W_4$ involves the longitudinal projection of the vectorial form factor of \eqref{eq:fvec_prime}, through the combination $B_4$, and enters the event rate through \eqref{eq:R_crystal_2D}. As in \autoref{fig:responses-QDs-W3}, all responses are normalized per silicon atom, and the dashed line marks the kinematic boundary in the large-mass limit.

\begin{figure}[!ht]
\centering
  \centering
  \includegraphics[width=0.49\textwidth]{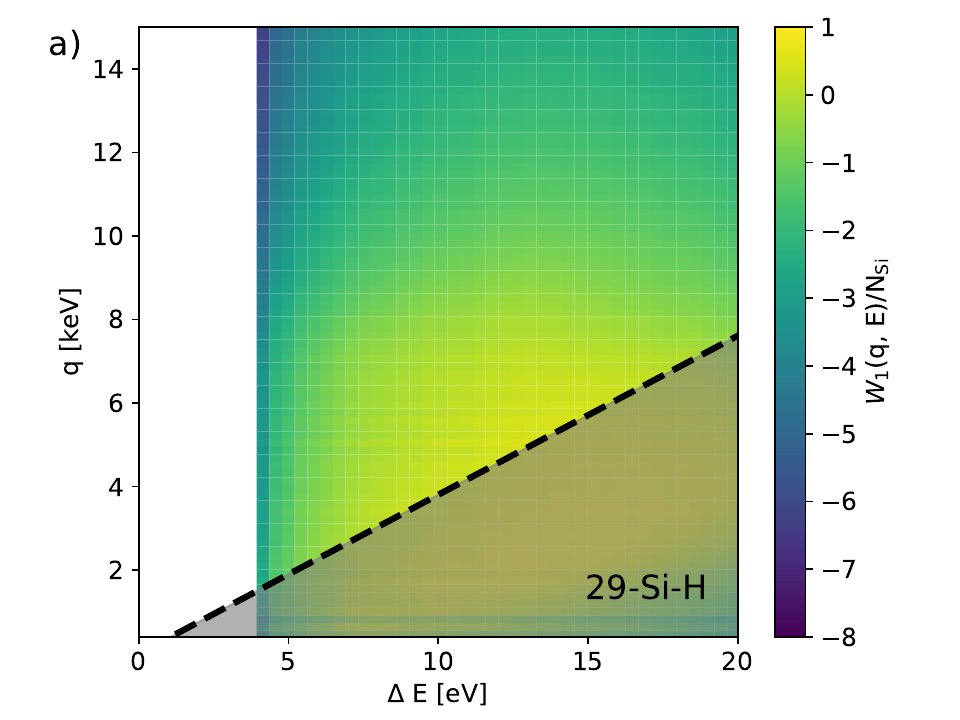}
  \includegraphics[width=0.49\textwidth]{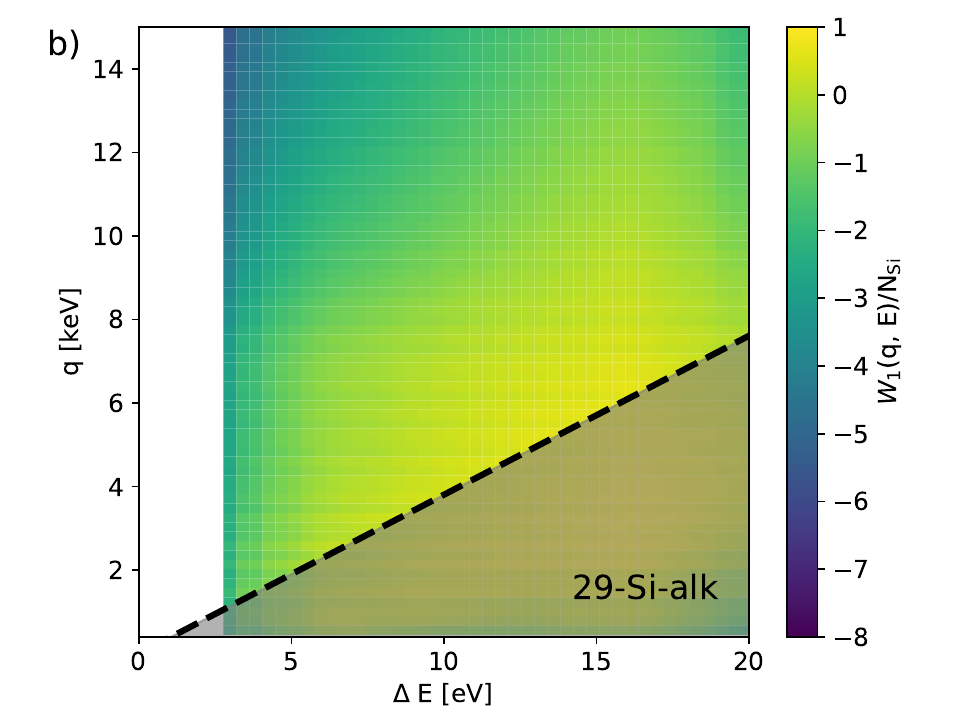}
  \includegraphics[width=0.49\textwidth]{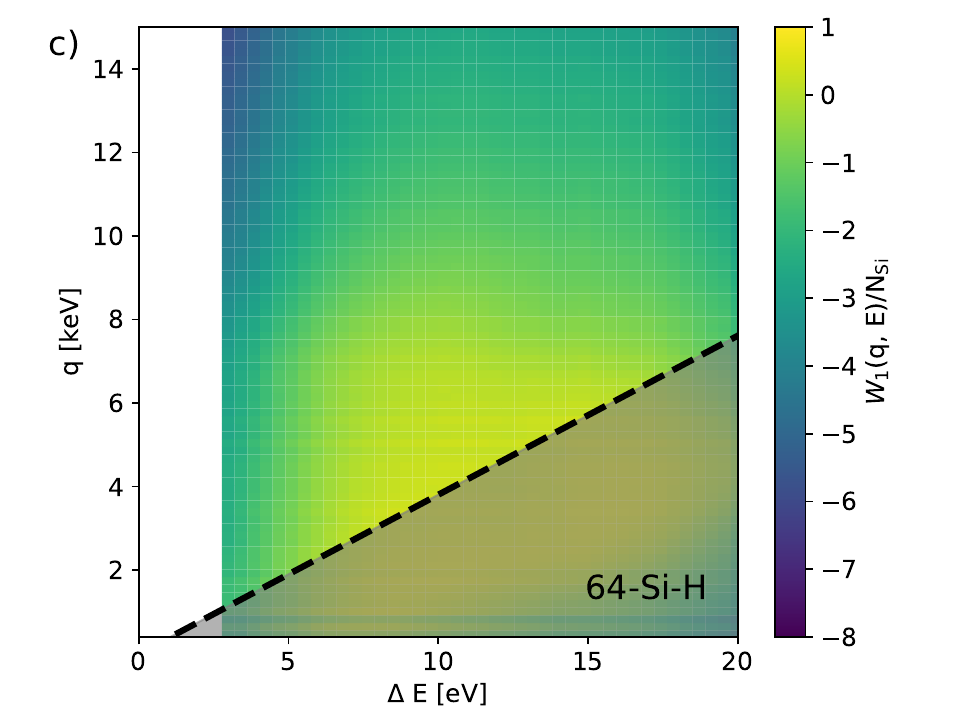}
  \includegraphics[width=0.49\textwidth]{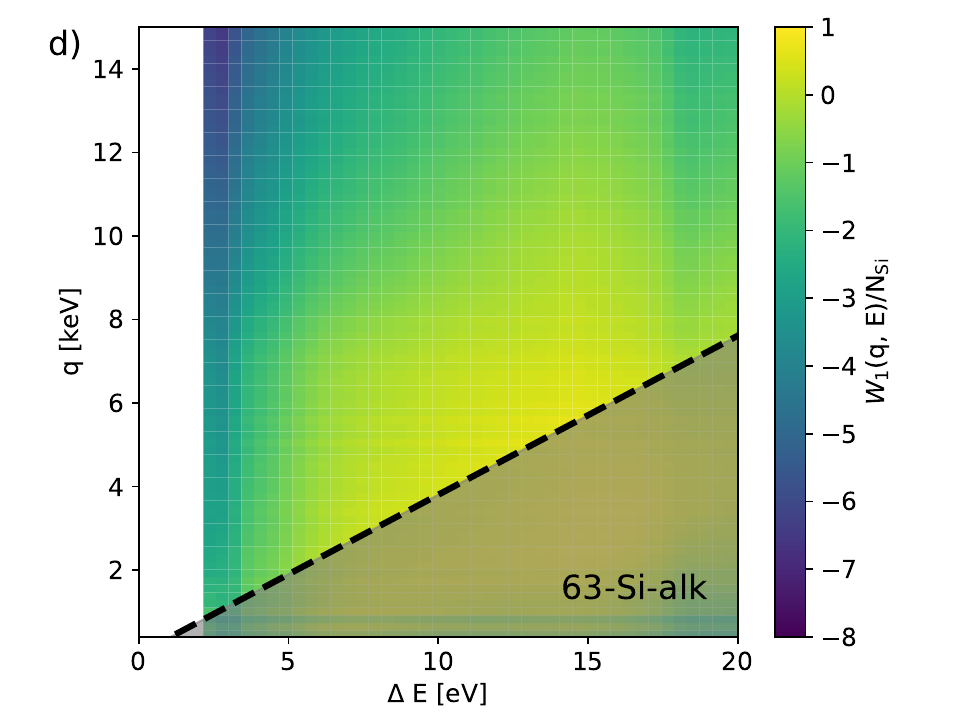}
  \includegraphics[width=0.49\textwidth]{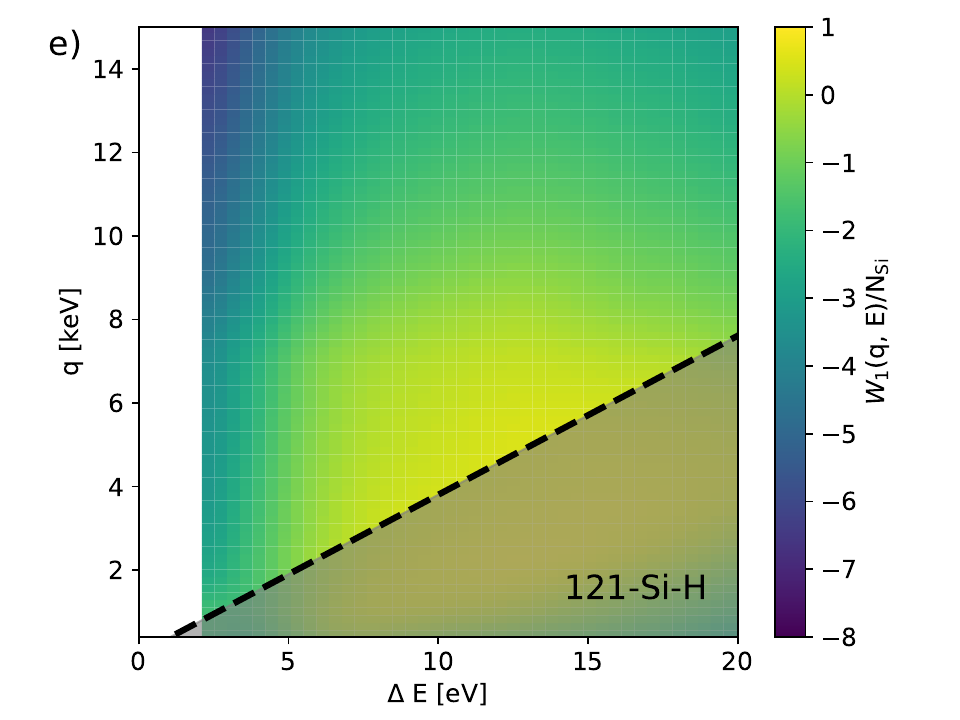}
  \includegraphics[width=0.49\textwidth]{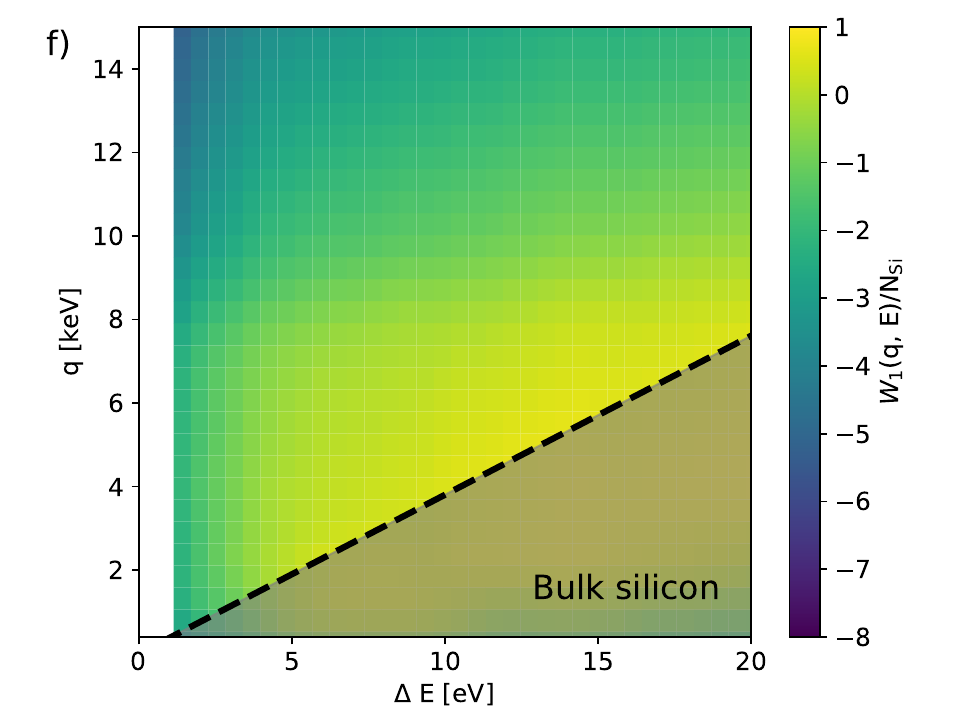}
\caption{Material response function $W_1$ normalized by the number of silicon atoms in a unit cell for all considered QDs, as well as for the bulk Si crystal. The shaded region below the dashed black line shows the kinematically inaccessible region. In the shaded region, the minimal DM velocity required to induce an excitation exceeds the galactic escape velocity in the large-mass limit.}
\label{fig:responses-QDs-W1}
\end{figure}

\autoref{fig:responses-QDs-W1} shows $W_1$. Its onset traces the optical gap of each target, from 1.12\,eV for bulk silicon up to 4.2\,eV for 29-Si-H, for which the empty band below threshold is clearly visible. In all targets, the support is concentrated at momentum transfers of $q\sim2$--$8$\,keV and falls steeply toward larger $q$. The H-terminated dots closely resemble the bulk response up to a rigid shift of the onset, while the alkyl-terminated dots carry additional weight at $\Delta E \sim 10$--$20$\,eV, originating from the deep Si--C surface states discussed in \autoref{sec:materialresponse}. The peak values per silicon atom are comparable across all six targets, which helps explain the factor-of-a-few rate comparison in the main text.

\begin{figure}[!ht]
\centering
  \centering
  \includegraphics[width=0.49\textwidth]{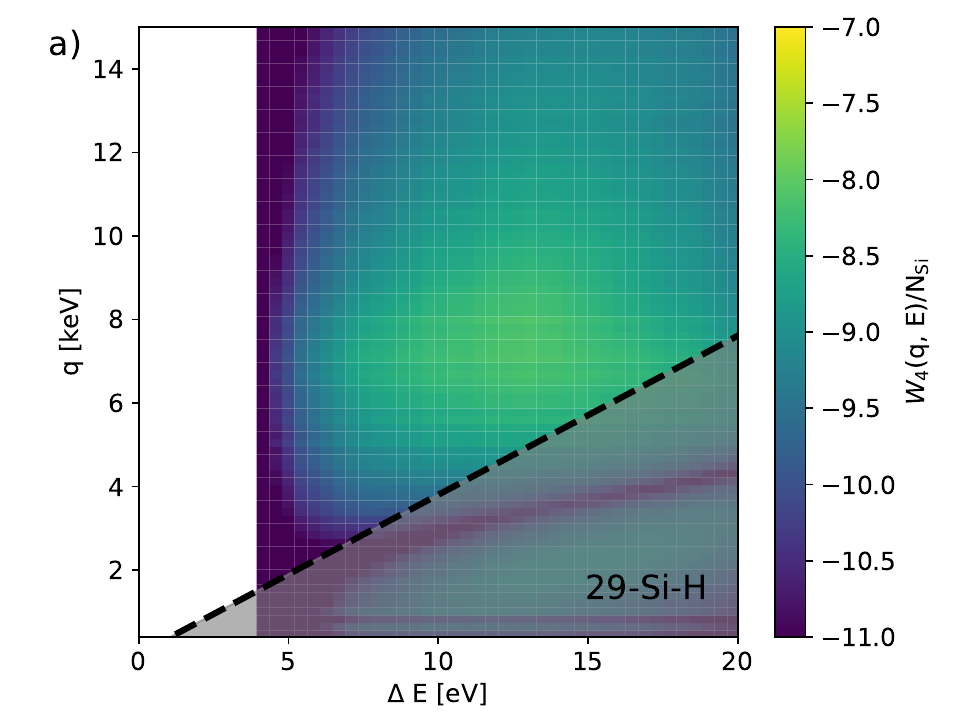}
  \includegraphics[width=0.49\textwidth]{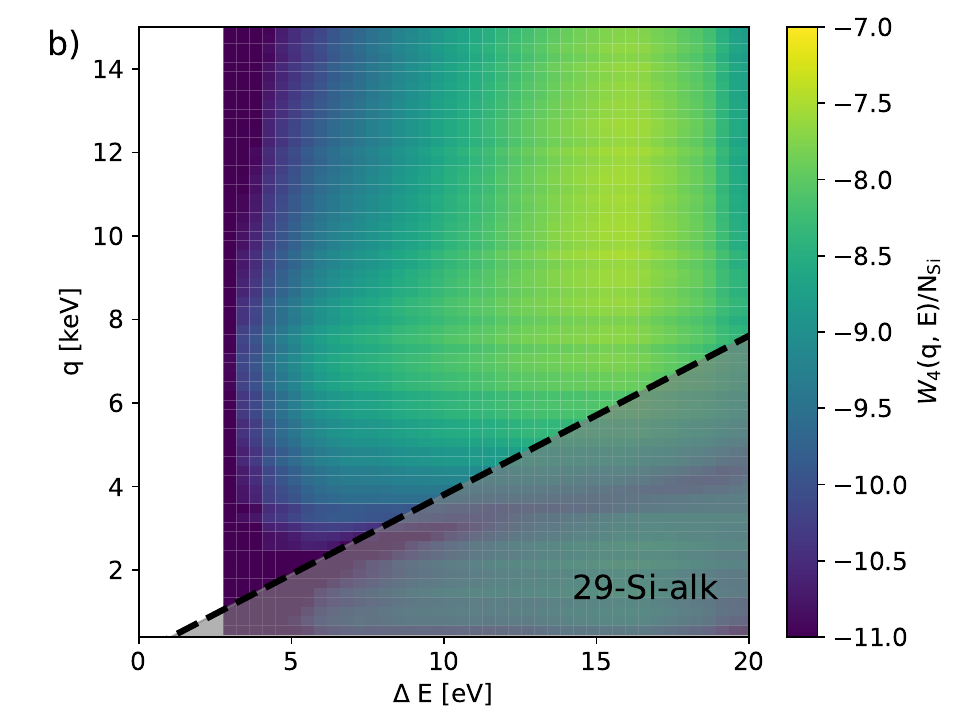}
  \includegraphics[width=0.49\textwidth]{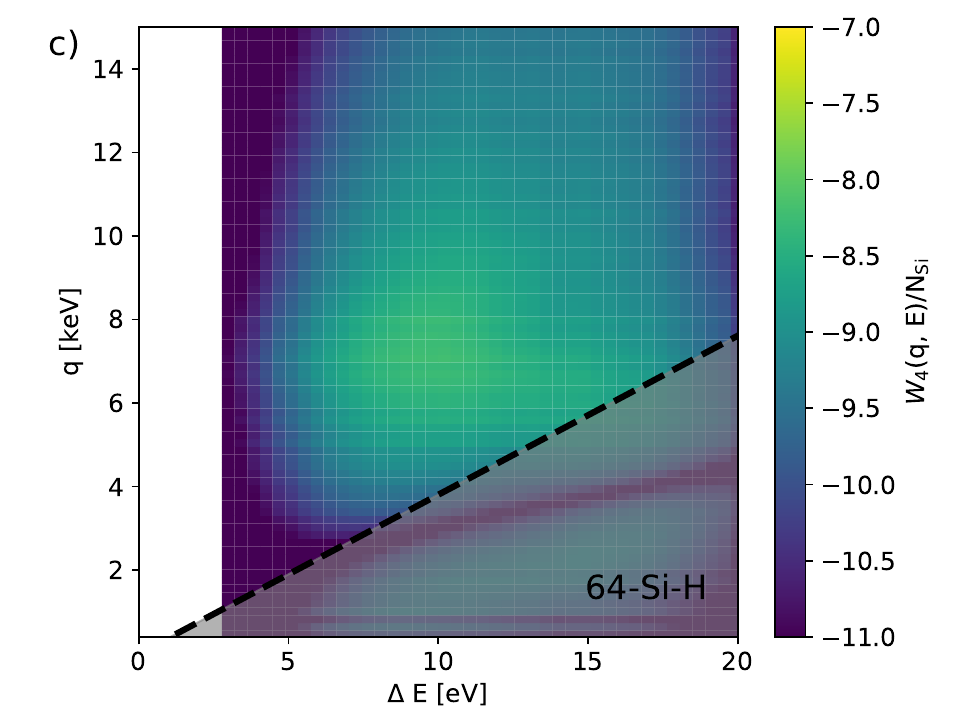}
  \includegraphics[width=0.49\textwidth]{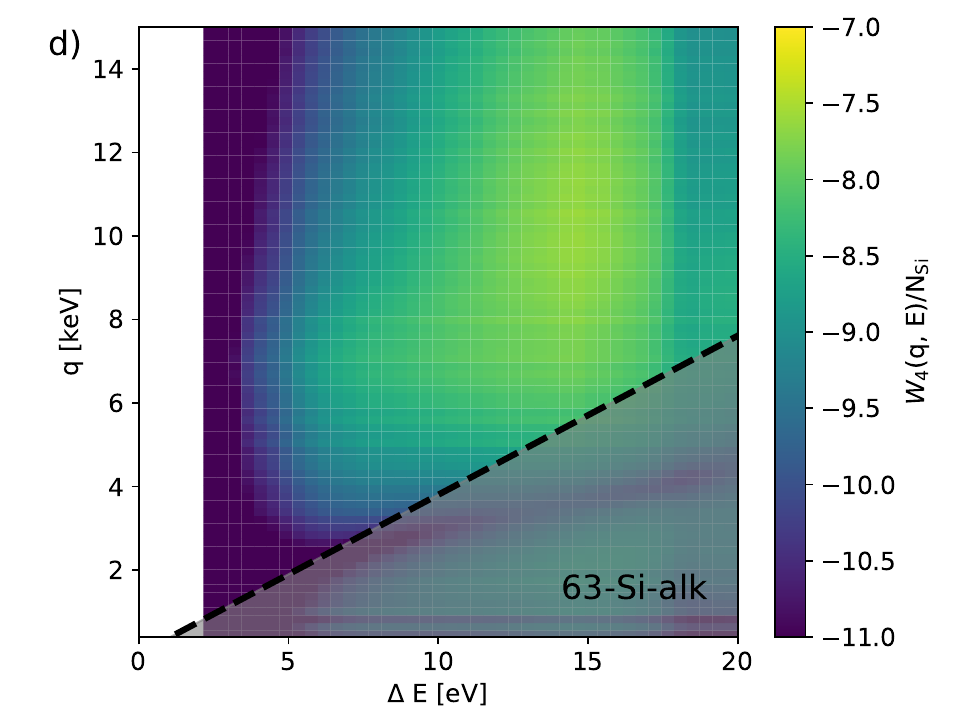}
  \includegraphics[width=0.49\textwidth]{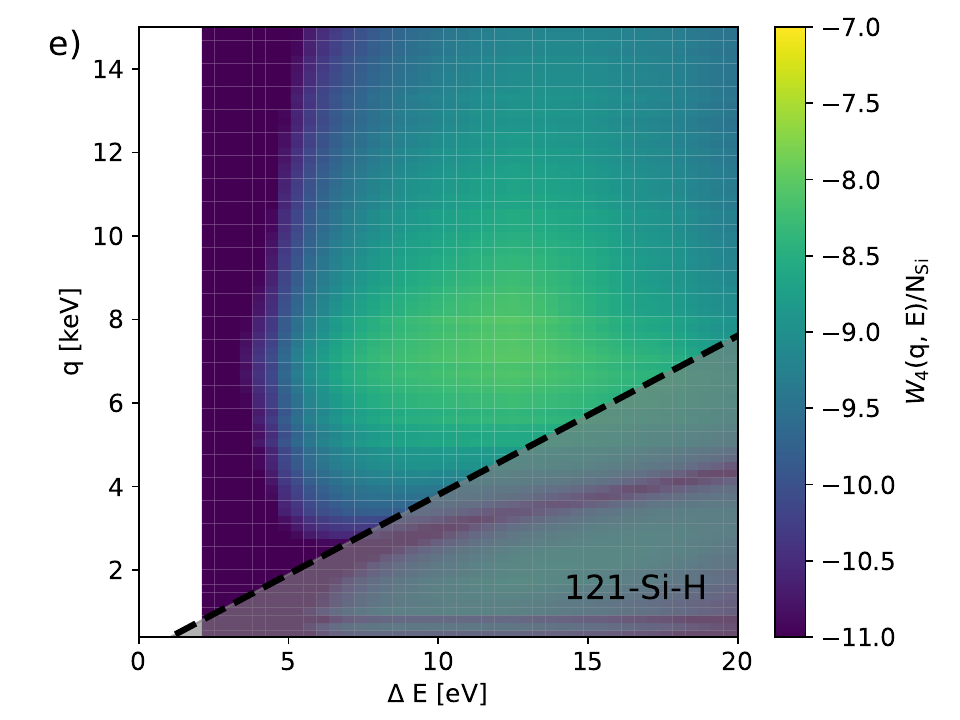}
  \includegraphics[width=0.49\textwidth]{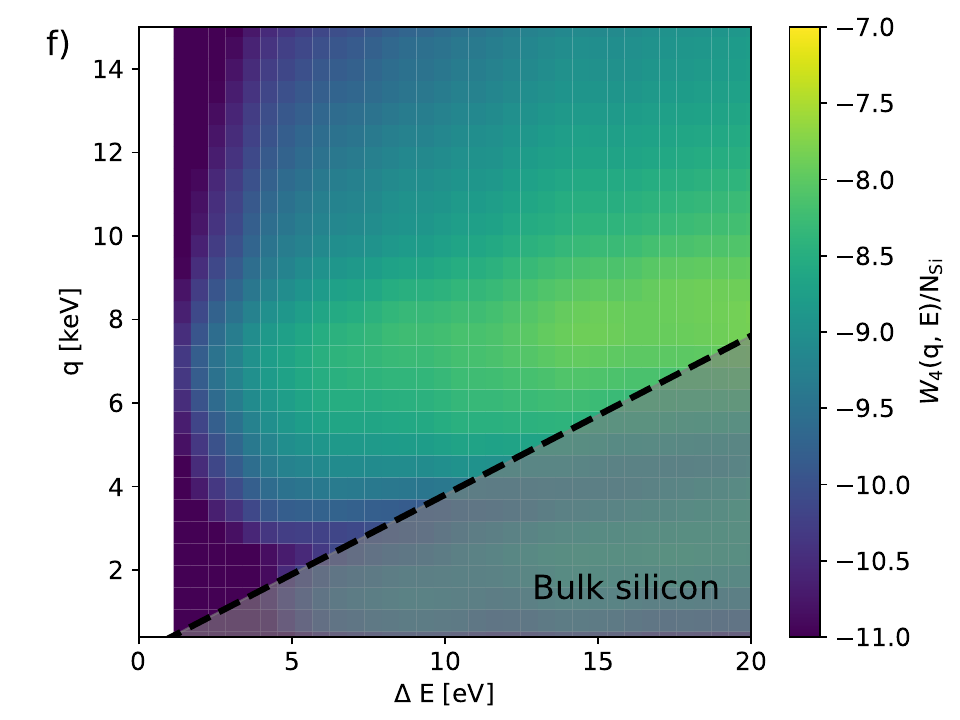}
\caption{Material response function $W_4$ normalized by the number of silicon atoms in a unit cell for all considered QDs, as well as for the bulk Si crystal. The shaded region below the dashed black line shows the kinematically inaccessible region. In the shaded region, the minimal DM velocity required to induce an excitation exceeds the galactic escape velocity in the large-mass limit.}
\label{fig:responses-QDs-W4}
\end{figure}

\autoref{fig:responses-QDs-W4} shows $W_4$. The explicit factor of $\mathbf{q}/m_e$ moves its support to larger momentum transfers, with the maximum near $q\sim6$--$10$\,keV, and suppresses the response at small $q$, in contrast to $W_1$. Its overall magnitude lies roughly eight orders of magnitude below $W_1$, reflecting the additional factors of $q/m_e$ and of the bound-electron velocity in the underlying amplitude. For the alkyl-terminated dots, the maximum is additionally pushed toward larger energy transfers.

A distinctive feature of $W_4$ is the dark valley running from the onset toward larger energy transfers, most clearly visible in the QD panels. Its origin is current conservation, as we now show. Consider the single-particle Hamiltonian $H = \mathbf{p}^2/2m_e + V(\mathbf{x})$ with a local potential $V(\mathbf{x})$, whose eigenstates satisfy $H|i\rangle = E_i |i\rangle$ and $H|i'\rangle = E_{i'} |i'\rangle$. The matrix element of the commutator $[H, e^{i\mathbf{q}\cdot\mathbf{x}}]$ between these states can be evaluated in two ways. First, letting $H$ act directly on the eigenstates on either side,
\begin{equation}
\langle i'| \left[H, e^{i\mathbf{q}\cdot\mathbf{x}}\right] |i\rangle = \left(E_{i'} - E_i\right) \langle i'| e^{i\mathbf{q}\cdot\mathbf{x}} |i\rangle = \Delta E \, f_{i\rightarrow i'}(\mathbf{q})\,,
\end{equation}
where the last step uses the definition of the scalar form factor, \autoref{eq:f}. Second, evaluating the commutator explicitly: the local potential is a function of position and commutes with $e^{i\mathbf{q}\cdot\mathbf{x}}$, so only the kinetic term contributes. Acting on any wavefunction, $\mathbf{p}\, e^{i\mathbf{q}\cdot\mathbf{x}} \psi = -i\nabla\left(e^{i\mathbf{q}\cdot\mathbf{x}} \psi\right) = e^{i\mathbf{q}\cdot\mathbf{x}} \left(\mathbf{p}+\mathbf{q}\right)\psi$, i.e., commuting the plane wave through the momentum operator shifts the latter by $\mathbf{q}$. Applying this twice,
\begin{equation}
\left[\mathbf{p}^2, e^{i\mathbf{q}\cdot\mathbf{x}}\right] = e^{i\mathbf{q}\cdot\mathbf{x}} \left[\left(\mathbf{p}+\mathbf{q}\right)^2 - \mathbf{p}^2\right] = e^{i\mathbf{q}\cdot\mathbf{x}} \left(2\,\mathbf{q}\cdot\mathbf{p} + q^2\right)\,,
\end{equation}
so that
\begin{equation}
\langle i'| \left[H, e^{i\mathbf{q}\cdot\mathbf{x}}\right] |i\rangle = \frac{1}{m_e}\langle i'| e^{i\mathbf{q}\cdot\mathbf{x}}\, \mathbf{q}\cdot\mathbf{p} |i\rangle + \frac{q^2}{2m_e} \, f_{i\rightarrow i'}(\mathbf{q})\,.
\end{equation}
The first term is the longitudinal projection of the vectorial form factor: since \autoref{eq:fvec} defines $\mathbf{f}_{i\rightarrow i'}$ with $i\nabla/m_e = -\mathbf{p}/m_e$, we have $\langle i'| e^{i\mathbf{q}\cdot\mathbf{x}}\,\mathbf{q}\cdot\mathbf{p} |i\rangle / m_e = -\,\mathbf{q}\cdot\mathbf{f}_{i\rightarrow i'}(\mathbf{q})$. Equating the two evaluations of the matrix element therefore yields
\begin{equation}\label{eq:continuity}
\mathbf{q}\cdot\mathbf{f}_{i\rightarrow i'}(\mathbf{q}) = \left(\frac{q^2}{2m_e} - \Delta E\right) f_{i\rightarrow i'}(\mathbf{q})\,,
\end{equation}
which is the matrix-element form of the continuity equation: the longitudinal current is fixed entirely by the density. Squaring \autoref{eq:continuity} and comparing with the definitions of $B_1$ and $B_4$ in \autoref{sec:formalism} gives
\begin{equation}\label{eq:B4fromB1}
B_4 = \left|\frac{\mathbf{q}}{m_e}\cdot\mathbf{f}_{i\rightarrow i'}\right|^{2} = \frac{1}{m_e^{2}}\left(\Delta E - \frac{q^{2}}{2m_e}\right)^{2} B_1\,.
\end{equation}
This correspondence has also been noted in recent literature~\cite{Giffin:2025hdx}. The longitudinal response thus vanishes on the free-electron line $\Delta E = q^{2}/2m_e$, for every transition and independently of the wavefunctions, which produces the valley. Since this line node lies mostly within the kinematically inaccessible region, it has little effect on the integrated rates. \autoref{eq:B4fromB1} also fixes the overall magnitude of $W_4$ relative to $W_1$, and the plotted responses are consistent with it. The identity holds exactly only for local potentials and unshifted eigenvalues; the nonlocal part of the pseudopotentials and the scissor correction displace it slightly. 

\clearpage

\twocolumngrid

\bibliography{biblio}

\begin{thebibliography}{72}%
\makeatletter
\providecommand \@ifxundefined [1]{%
 \@ifx{#1\undefined}
}%
\providecommand \@ifnum [1]{%
 \ifnum #1\expandafter \@firstoftwo
 \else \expandafter \@secondoftwo
 \fi
}%
\providecommand \@ifx [1]{%
 \ifx #1\expandafter \@firstoftwo
 \else \expandafter \@secondoftwo
 \fi
}%
\providecommand \natexlab [1]{#1}%
\providecommand \enquote  [1]{``#1''}%
\providecommand \bibnamefont  [1]{#1}%
\providecommand \bibfnamefont [1]{#1}%
\providecommand \citenamefont [1]{#1}%
\providecommand \href@noop [0]{\@secondoftwo}%
\providecommand \href [0]{\begingroup \@sanitize@url \@href}%
\providecommand \@href[1]{\@@startlink{#1}\@@href}%
\providecommand \@@href[1]{\endgroup#1\@@endlink}%
\providecommand \@sanitize@url [0]{\catcode `\\12\catcode `\$12\catcode
  `\&12\catcode `\#12\catcode `\^12\catcode `\_12\catcode `\%12\relax}%
\providecommand \@@startlink[1]{}%
\providecommand \@@endlink[0]{}%
\providecommand \url  [0]{\begingroup\@sanitize@url \@url }%
\providecommand \@url [1]{\endgroup\@href {#1}{\urlprefix }}%
\providecommand \urlprefix  [0]{URL }%
\providecommand \Eprint [0]{\href }%
\providecommand \doibase [0]{https://doi.org/}%
\providecommand \selectlanguage [0]{\@gobble}%
\providecommand \bibinfo  [0]{\@secondoftwo}%
\providecommand \bibfield  [0]{\@secondoftwo}%
\providecommand \translation [1]{[#1]}%
\providecommand \BibitemOpen [0]{}%
\providecommand \bibitemStop [0]{}%
\providecommand \bibitemNoStop [0]{.\EOS\space}%
\providecommand \EOS [0]{\spacefactor3000\relax}%
\providecommand \BibitemShut  [1]{\csname bibitem#1\endcsname}%
\let\auto@bib@innerbib\@empty
\bibitem [{\citenamefont {Aprile}\ \emph {et~al.}(2025)\citenamefont {Aprile}
  \emph {et~al.}}]{XENON:2025vwd}%
  \BibitemOpen
  \bibfield  {author} {\bibinfo {author} {\bibfnamefont {E.}~\bibnamefont
  {Aprile}} \emph {et~al.} (\bibinfo {collaboration} {XENON}),\ }\href
  {https://doi.org/10.1103/msw4-t342} {\bibfield  {journal} {\bibinfo
  {journal} {Phys. Rev. Lett.}\ }\textbf {\bibinfo {volume} {135}},\ \bibinfo
  {pages} {221003} (\bibinfo {year} {2025})},\ \Eprint
  {https://arxiv.org/abs/2502.18005} {arXiv:2502.18005 [hep-ex]} \BibitemShut
  {NoStop}%
\bibitem [{\citenamefont {Aalbers}\ \emph {et~al.}(2025)\citenamefont {Aalbers}
  \emph {et~al.}}]{LZ:2024zvo}%
  \BibitemOpen
  \bibfield  {author} {\bibinfo {author} {\bibfnamefont {J.}~\bibnamefont
  {Aalbers}} \emph {et~al.} (\bibinfo {collaboration} {LZ}),\ }\href
  {https://doi.org/10.1103/4dyc-z8zf} {\bibfield  {journal} {\bibinfo
  {journal} {Phys. Rev. Lett.}\ }\textbf {\bibinfo {volume} {135}},\ \bibinfo
  {pages} {011802} (\bibinfo {year} {2025})},\ \Eprint
  {https://arxiv.org/abs/2410.17036} {arXiv:2410.17036 [hep-ex]} \BibitemShut
  {NoStop}%
\bibitem [{\citenamefont {Bo}\ \emph {et~al.}(2025)\citenamefont {Bo} \emph
  {et~al.}}]{PandaX:2024qfu}%
  \BibitemOpen
  \bibfield  {author} {\bibinfo {author} {\bibfnamefont {Z.}~\bibnamefont {Bo}}
  \emph {et~al.} (\bibinfo {collaboration} {PandaX}),\ }\href
  {https://doi.org/10.1103/PhysRevLett.134.011805} {\bibfield  {journal}
  {\bibinfo  {journal} {Phys. Rev. Lett.}\ }\textbf {\bibinfo {volume} {134}},\
  \bibinfo {pages} {011805} (\bibinfo {year} {2025})},\ \Eprint
  {https://arxiv.org/abs/2408.00664} {arXiv:2408.00664 [hep-ex]} \BibitemShut
  {NoStop}%
\bibitem [{\citenamefont {Battaglieri}\ \emph {et~al.}(2017)\citenamefont
  {Battaglieri}, \citenamefont {Belloni}, \citenamefont {Chou}, \citenamefont
  {Cushman}, \citenamefont {Echenard}, \citenamefont {Essig},\ and\
  \citenamefont {Others}}]{Battaglieri:2017aum}%
  \BibitemOpen
  \bibfield  {author} {\bibinfo {author} {\bibfnamefont {M.}~\bibnamefont
  {Battaglieri}}, \bibinfo {author} {\bibfnamefont {A.}~\bibnamefont
  {Belloni}}, \bibinfo {author} {\bibfnamefont {A.}~\bibnamefont {Chou}},
  \bibinfo {author} {\bibfnamefont {P.}~\bibnamefont {Cushman}}, \bibinfo
  {author} {\bibfnamefont {B.}~\bibnamefont {Echenard}}, \bibinfo {author}
  {\bibfnamefont {R.}~\bibnamefont {Essig}},\ and\ \bibinfo {author}
  {\bibnamefont {Others}},\ }\href@noop {} {\bibfield  {journal} {\bibinfo
  {journal} {arXiv:1707.04591 [hep-ph]}\ } (\bibinfo {year} {2017})},\ \Eprint
  {https://arxiv.org/abs/1707.04591} {arXiv:1707.04591 [hep-ph]} \BibitemShut
  {NoStop}%
\bibitem [{\citenamefont {Essig}\ \emph {et~al.}(2022)\citenamefont {Essig}
  \emph {et~al.}}]{Essig:2022dfa}%
  \BibitemOpen
  \bibfield  {author} {\bibinfo {author} {\bibfnamefont {R.}~\bibnamefont
  {Essig}} \emph {et~al.},\ }in\ \href@noop {} {\emph {\bibinfo {booktitle}
  {{Snowmass 2021}}}}\ (\bibinfo {year} {2022})\ \Eprint
  {https://arxiv.org/abs/2203.08297} {arXiv:2203.08297 [hep-ph]} \BibitemShut
  {NoStop}%
\bibitem [{\citenamefont {Essig}\ \emph {et~al.}(2012)\citenamefont {Essig},
  \citenamefont {Mardon},\ and\ \citenamefont {Volansky}}]{Essig:2011nj}%
  \BibitemOpen
  \bibfield  {author} {\bibinfo {author} {\bibfnamefont {R.}~\bibnamefont
  {Essig}}, \bibinfo {author} {\bibfnamefont {J.}~\bibnamefont {Mardon}},\ and\
  \bibinfo {author} {\bibfnamefont {T.}~\bibnamefont {Volansky}},\ }\href
  {https://doi.org/10.1103/PhysRevD.85.076007} {\bibfield  {journal} {\bibinfo
  {journal} {Phys. Rev. D}\ }\textbf {\bibinfo {volume} {85}},\ \bibinfo
  {pages} {076007} (\bibinfo {year} {2012})},\ \Eprint
  {https://arxiv.org/abs/1108.5383} {arXiv:1108.5383 [hep-ph]} \BibitemShut
  {NoStop}%
\bibitem [{\citenamefont {Essig}\ \emph {et~al.}(2016)\citenamefont {Essig},
  \citenamefont {Fernandez-Serra}, \citenamefont {Mardon}, \citenamefont
  {Soto}, \citenamefont {Volansky},\ and\ \citenamefont {Yu}}]{Essig:2015cda}%
  \BibitemOpen
  \bibfield  {author} {\bibinfo {author} {\bibfnamefont {R.}~\bibnamefont
  {Essig}}, \bibinfo {author} {\bibfnamefont {M.}~\bibnamefont
  {Fernandez-Serra}}, \bibinfo {author} {\bibfnamefont {J.}~\bibnamefont
  {Mardon}}, \bibinfo {author} {\bibfnamefont {A.}~\bibnamefont {Soto}},
  \bibinfo {author} {\bibfnamefont {T.}~\bibnamefont {Volansky}},\ and\
  \bibinfo {author} {\bibfnamefont {T.-T.}\ \bibnamefont {Yu}},\ }\href
  {https://doi.org/10.1007/JHEP05(2016)046} {\bibfield  {journal} {\bibinfo
  {journal} {JHEP}\ }\textbf {\bibinfo {volume} {05}},\ \bibinfo {pages}
  {046}},\ \Eprint {https://arxiv.org/abs/1509.01598} {arXiv:1509.01598
  [hep-ph]} \BibitemShut {NoStop}%
\bibitem [{\citenamefont {Aggarwal}\ \emph {et~al.}(2025)\citenamefont
  {Aggarwal} \emph {et~al.}}]{DAMIC-M:2025luv}%
  \BibitemOpen
  \bibfield  {author} {\bibinfo {author} {\bibfnamefont {K.}~\bibnamefont
  {Aggarwal}} \emph {et~al.} (\bibinfo {collaboration} {DAMIC-M}),\ }\href
  {https://doi.org/10.1103/2tcc-bqck} {\bibfield  {journal} {\bibinfo
  {journal} {Phys. Rev. Lett.}\ }\textbf {\bibinfo {volume} {135}},\ \bibinfo
  {pages} {071002} (\bibinfo {year} {2025})},\ \Eprint
  {https://arxiv.org/abs/2503.14617} {arXiv:2503.14617 [hep-ex]} \BibitemShut
  {NoStop}%
\bibitem [{\citenamefont {Bloch}\ \emph {et~al.}(2025)\citenamefont {Bloch}
  \emph {et~al.}}]{SENSEI:2024yyt}%
  \BibitemOpen
  \bibfield  {author} {\bibinfo {author} {\bibfnamefont {I.~M.}\ \bibnamefont
  {Bloch}} \emph {et~al.} (\bibinfo {collaboration} {SENSEI}),\ }\href
  {https://doi.org/10.1103/PhysRevLett.134.161002} {\bibfield  {journal}
  {\bibinfo  {journal} {Phys. Rev. Lett.}\ }\textbf {\bibinfo {volume} {134}},\
  \bibinfo {pages} {161002} (\bibinfo {year} {2025})},\ \Eprint
  {https://arxiv.org/abs/2410.18716} {arXiv:2410.18716 [astro-ph.CO]}
  \BibitemShut {NoStop}%
\bibitem [{\citenamefont {Adari}\ \emph {et~al.}(2025)\citenamefont {Adari}
  \emph {et~al.}}]{SENSEI:2023zdf}%
  \BibitemOpen
  \bibfield  {author} {\bibinfo {author} {\bibfnamefont {P.}~\bibnamefont
  {Adari}} \emph {et~al.} (\bibinfo {collaboration} {SENSEI}),\ }\href
  {https://doi.org/10.1103/PhysRevLett.134.011804} {\bibfield  {journal}
  {\bibinfo  {journal} {Phys. Rev. Lett.}\ }\textbf {\bibinfo {volume} {134}},\
  \bibinfo {pages} {011804} (\bibinfo {year} {2025})},\ \Eprint
  {https://arxiv.org/abs/2312.13342} {arXiv:2312.13342 [astro-ph.CO]}
  \BibitemShut {NoStop}%
\bibitem [{\citenamefont {Aguilar-Arevalo}\ \emph {et~al.}(2022)\citenamefont
  {Aguilar-Arevalo} \emph {et~al.}}]{Oscura:2022vmi}%
  \BibitemOpen
  \bibfield  {author} {\bibinfo {author} {\bibfnamefont {A.}~\bibnamefont
  {Aguilar-Arevalo}} \emph {et~al.} (\bibinfo {collaboration} {Oscura}),\
  }\href@noop {} {\  (\bibinfo {year} {2022})},\ \Eprint
  {https://arxiv.org/abs/2202.10518} {arXiv:2202.10518 [astro-ph.IM]}
  \BibitemShut {NoStop}%
\bibitem [{\citenamefont {Derenzo}\ \emph {et~al.}(2017)\citenamefont
  {Derenzo}, \citenamefont {Essig}, \citenamefont {Massari}, \citenamefont
  {Soto},\ and\ \citenamefont {Yu}}]{Derenzo:2016fse}%
  \BibitemOpen
  \bibfield  {author} {\bibinfo {author} {\bibfnamefont {S.}~\bibnamefont
  {Derenzo}}, \bibinfo {author} {\bibfnamefont {R.}~\bibnamefont {Essig}},
  \bibinfo {author} {\bibfnamefont {A.}~\bibnamefont {Massari}}, \bibinfo
  {author} {\bibfnamefont {A.}~\bibnamefont {Soto}},\ and\ \bibinfo {author}
  {\bibfnamefont {T.-T.}\ \bibnamefont {Yu}},\ }\href
  {https://doi.org/10.1103/PhysRevD.96.016026} {\bibfield  {journal} {\bibinfo
  {journal} {Phys. Rev. D}\ }\textbf {\bibinfo {volume} {96}},\ \bibinfo
  {pages} {016026} (\bibinfo {year} {2017})},\ \Eprint
  {https://arxiv.org/abs/1607.01009} {arXiv:1607.01009 [hep-ph]} \BibitemShut
  {NoStop}%
\bibitem [{\citenamefont {Blanco}\ \emph {et~al.}(2020)\citenamefont {Blanco},
  \citenamefont {Collar}, \citenamefont {Kahn},\ and\ \citenamefont
  {Lillard}}]{Blanco:2019lrf}%
  \BibitemOpen
  \bibfield  {author} {\bibinfo {author} {\bibfnamefont {C.}~\bibnamefont
  {Blanco}}, \bibinfo {author} {\bibfnamefont {J.~I.}\ \bibnamefont {Collar}},
  \bibinfo {author} {\bibfnamefont {Y.}~\bibnamefont {Kahn}},\ and\ \bibinfo
  {author} {\bibfnamefont {B.}~\bibnamefont {Lillard}},\ }\href
  {https://doi.org/10.1103/PhysRevD.101.056001} {\bibfield  {journal} {\bibinfo
   {journal} {Phys. Rev. D}\ }\textbf {\bibinfo {volume} {101}},\ \bibinfo
  {pages} {056001} (\bibinfo {year} {2020})},\ \Eprint
  {https://arxiv.org/abs/1912.02822} {arXiv:1912.02822 [hep-ph]} \BibitemShut
  {NoStop}%
\bibitem [{\citenamefont {Arnaud}\ \emph {et~al.}(2020)\citenamefont {Arnaud}
  \emph {et~al.}}]{EDELWEISS:2020fxc}%
  \BibitemOpen
  \bibfield  {author} {\bibinfo {author} {\bibfnamefont {Q.}~\bibnamefont
  {Arnaud}} \emph {et~al.} (\bibinfo {collaboration} {EDELWEISS}),\ }\href
  {https://doi.org/10.1103/PhysRevLett.125.141301} {\bibfield  {journal}
  {\bibinfo  {journal} {Phys. Rev. Lett.}\ }\textbf {\bibinfo {volume} {125}},\
  \bibinfo {pages} {141301} (\bibinfo {year} {2020})},\ \Eprint
  {https://arxiv.org/abs/2003.01046} {arXiv:2003.01046 [astro-ph.GA]}
  \BibitemShut {NoStop}%
\bibitem [{\citenamefont {Albakry}\ \emph {et~al.}(2025)\citenamefont {Albakry}
  \emph {et~al.}}]{SuperCDMS:2025dha}%
  \BibitemOpen
  \bibfield  {author} {\bibinfo {author} {\bibfnamefont {M.~F.}\ \bibnamefont
  {Albakry}} \emph {et~al.} (\bibinfo {collaboration} {SuperCDMS}),\
  }\href@noop {} {\  (\bibinfo {year} {2025})},\ \Eprint
  {https://arxiv.org/abs/2509.03608} {arXiv:2509.03608 [hep-ex]} \BibitemShut
  {NoStop}%
\bibitem [{\citenamefont {Aprile}\ \emph {et~al.}(2026)\citenamefont {Aprile}
  \emph {et~al.}}]{XENON:2026qow}%
  \BibitemOpen
  \bibfield  {author} {\bibinfo {author} {\bibfnamefont {E.}~\bibnamefont
  {Aprile}} \emph {et~al.} (\bibinfo {collaboration} {XENON}),\ }\href@noop {}
  {\  (\bibinfo {year} {2026})},\ \Eprint {https://arxiv.org/abs/2601.11296}
  {arXiv:2601.11296 [hep-ex]} \BibitemShut {NoStop}%
\bibitem [{\citenamefont {Li}\ \emph {et~al.}(2023)\citenamefont {Li},
  \citenamefont {Wu}, \citenamefont {Abdukerim}, \citenamefont {Bo},
  \citenamefont {Chen}, \citenamefont {Chen}, \citenamefont {Chen},
  \citenamefont {Cheng}, \citenamefont {Cheng}, \citenamefont {Cui},
  \citenamefont {Fan}, \citenamefont {Fang}, \citenamefont {Fu}, \citenamefont
  {Fu}, \citenamefont {Geng}, \citenamefont {Giboni}, \citenamefont {Gu},
  \citenamefont {Guo}, \citenamefont {Han}, \citenamefont {Han}, \citenamefont
  {He}, \citenamefont {He}, \citenamefont {Huang}, \citenamefont {Huang},
  \citenamefont {Huang}, \citenamefont {Hou}, \citenamefont {Ji}, \citenamefont
  {Ju}, \citenamefont {Li}, \citenamefont {Li}, \citenamefont {Li},
  \citenamefont {Li}, \citenamefont {Lin}, \citenamefont {Liu}, \citenamefont
  {Lu}, \citenamefont {Luo}, \citenamefont {Luo}, \citenamefont {Ma},
  \citenamefont {Ma}, \citenamefont {Mao}, \citenamefont {Meng}, \citenamefont
  {Ning}, \citenamefont {Qi}, \citenamefont {Qian}, \citenamefont {Ren},
  \citenamefont {Shaheed}, \citenamefont {Shang}, \citenamefont {Shang},
  \citenamefont {Shen}, \citenamefont {Si}, \citenamefont {Sun}, \citenamefont
  {Tan}, \citenamefont {Tao}, \citenamefont {Wang}, \citenamefont {Wang},
  \citenamefont {Wang}, \citenamefont {Wang}, \citenamefont {Wang},
  \citenamefont {Wang}, \citenamefont {Wang}, \citenamefont {Wang},
  \citenamefont {Wei}, \citenamefont {Wu}, \citenamefont {Xia}, \citenamefont
  {Xiao}, \citenamefont {Xiao}, \citenamefont {Xie}, \citenamefont {Yan},
  \citenamefont {Yan}, \citenamefont {Yang}, \citenamefont {Yang},
  \citenamefont {Yao}, \citenamefont {You}, \citenamefont {Yu}, \citenamefont
  {Yuan}, \citenamefont {Yuan}, \citenamefont {Yuan}, \citenamefont {Zeng},
  \citenamefont {Zhang}, \citenamefont {Zhang}, \citenamefont {Zhang},
  \citenamefont {Zhang}, \citenamefont {Zhang}, \citenamefont {Zhang},
  \citenamefont {Zhang}, \citenamefont {Zhang}, \citenamefont {Zhang},
  \citenamefont {Zhao}, \citenamefont {Zheng}, \citenamefont {Zhou},
  \citenamefont {Zhou}, \citenamefont {Zhou}, \citenamefont {Zhou},\ and\
  \citenamefont {Zhou}}]{PhysRevLett.130.261001}%
  \BibitemOpen
  \bibfield  {author} {\bibinfo {author} {\bibfnamefont {S.}~\bibnamefont
  {Li}}, \bibinfo {author} {\bibfnamefont {M.}~\bibnamefont {Wu}}, \bibinfo
  {author} {\bibfnamefont {A.}~\bibnamefont {Abdukerim}}, \bibinfo {author}
  {\bibfnamefont {Z.}~\bibnamefont {Bo}}, \bibinfo {author} {\bibfnamefont
  {W.}~\bibnamefont {Chen}}, \bibinfo {author} {\bibfnamefont {X.}~\bibnamefont
  {Chen}}, \bibinfo {author} {\bibfnamefont {Y.}~\bibnamefont {Chen}}, \bibinfo
  {author} {\bibfnamefont {C.}~\bibnamefont {Cheng}}, \bibinfo {author}
  {\bibfnamefont {Z.}~\bibnamefont {Cheng}}, \bibinfo {author} {\bibfnamefont
  {X.}~\bibnamefont {Cui}}, \bibinfo {author} {\bibfnamefont {Y.}~\bibnamefont
  {Fan}}, \bibinfo {author} {\bibfnamefont {D.}~\bibnamefont {Fang}}, \bibinfo
  {author} {\bibfnamefont {C.}~\bibnamefont {Fu}}, \bibinfo {author}
  {\bibfnamefont {M.}~\bibnamefont {Fu}}, \bibinfo {author} {\bibfnamefont
  {L.}~\bibnamefont {Geng}}, \bibinfo {author} {\bibfnamefont {K.}~\bibnamefont
  {Giboni}}, \bibinfo {author} {\bibfnamefont {L.}~\bibnamefont {Gu}}, \bibinfo
  {author} {\bibfnamefont {X.}~\bibnamefont {Guo}}, \bibinfo {author}
  {\bibfnamefont {C.}~\bibnamefont {Han}}, \bibinfo {author} {\bibfnamefont
  {K.}~\bibnamefont {Han}}, \bibinfo {author} {\bibfnamefont {C.}~\bibnamefont
  {He}}, \bibinfo {author} {\bibfnamefont {J.}~\bibnamefont {He}}, \bibinfo
  {author} {\bibfnamefont {D.}~\bibnamefont {Huang}}, \bibinfo {author}
  {\bibfnamefont {Y.}~\bibnamefont {Huang}}, \bibinfo {author} {\bibfnamefont
  {Z.}~\bibnamefont {Huang}}, \bibinfo {author} {\bibfnamefont
  {R.}~\bibnamefont {Hou}}, \bibinfo {author} {\bibfnamefont {X.}~\bibnamefont
  {Ji}}, \bibinfo {author} {\bibfnamefont {Y.}~\bibnamefont {Ju}}, \bibinfo
  {author} {\bibfnamefont {C.}~\bibnamefont {Li}}, \bibinfo {author}
  {\bibfnamefont {J.}~\bibnamefont {Li}}, \bibinfo {author} {\bibfnamefont
  {M.}~\bibnamefont {Li}}, \bibinfo {author} {\bibfnamefont {S.}~\bibnamefont
  {Li}}, \bibinfo {author} {\bibfnamefont {Q.}~\bibnamefont {Lin}}, \bibinfo
  {author} {\bibfnamefont {J.}~\bibnamefont {Liu}}, \bibinfo {author}
  {\bibfnamefont {X.}~\bibnamefont {Lu}}, \bibinfo {author} {\bibfnamefont
  {L.}~\bibnamefont {Luo}}, \bibinfo {author} {\bibfnamefont {Y.}~\bibnamefont
  {Luo}}, \bibinfo {author} {\bibfnamefont {W.}~\bibnamefont {Ma}}, \bibinfo
  {author} {\bibfnamefont {Y.}~\bibnamefont {Ma}}, \bibinfo {author}
  {\bibfnamefont {Y.}~\bibnamefont {Mao}}, \bibinfo {author} {\bibfnamefont
  {Y.}~\bibnamefont {Meng}}, \bibinfo {author} {\bibfnamefont {X.}~\bibnamefont
  {Ning}}, \bibinfo {author} {\bibfnamefont {N.}~\bibnamefont {Qi}}, \bibinfo
  {author} {\bibfnamefont {Z.}~\bibnamefont {Qian}}, \bibinfo {author}
  {\bibfnamefont {X.}~\bibnamefont {Ren}}, \bibinfo {author} {\bibfnamefont
  {N.}~\bibnamefont {Shaheed}}, \bibinfo {author} {\bibfnamefont
  {C.}~\bibnamefont {Shang}}, \bibinfo {author} {\bibfnamefont
  {X.}~\bibnamefont {Shang}}, \bibinfo {author} {\bibfnamefont
  {G.}~\bibnamefont {Shen}}, \bibinfo {author} {\bibfnamefont {L.}~\bibnamefont
  {Si}}, \bibinfo {author} {\bibfnamefont {W.}~\bibnamefont {Sun}}, \bibinfo
  {author} {\bibfnamefont {A.}~\bibnamefont {Tan}}, \bibinfo {author}
  {\bibfnamefont {Y.}~\bibnamefont {Tao}}, \bibinfo {author} {\bibfnamefont
  {A.}~\bibnamefont {Wang}}, \bibinfo {author} {\bibfnamefont {M.}~\bibnamefont
  {Wang}}, \bibinfo {author} {\bibfnamefont {Q.}~\bibnamefont {Wang}}, \bibinfo
  {author} {\bibfnamefont {S.}~\bibnamefont {Wang}}, \bibinfo {author}
  {\bibfnamefont {S.}~\bibnamefont {Wang}}, \bibinfo {author} {\bibfnamefont
  {W.}~\bibnamefont {Wang}}, \bibinfo {author} {\bibfnamefont {X.}~\bibnamefont
  {Wang}}, \bibinfo {author} {\bibfnamefont {Z.}~\bibnamefont {Wang}}, \bibinfo
  {author} {\bibfnamefont {Y.}~\bibnamefont {Wei}}, \bibinfo {author}
  {\bibfnamefont {W.}~\bibnamefont {Wu}}, \bibinfo {author} {\bibfnamefont
  {J.}~\bibnamefont {Xia}}, \bibinfo {author} {\bibfnamefont {M.}~\bibnamefont
  {Xiao}}, \bibinfo {author} {\bibfnamefont {X.}~\bibnamefont {Xiao}}, \bibinfo
  {author} {\bibfnamefont {P.}~\bibnamefont {Xie}}, \bibinfo {author}
  {\bibfnamefont {B.}~\bibnamefont {Yan}}, \bibinfo {author} {\bibfnamefont
  {X.}~\bibnamefont {Yan}}, \bibinfo {author} {\bibfnamefont {J.}~\bibnamefont
  {Yang}}, \bibinfo {author} {\bibfnamefont {Y.}~\bibnamefont {Yang}}, \bibinfo
  {author} {\bibfnamefont {Y.}~\bibnamefont {Yao}}, \bibinfo {author}
  {\bibfnamefont {Z.}~\bibnamefont {You}}, \bibinfo {author} {\bibfnamefont
  {C.}~\bibnamefont {Yu}}, \bibinfo {author} {\bibfnamefont {J.}~\bibnamefont
  {Yuan}}, \bibinfo {author} {\bibfnamefont {Y.}~\bibnamefont {Yuan}}, \bibinfo
  {author} {\bibfnamefont {Z.}~\bibnamefont {Yuan}}, \bibinfo {author}
  {\bibfnamefont {X.}~\bibnamefont {Zeng}}, \bibinfo {author} {\bibfnamefont
  {D.}~\bibnamefont {Zhang}}, \bibinfo {author} {\bibfnamefont
  {M.}~\bibnamefont {Zhang}}, \bibinfo {author} {\bibfnamefont
  {P.}~\bibnamefont {Zhang}}, \bibinfo {author} {\bibfnamefont
  {S.}~\bibnamefont {Zhang}}, \bibinfo {author} {\bibfnamefont
  {S.}~\bibnamefont {Zhang}}, \bibinfo {author} {\bibfnamefont
  {T.}~\bibnamefont {Zhang}}, \bibinfo {author} {\bibfnamefont
  {Y.}~\bibnamefont {Zhang}}, \bibinfo {author} {\bibfnamefont
  {Y.}~\bibnamefont {Zhang}}, \bibinfo {author} {\bibfnamefont
  {Y.}~\bibnamefont {Zhang}}, \bibinfo {author} {\bibfnamefont
  {L.}~\bibnamefont {Zhao}}, \bibinfo {author} {\bibfnamefont {Q.}~\bibnamefont
  {Zheng}}, \bibinfo {author} {\bibfnamefont {J.}~\bibnamefont {Zhou}},
  \bibinfo {author} {\bibfnamefont {N.}~\bibnamefont {Zhou}}, \bibinfo {author}
  {\bibfnamefont {X.}~\bibnamefont {Zhou}}, \bibinfo {author} {\bibfnamefont
  {Y.}~\bibnamefont {Zhou}},\ and\ \bibinfo {author} {\bibfnamefont
  {Y.}~\bibnamefont {Zhou}} (\bibinfo {collaboration} {PandaX Collaboration}),\
  }\href {https://doi.org/10.1103/PhysRevLett.130.261001} {\bibfield  {journal}
  {\bibinfo  {journal} {Phys. Rev. Lett.}\ }\textbf {\bibinfo {volume} {130}},\
  \bibinfo {pages} {261001} (\bibinfo {year} {2023})}\BibitemShut {NoStop}%
\bibitem [{\citenamefont {Zhang}\ \emph {et~al.}(2025)\citenamefont {Zhang}
  \emph {et~al.}}]{PandaX:2025rrz}%
  \BibitemOpen
  \bibfield  {author} {\bibinfo {author} {\bibfnamefont {M.}~\bibnamefont
  {Zhang}} \emph {et~al.} (\bibinfo {collaboration} {PandaX}),\ }\href
  {https://doi.org/10.1103/rtnh-jn8s} {\bibfield  {journal} {\bibinfo
  {journal} {Phys. Rev. Lett.}\ }\textbf {\bibinfo {volume} {135}},\ \bibinfo
  {pages} {211001} (\bibinfo {year} {2025})},\ \bibinfo {note} {[Erratum:
  Phys.Rev.Lett. 136, 069901 (2026)]},\ \Eprint
  {https://arxiv.org/abs/2507.11930} {arXiv:2507.11930 [hep-ex]} \BibitemShut
  {NoStop}%
\bibitem [{\citenamefont {Blanco}\ \emph {et~al.}(2021)\citenamefont {Blanco},
  \citenamefont {Kahn}, \citenamefont {Lillard},\ and\ \citenamefont
  {McDermott}}]{Blanco:2021hlm}%
  \BibitemOpen
  \bibfield  {author} {\bibinfo {author} {\bibfnamefont {C.}~\bibnamefont
  {Blanco}}, \bibinfo {author} {\bibfnamefont {Y.}~\bibnamefont {Kahn}},
  \bibinfo {author} {\bibfnamefont {B.}~\bibnamefont {Lillard}},\ and\ \bibinfo
  {author} {\bibfnamefont {S.~D.}\ \bibnamefont {McDermott}},\ }\href
  {https://doi.org/10.1103/PhysRevD.104.036011} {\bibfield  {journal} {\bibinfo
   {journal} {Phys. Rev. D}\ }\textbf {\bibinfo {volume} {104}},\ \bibinfo
  {pages} {036011} (\bibinfo {year} {2021})},\ \Eprint
  {https://arxiv.org/abs/2103.08601} {arXiv:2103.08601 [hep-ph]} \BibitemShut
  {NoStop}%
\bibitem [{\citenamefont {Blanco}\ \emph {et~al.}(2022)\citenamefont {Blanco},
  \citenamefont {Harris}, \citenamefont {Kahn}, \citenamefont {Lillard},\ and\
  \citenamefont {P{\'e}rez-R{\'\i}os}}]{Blanco:2022pkt}%
  \BibitemOpen
  \bibfield  {author} {\bibinfo {author} {\bibfnamefont {C.}~\bibnamefont
  {Blanco}}, \bibinfo {author} {\bibfnamefont {I.}~\bibnamefont {Harris}},
  \bibinfo {author} {\bibfnamefont {Y.}~\bibnamefont {Kahn}}, \bibinfo {author}
  {\bibfnamefont {B.}~\bibnamefont {Lillard}},\ and\ \bibinfo {author}
  {\bibfnamefont {J.}~\bibnamefont {P{\'e}rez-R{\'\i}os}},\ }\href
  {https://doi.org/10.1103/PhysRevD.106.115015} {\bibfield  {journal} {\bibinfo
   {journal} {Phys. Rev. D}\ }\textbf {\bibinfo {volume} {106}},\ \bibinfo
  {pages} {115015} (\bibinfo {year} {2022})},\ \Eprint
  {https://arxiv.org/abs/2208.09002} {arXiv:2208.09002 [hep-ph]} \BibitemShut
  {NoStop}%
\bibitem [{\citenamefont {Blanco}\ \emph {et~al.}(2023)\citenamefont {Blanco},
  \citenamefont {Essig}, \citenamefont {Fernandez-Serra}, \citenamefont
  {Ramani},\ and\ \citenamefont {Slone}}]{Blanco:2022cel}%
  \BibitemOpen
  \bibfield  {author} {\bibinfo {author} {\bibfnamefont {C.}~\bibnamefont
  {Blanco}}, \bibinfo {author} {\bibfnamefont {R.}~\bibnamefont {Essig}},
  \bibinfo {author} {\bibfnamefont {M.}~\bibnamefont {Fernandez-Serra}},
  \bibinfo {author} {\bibfnamefont {H.}~\bibnamefont {Ramani}},\ and\ \bibinfo
  {author} {\bibfnamefont {O.}~\bibnamefont {Slone}},\ }\href
  {https://doi.org/10.1103/PhysRevD.107.095035} {\bibfield  {journal} {\bibinfo
   {journal} {Phys. Rev. D}\ }\textbf {\bibinfo {volume} {107}},\ \bibinfo
  {pages} {095035} (\bibinfo {year} {2023})},\ \Eprint
  {https://arxiv.org/abs/2208.05967} {arXiv:2208.05967 [hep-ph]} \BibitemShut
  {NoStop}%
\bibitem [{\citenamefont {Tiffenberg}\ \emph {et~al.}(2017)\citenamefont
  {Tiffenberg}, \citenamefont {Sofo-Haro}, \citenamefont {Drlica-Wagner},
  \citenamefont {Essig}, \citenamefont {Guardincerri}, \citenamefont {Holland},
  \citenamefont {Volansky},\ and\ \citenamefont {Yu}}]{Tiffenberg:2017aac}%
  \BibitemOpen
  \bibfield  {author} {\bibinfo {author} {\bibfnamefont {J.}~\bibnamefont
  {Tiffenberg}}, \bibinfo {author} {\bibfnamefont {M.}~\bibnamefont
  {Sofo-Haro}}, \bibinfo {author} {\bibfnamefont {A.}~\bibnamefont
  {Drlica-Wagner}}, \bibinfo {author} {\bibfnamefont {R.}~\bibnamefont
  {Essig}}, \bibinfo {author} {\bibfnamefont {Y.}~\bibnamefont {Guardincerri}},
  \bibinfo {author} {\bibfnamefont {S.}~\bibnamefont {Holland}}, \bibinfo
  {author} {\bibfnamefont {T.}~\bibnamefont {Volansky}},\ and\ \bibinfo
  {author} {\bibfnamefont {T.-T.}\ \bibnamefont {Yu}} (\bibinfo {collaboration}
  {SENSEI}),\ }\href {https://doi.org/10.1103/PhysRevLett.119.131802}
  {\bibfield  {journal} {\bibinfo  {journal} {Phys. Rev. Lett.}\ }\textbf
  {\bibinfo {volume} {119}},\ \bibinfo {pages} {131802} (\bibinfo {year}
  {2017})},\ \Eprint {https://arxiv.org/abs/1706.00028} {arXiv:1706.00028
  [physics.ins-det]} \BibitemShut {NoStop}%
\bibitem [{\citenamefont {Brus}(1984)}]{Brus1984}%
  \BibitemOpen
  \bibfield  {author} {\bibinfo {author} {\bibfnamefont {L.~E.}\ \bibnamefont
  {Brus}},\ }\href {https://doi.org/10.1063/1.447218} {\bibfield  {journal}
  {\bibinfo  {journal} {The Journal of Chemical Physics}\ }\textbf {\bibinfo
  {volume} {80}},\ \bibinfo {pages} {4403} (\bibinfo {year}
  {1984})}\BibitemShut {NoStop}%
\bibitem [{\citenamefont {Alivisatos}(1996)}]{Alivisatos1996}%
  \BibitemOpen
  \bibfield  {author} {\bibinfo {author} {\bibfnamefont {A.~P.}\ \bibnamefont
  {Alivisatos}},\ }\href {https://doi.org/10.1126/science.271.5251.933}
  {\bibfield  {journal} {\bibinfo  {journal} {Science}\ }\textbf {\bibinfo
  {volume} {271}},\ \bibinfo {pages} {933} (\bibinfo {year}
  {1996})}\BibitemShut {NoStop}%
\bibitem [{\citenamefont {Efros}\ and\ \citenamefont
  {Brus}(2021)}]{EfrosBrus2021}%
  \BibitemOpen
  \bibfield  {author} {\bibinfo {author} {\bibfnamefont {A.~L.}\ \bibnamefont
  {Efros}}\ and\ \bibinfo {author} {\bibfnamefont {L.~E.}\ \bibnamefont
  {Brus}},\ }\href {https://doi.org/10.1021/acsnano.1c01399} {\bibfield
  {journal} {\bibinfo  {journal} {ACS Nano}\ }\textbf {\bibinfo {volume}
  {15}},\ \bibinfo {pages} {6192} (\bibinfo {year} {2021})}\BibitemShut
  {NoStop}%
\bibitem [{\citenamefont {Murray}\ \emph {et~al.}(1993)\citenamefont {Murray},
  \citenamefont {Norris},\ and\ \citenamefont {Bawendi}}]{Murray1993}%
  \BibitemOpen
  \bibfield  {author} {\bibinfo {author} {\bibfnamefont {C.~B.}\ \bibnamefont
  {Murray}}, \bibinfo {author} {\bibfnamefont {D.~J.}\ \bibnamefont {Norris}},\
  and\ \bibinfo {author} {\bibfnamefont {M.~G.}\ \bibnamefont {Bawendi}},\
  }\href {https://doi.org/10.1021/ja00072a025} {\bibfield  {journal} {\bibinfo
  {journal} {Journal of the American Chemical Society}\ }\textbf {\bibinfo
  {volume} {115}},\ \bibinfo {pages} {8706} (\bibinfo {year}
  {1993})}\BibitemShut {NoStop}%
\bibitem [{\citenamefont {Wolkin}\ \emph {et~al.}(1999)\citenamefont {Wolkin},
  \citenamefont {Jorne}, \citenamefont {Fauchet}, \citenamefont {Allan},\ and\
  \citenamefont {Delerue}}]{wolkin1999electronic}%
  \BibitemOpen
  \bibfield  {author} {\bibinfo {author} {\bibfnamefont {M.}~\bibnamefont
  {Wolkin}}, \bibinfo {author} {\bibfnamefont {J.}~\bibnamefont {Jorne}},
  \bibinfo {author} {\bibfnamefont {P.}~\bibnamefont {Fauchet}}, \bibinfo
  {author} {\bibfnamefont {G.}~\bibnamefont {Allan}},\ and\ \bibinfo {author}
  {\bibfnamefont {C.}~\bibnamefont {Delerue}},\ }\href@noop {} {\bibfield
  {journal} {\bibinfo  {journal} {Physical review letters}\ }\textbf {\bibinfo
  {volume} {82}},\ \bibinfo {pages} {197} (\bibinfo {year} {1999})}\BibitemShut
  {NoStop}%
\bibitem [{\citenamefont {Kanemitsu}\ and\ \citenamefont
  {Okamoto}(1997)}]{kanemitsu1997visible}%
  \BibitemOpen
  \bibfield  {author} {\bibinfo {author} {\bibfnamefont {Y.}~\bibnamefont
  {Kanemitsu}}\ and\ \bibinfo {author} {\bibfnamefont {S.}~\bibnamefont
  {Okamoto}},\ }\href@noop {} {\bibfield  {journal} {\bibinfo  {journal}
  {Materials Science and Engineering: B}\ }\textbf {\bibinfo {volume} {48}},\
  \bibinfo {pages} {108} (\bibinfo {year} {1997})}\BibitemShut {NoStop}%
\bibitem [{\citenamefont {Hannah}\ \emph {et~al.}(2012)\citenamefont {Hannah},
  \citenamefont {Yang}, \citenamefont {Podsiadlo}, \citenamefont {Chan},
  \citenamefont {Demortiere}, \citenamefont {Gosztola}, \citenamefont
  {Prakapenka}, \citenamefont {Schatz}, \citenamefont {Kortshagen},\ and\
  \citenamefont {Schaller}}]{hannah2012origin}%
  \BibitemOpen
  \bibfield  {author} {\bibinfo {author} {\bibfnamefont {D.~C.}\ \bibnamefont
  {Hannah}}, \bibinfo {author} {\bibfnamefont {J.}~\bibnamefont {Yang}},
  \bibinfo {author} {\bibfnamefont {P.}~\bibnamefont {Podsiadlo}}, \bibinfo
  {author} {\bibfnamefont {M.~K.}\ \bibnamefont {Chan}}, \bibinfo {author}
  {\bibfnamefont {A.}~\bibnamefont {Demortiere}}, \bibinfo {author}
  {\bibfnamefont {D.~J.}\ \bibnamefont {Gosztola}}, \bibinfo {author}
  {\bibfnamefont {V.~B.}\ \bibnamefont {Prakapenka}}, \bibinfo {author}
  {\bibfnamefont {G.~C.}\ \bibnamefont {Schatz}}, \bibinfo {author}
  {\bibfnamefont {U.}~\bibnamefont {Kortshagen}},\ and\ \bibinfo {author}
  {\bibfnamefont {R.~D.}\ \bibnamefont {Schaller}},\ }\href@noop {} {\bibfield
  {journal} {\bibinfo  {journal} {Nano letters}\ }\textbf {\bibinfo {volume}
  {12}},\ \bibinfo {pages} {4200} (\bibinfo {year} {2012})}\BibitemShut
  {NoStop}%
\bibitem [{\citenamefont {Takai}\ \emph {et~al.}(2017)\citenamefont {Takai},
  \citenamefont {Ikeda}, \citenamefont {Yamasaki},\ and\ \citenamefont
  {Kaneta}}]{takai2017size}%
  \BibitemOpen
  \bibfield  {author} {\bibinfo {author} {\bibfnamefont {K.}~\bibnamefont
  {Takai}}, \bibinfo {author} {\bibfnamefont {M.}~\bibnamefont {Ikeda}},
  \bibinfo {author} {\bibfnamefont {T.}~\bibnamefont {Yamasaki}},\ and\
  \bibinfo {author} {\bibfnamefont {C.}~\bibnamefont {Kaneta}},\ }\href@noop {}
  {\bibfield  {journal} {\bibinfo  {journal} {Journal of Physics
  Communications}\ }\textbf {\bibinfo {volume} {1}},\ \bibinfo {pages} {045010}
  (\bibinfo {year} {2017})}\BibitemShut {NoStop}%
\bibitem [{\citenamefont {Gert}\ \emph {et~al.}(2017)\citenamefont {Gert},
  \citenamefont {Nestoklon}, \citenamefont {Prokofiev},\ and\ \citenamefont
  {Yassievich}}]{gert2017tight}%
  \BibitemOpen
  \bibfield  {author} {\bibinfo {author} {\bibfnamefont {A.~V.}\ \bibnamefont
  {Gert}}, \bibinfo {author} {\bibfnamefont {M.~O.}\ \bibnamefont {Nestoklon}},
  \bibinfo {author} {\bibfnamefont {A.}~\bibnamefont {Prokofiev}},\ and\
  \bibinfo {author} {\bibfnamefont {I.~N.}\ \bibnamefont {Yassievich}},\
  }\href@noop {} {\bibfield  {journal} {\bibinfo  {journal} {Semiconductors}\
  }\textbf {\bibinfo {volume} {51}},\ \bibinfo {pages} {1274} (\bibinfo {year}
  {2017})}\BibitemShut {NoStop}%
\bibitem [{\citenamefont {Dohnalov{\'a}}\ \emph {et~al.}(2013)\citenamefont
  {Dohnalov{\'a}}, \citenamefont {Poddubny}, \citenamefont {Prokofiev},
  \citenamefont {De~Boer}, \citenamefont {Umesh}, \citenamefont {Paulusse},
  \citenamefont {Zuilhof},\ and\ \citenamefont
  {Gregorkiewicz}}]{dohnalova2013surface}%
  \BibitemOpen
  \bibfield  {author} {\bibinfo {author} {\bibfnamefont {K.}~\bibnamefont
  {Dohnalov{\'a}}}, \bibinfo {author} {\bibfnamefont {A.~N.}\ \bibnamefont
  {Poddubny}}, \bibinfo {author} {\bibfnamefont {A.~A.}\ \bibnamefont
  {Prokofiev}}, \bibinfo {author} {\bibfnamefont {W.~D.}\ \bibnamefont
  {De~Boer}}, \bibinfo {author} {\bibfnamefont {C.~P.}\ \bibnamefont {Umesh}},
  \bibinfo {author} {\bibfnamefont {J.~M.}\ \bibnamefont {Paulusse}}, \bibinfo
  {author} {\bibfnamefont {H.}~\bibnamefont {Zuilhof}},\ and\ \bibinfo {author}
  {\bibfnamefont {T.}~\bibnamefont {Gregorkiewicz}},\ }\href
  {https://doi.org/10.1038/lsa.2013.3} {\bibfield  {journal} {\bibinfo
  {journal} {Light: science \& applications}\ }\textbf {\bibinfo {volume}
  {2}},\ \bibinfo {pages} {e47} (\bibinfo {year} {2013})}\BibitemShut {NoStop}%
\bibitem [{\citenamefont {Reboredo}\ and\ \citenamefont
  {Galli}(2005)}]{reboredo2005theory}%
  \BibitemOpen
  \bibfield  {author} {\bibinfo {author} {\bibfnamefont {F.~A.}\ \bibnamefont
  {Reboredo}}\ and\ \bibinfo {author} {\bibfnamefont {G.}~\bibnamefont
  {Galli}},\ }\href@noop {} {\bibfield  {journal} {\bibinfo  {journal} {The
  Journal of Physical Chemistry B}\ }\textbf {\bibinfo {volume} {109}},\
  \bibinfo {pages} {1072} (\bibinfo {year} {2005})}\BibitemShut {NoStop}%
\bibitem [{\citenamefont {Sangghaleh}\ \emph {et~al.}(2015)\citenamefont
  {Sangghaleh}, \citenamefont {Sychugov}, \citenamefont {Yang}, \citenamefont
  {Veinot},\ and\ \citenamefont {Linnros}}]{Sangghaleh_2015}%
  \BibitemOpen
  \bibfield  {author} {\bibinfo {author} {\bibfnamefont {F.}~\bibnamefont
  {Sangghaleh}}, \bibinfo {author} {\bibfnamefont {I.}~\bibnamefont
  {Sychugov}}, \bibinfo {author} {\bibfnamefont {Z.}~\bibnamefont {Yang}},
  \bibinfo {author} {\bibfnamefont {J.~G.~C.}\ \bibnamefont {Veinot}},\ and\
  \bibinfo {author} {\bibfnamefont {J.}~\bibnamefont {Linnros}},\ }\href
  {https://doi.org/10.1021/acsnano.5b01717} {\bibfield  {journal} {\bibinfo
  {journal} {ACS Nano}\ }\textbf {\bibinfo {volume} {9}},\ \bibinfo {pages}
  {7097} (\bibinfo {year} {2015})}\BibitemShut {NoStop}%
\bibitem [{\citenamefont {Catena}\ \emph {et~al.}(2021)\citenamefont {Catena},
  \citenamefont {Emken}, \citenamefont {Matas}, \citenamefont {Spaldin},\ and\
  \citenamefont {Urdshals}}]{Catena:2021qsr}%
  \BibitemOpen
  \bibfield  {author} {\bibinfo {author} {\bibfnamefont {R.}~\bibnamefont
  {Catena}}, \bibinfo {author} {\bibfnamefont {T.}~\bibnamefont {Emken}},
  \bibinfo {author} {\bibfnamefont {M.}~\bibnamefont {Matas}}, \bibinfo
  {author} {\bibfnamefont {N.~A.}\ \bibnamefont {Spaldin}},\ and\ \bibinfo
  {author} {\bibfnamefont {E.}~\bibnamefont {Urdshals}},\ }\href
  {https://doi.org/10.1103/PhysRevResearch.3.033149} {\bibfield  {journal}
  {\bibinfo  {journal} {Phys. Rev. Res.}\ }\textbf {\bibinfo {volume} {3}},\
  \bibinfo {pages} {033149} (\bibinfo {year} {2021})},\ \Eprint
  {https://arxiv.org/abs/2105.02233} {arXiv:2105.02233 [hep-ph]} \BibitemShut
  {NoStop}%
\bibitem [{\citenamefont {Catena}\ \emph {et~al.}(2020)\citenamefont {Catena},
  \citenamefont {Emken}, \citenamefont {Spaldin},\ and\ \citenamefont
  {Tarantino}}]{Catena:2019gfa}%
  \BibitemOpen
  \bibfield  {author} {\bibinfo {author} {\bibfnamefont {R.}~\bibnamefont
  {Catena}}, \bibinfo {author} {\bibfnamefont {T.}~\bibnamefont {Emken}},
  \bibinfo {author} {\bibfnamefont {N.~A.}\ \bibnamefont {Spaldin}},\ and\
  \bibinfo {author} {\bibfnamefont {W.}~\bibnamefont {Tarantino}},\ }\href
  {https://doi.org/10.1103/PhysRevResearch.2.033195} {\bibfield  {journal}
  {\bibinfo  {journal} {Phys. Rev. Res.}\ }\textbf {\bibinfo {volume} {2}},\
  \bibinfo {pages} {033195} (\bibinfo {year} {2020})},\ \bibinfo {note}
  {[Erratum: Phys.Rev.Res. 7, 019001 (2025)]},\ \Eprint
  {https://arxiv.org/abs/1912.08204} {arXiv:1912.08204 [hep-ph]} \BibitemShut
  {NoStop}%
\bibitem [{\citenamefont {Catena}\ \emph {et~al.}(2023)\citenamefont {Catena},
  \citenamefont {Cole}, \citenamefont {Emken}, \citenamefont {Matas},
  \citenamefont {Spaldin}, \citenamefont {Tarantino},\ and\ \citenamefont
  {Urdshals}}]{Catena:2022fnk}%
  \BibitemOpen
  \bibfield  {author} {\bibinfo {author} {\bibfnamefont {R.}~\bibnamefont
  {Catena}}, \bibinfo {author} {\bibfnamefont {D.}~\bibnamefont {Cole}},
  \bibinfo {author} {\bibfnamefont {T.}~\bibnamefont {Emken}}, \bibinfo
  {author} {\bibfnamefont {M.}~\bibnamefont {Matas}}, \bibinfo {author}
  {\bibfnamefont {N.}~\bibnamefont {Spaldin}}, \bibinfo {author} {\bibfnamefont
  {W.}~\bibnamefont {Tarantino}},\ and\ \bibinfo {author} {\bibfnamefont
  {E.}~\bibnamefont {Urdshals}},\ }\href
  {https://doi.org/10.1088/1475-7516/2023/03/052} {\bibfield  {journal}
  {\bibinfo  {journal} {JCAP}\ }\textbf {\bibinfo {volume} {03}},\ \bibinfo
  {pages} {052}},\ \Eprint {https://arxiv.org/abs/2210.07305} {arXiv:2210.07305
  [hep-ph]} \BibitemShut {NoStop}%
\bibitem [{\citenamefont {Del~Nobile}(2018)}]{PhysRevD.98.123003}%
  \BibitemOpen
  \bibfield  {author} {\bibinfo {author} {\bibfnamefont {E.}~\bibnamefont
  {Del~Nobile}},\ }\href {https://doi.org/10.1103/PhysRevD.98.123003}
  {\bibfield  {journal} {\bibinfo  {journal} {Phys. Rev. D}\ }\textbf {\bibinfo
  {volume} {98}},\ \bibinfo {pages} {123003} (\bibinfo {year}
  {2018})}\BibitemShut {NoStop}%
\bibitem [{\citenamefont {Catena}\ \emph {et~al.}(2019)\citenamefont {Catena},
  \citenamefont {Fridell},\ and\ \citenamefont {Krauss}}]{Catena:2019hzw}%
  \BibitemOpen
  \bibfield  {author} {\bibinfo {author} {\bibfnamefont {R.}~\bibnamefont
  {Catena}}, \bibinfo {author} {\bibfnamefont {K.}~\bibnamefont {Fridell}},\
  and\ \bibinfo {author} {\bibfnamefont {M.~B.}\ \bibnamefont {Krauss}},\
  }\href {https://doi.org/10.1007/JHEP08(2019)030} {\bibfield  {journal}
  {\bibinfo  {journal} {JHEP}\ }\textbf {\bibinfo {volume} {08}},\ \bibinfo
  {pages} {030}},\ \Eprint {https://arxiv.org/abs/1907.02910} {arXiv:1907.02910
  [hep-ph]} \BibitemShut {NoStop}%
\bibitem [{\citenamefont {Catena}\ and\ \citenamefont
  {Spaldin}(2024)}]{Catena:2024rym}%
  \BibitemOpen
  \bibfield  {author} {\bibinfo {author} {\bibfnamefont {R.}~\bibnamefont
  {Catena}}\ and\ \bibinfo {author} {\bibfnamefont {N.~A.}\ \bibnamefont
  {Spaldin}},\ }\href {https://doi.org/10.1103/PhysRevResearch.6.033230}
  {\bibfield  {journal} {\bibinfo  {journal} {Phys. Rev. Res.}\ }\textbf
  {\bibinfo {volume} {6}},\ \bibinfo {pages} {033230} (\bibinfo {year}
  {2024})},\ \Eprint {https://arxiv.org/abs/2402.06817} {arXiv:2402.06817
  [hep-ph]} \BibitemShut {NoStop}%
\bibitem [{\citenamefont {Catena}\ and\ \citenamefont
  {Ullio}(2010)}]{Catena:2009mf}%
  \BibitemOpen
  \bibfield  {author} {\bibinfo {author} {\bibfnamefont {R.}~\bibnamefont
  {Catena}}\ and\ \bibinfo {author} {\bibfnamefont {P.}~\bibnamefont {Ullio}},\
  }\href {https://doi.org/10.1088/1475-7516/2010/08/004} {\bibfield  {journal}
  {\bibinfo  {journal} {JCAP}\ }\textbf {\bibinfo {volume} {08}},\ \bibinfo
  {pages} {004}},\ \Eprint {https://arxiv.org/abs/0907.0018} {arXiv:0907.0018
  [astro-ph.CO]} \BibitemShut {NoStop}%
\bibitem [{\citenamefont {Kerr}\ and\ \citenamefont
  {Lynden-Bell}(1986)}]{Kerr:1986hz}%
  \BibitemOpen
  \bibfield  {author} {\bibinfo {author} {\bibfnamefont {F.~J.}\ \bibnamefont
  {Kerr}}\ and\ \bibinfo {author} {\bibfnamefont {D.}~\bibnamefont
  {Lynden-Bell}},\ }\href@noop {} {\bibfield  {journal} {\bibinfo  {journal}
  {Mon. Not. Roy. Astron. Soc.}\ }\textbf {\bibinfo {volume} {221}},\ \bibinfo
  {pages} {1023} (\bibinfo {year} {1986})}\BibitemShut {NoStop}%
\bibitem [{\citenamefont {Smith}\ \emph {et~al.}(2007)\citenamefont {Smith}
  \emph {et~al.}}]{Smith:2006ym}%
  \BibitemOpen
  \bibfield  {author} {\bibinfo {author} {\bibfnamefont {M.~C.}\ \bibnamefont
  {Smith}} \emph {et~al.},\ }\href
  {https://doi.org/10.1111/j.1365-2966.2007.11964.x} {\bibfield  {journal}
  {\bibinfo  {journal} {Mon. Not. Roy. Astron. Soc.}\ }\textbf {\bibinfo
  {volume} {379}},\ \bibinfo {pages} {755} (\bibinfo {year} {2007})},\ \Eprint
  {https://arxiv.org/abs/astro-ph/0611671} {arXiv:astro-ph/0611671}
  \BibitemShut {NoStop}%
\bibitem [{\citenamefont {Fan}\ \emph {et~al.}(2010)\citenamefont {Fan},
  \citenamefont {Reece},\ and\ \citenamefont {Wang}}]{Fan:2010gt}%
  \BibitemOpen
  \bibfield  {author} {\bibinfo {author} {\bibfnamefont {J.}~\bibnamefont
  {Fan}}, \bibinfo {author} {\bibfnamefont {M.}~\bibnamefont {Reece}},\ and\
  \bibinfo {author} {\bibfnamefont {L.-T.}\ \bibnamefont {Wang}},\ }\href
  {https://doi.org/10.1088/1475-7516/2010/11/042} {\bibfield  {journal}
  {\bibinfo  {journal} {JCAP}\ }\textbf {\bibinfo {volume} {11}},\ \bibinfo
  {pages} {042}},\ \Eprint {https://arxiv.org/abs/1008.1591} {arXiv:1008.1591
  [hep-ph]} \BibitemShut {NoStop}%
\bibitem [{\citenamefont {Fitzpatrick}\ \emph {et~al.}(2013)\citenamefont
  {Fitzpatrick}, \citenamefont {Haxton}, \citenamefont {Katz}, \citenamefont
  {Lubbers},\ and\ \citenamefont {Xu}}]{Fitzpatrick:2012ix}%
  \BibitemOpen
  \bibfield  {author} {\bibinfo {author} {\bibfnamefont {A.~L.}\ \bibnamefont
  {Fitzpatrick}}, \bibinfo {author} {\bibfnamefont {W.}~\bibnamefont {Haxton}},
  \bibinfo {author} {\bibfnamefont {E.}~\bibnamefont {Katz}}, \bibinfo {author}
  {\bibfnamefont {N.}~\bibnamefont {Lubbers}},\ and\ \bibinfo {author}
  {\bibfnamefont {Y.}~\bibnamefont {Xu}},\ }\href
  {https://doi.org/10.1088/1475-7516/2013/02/004} {\bibfield  {journal}
  {\bibinfo  {journal} {JCAP}\ }\textbf {\bibinfo {volume} {02}},\ \bibinfo
  {pages} {004}},\ \Eprint {https://arxiv.org/abs/1203.3542} {arXiv:1203.3542
  [hep-ph]} \BibitemShut {NoStop}%
\bibitem [{\citenamefont {Giannozzi}\ \emph {et~al.}(2009)\citenamefont
  {Giannozzi}, \citenamefont {Baroni}, \citenamefont {Bonini}, \citenamefont
  {Calandra}, \citenamefont {Car}, \citenamefont {Cavazzoni}, \citenamefont
  {Ceresoli}, \citenamefont {Chiarotti}, \citenamefont {Cococcioni},
  \citenamefont {Dabo}, \citenamefont {Dal~Corso}, \citenamefont
  {de~Gironcoli}, \citenamefont {Fabris}, \citenamefont {Fratesi},
  \citenamefont {Gebauer}, \citenamefont {Gerstmann}, \citenamefont
  {Gougoussis}, \citenamefont {Kokalj}, \citenamefont {Lazzeri}, \citenamefont
  {Martin-Samos}, \citenamefont {Marzari}, \citenamefont {Mauri}, \citenamefont
  {Mazzarello}, \citenamefont {Paolini}, \citenamefont {Pasquarello},
  \citenamefont {Paulatto}, \citenamefont {Sbraccia}, \citenamefont {Scandolo},
  \citenamefont {Sclauzero}, \citenamefont {Seitsonen}, \citenamefont
  {Smogunov}, \citenamefont {Umari},\ and\ \citenamefont
  {Wentzcovitch}}]{Giannozzi_2009}%
  \BibitemOpen
  \bibfield  {author} {\bibinfo {author} {\bibfnamefont {P.}~\bibnamefont
  {Giannozzi}}, \bibinfo {author} {\bibfnamefont {S.}~\bibnamefont {Baroni}},
  \bibinfo {author} {\bibfnamefont {N.}~\bibnamefont {Bonini}}, \bibinfo
  {author} {\bibfnamefont {M.}~\bibnamefont {Calandra}}, \bibinfo {author}
  {\bibfnamefont {R.}~\bibnamefont {Car}}, \bibinfo {author} {\bibfnamefont
  {C.}~\bibnamefont {Cavazzoni}}, \bibinfo {author} {\bibfnamefont
  {D.}~\bibnamefont {Ceresoli}}, \bibinfo {author} {\bibfnamefont {G.~L.}\
  \bibnamefont {Chiarotti}}, \bibinfo {author} {\bibfnamefont {M.}~\bibnamefont
  {Cococcioni}}, \bibinfo {author} {\bibfnamefont {I.}~\bibnamefont {Dabo}},
  \bibinfo {author} {\bibfnamefont {A.}~\bibnamefont {Dal~Corso}}, \bibinfo
  {author} {\bibfnamefont {S.}~\bibnamefont {de~Gironcoli}}, \bibinfo {author}
  {\bibfnamefont {S.}~\bibnamefont {Fabris}}, \bibinfo {author} {\bibfnamefont
  {G.}~\bibnamefont {Fratesi}}, \bibinfo {author} {\bibfnamefont
  {R.}~\bibnamefont {Gebauer}}, \bibinfo {author} {\bibfnamefont
  {U.}~\bibnamefont {Gerstmann}}, \bibinfo {author} {\bibfnamefont
  {C.}~\bibnamefont {Gougoussis}}, \bibinfo {author} {\bibfnamefont
  {A.}~\bibnamefont {Kokalj}}, \bibinfo {author} {\bibfnamefont
  {M.}~\bibnamefont {Lazzeri}}, \bibinfo {author} {\bibfnamefont
  {L.}~\bibnamefont {Martin-Samos}}, \bibinfo {author} {\bibfnamefont
  {N.}~\bibnamefont {Marzari}}, \bibinfo {author} {\bibfnamefont
  {F.}~\bibnamefont {Mauri}}, \bibinfo {author} {\bibfnamefont
  {R.}~\bibnamefont {Mazzarello}}, \bibinfo {author} {\bibfnamefont
  {S.}~\bibnamefont {Paolini}}, \bibinfo {author} {\bibfnamefont
  {A.}~\bibnamefont {Pasquarello}}, \bibinfo {author} {\bibfnamefont
  {L.}~\bibnamefont {Paulatto}}, \bibinfo {author} {\bibfnamefont
  {C.}~\bibnamefont {Sbraccia}}, \bibinfo {author} {\bibfnamefont
  {S.}~\bibnamefont {Scandolo}}, \bibinfo {author} {\bibfnamefont
  {G.}~\bibnamefont {Sclauzero}}, \bibinfo {author} {\bibfnamefont {A.~P.}\
  \bibnamefont {Seitsonen}}, \bibinfo {author} {\bibfnamefont {A.}~\bibnamefont
  {Smogunov}}, \bibinfo {author} {\bibfnamefont {P.}~\bibnamefont {Umari}},\
  and\ \bibinfo {author} {\bibfnamefont {R.~M.}\ \bibnamefont {Wentzcovitch}},\
  }\href {https://doi.org/10.1088/0953-8984/21/39/395502} {\bibfield  {journal}
  {\bibinfo  {journal} {Journal of Physics: Condensed Matter}\ }\textbf
  {\bibinfo {volume} {21}},\ \bibinfo {pages} {395502} (\bibinfo {year}
  {2009})}\BibitemShut {NoStop}%
\bibitem [{\citenamefont {Giannozzi}\ \emph {et~al.}(2017)\citenamefont
  {Giannozzi}, \citenamefont {Andreussi}, \citenamefont {Brumme}, \citenamefont
  {Bunau}, \citenamefont {Buongiorno~Nardelli}, \citenamefont {Calandra},
  \citenamefont {Car}, \citenamefont {Cavazzoni}, \citenamefont {Ceresoli},
  \citenamefont {Cococcioni}, \citenamefont {Colonna}, \citenamefont
  {Carnimeo}, \citenamefont {Dal~Corso}, \citenamefont {de~Gironcoli},
  \citenamefont {Delugas}, \citenamefont {DiStasio}, \citenamefont {Ferretti},
  \citenamefont {Floris}, \citenamefont {Fratesi}, \citenamefont {Fugallo},
  \citenamefont {Gebauer}, \citenamefont {Gerstmann}, \citenamefont {Giustino},
  \citenamefont {Gorni}, \citenamefont {Jia}, \citenamefont {Kawamura},
  \citenamefont {Ko}, \citenamefont {Kokalj}, \citenamefont {Küçükbenli},
  \citenamefont {Lazzeri}, \citenamefont {Marsili}, \citenamefont {Marzari},
  \citenamefont {Mauri}, \citenamefont {Nguyen}, \citenamefont {Nguyen},
  \citenamefont {Otero-de-la Roza}, \citenamefont {Paulatto}, \citenamefont
  {Poncé}, \citenamefont {Rocca}, \citenamefont {Sabatini}, \citenamefont
  {Santra}, \citenamefont {Schlipf}, \citenamefont {Seitsonen}, \citenamefont
  {Smogunov}, \citenamefont {Timrov}, \citenamefont {Thonhauser}, \citenamefont
  {Umari}, \citenamefont {Vast}, \citenamefont {Wu},\ and\ \citenamefont
  {Baroni}}]{Giannozzi_2017}%
  \BibitemOpen
  \bibfield  {author} {\bibinfo {author} {\bibfnamefont {P.}~\bibnamefont
  {Giannozzi}}, \bibinfo {author} {\bibfnamefont {O.}~\bibnamefont
  {Andreussi}}, \bibinfo {author} {\bibfnamefont {T.}~\bibnamefont {Brumme}},
  \bibinfo {author} {\bibfnamefont {O.}~\bibnamefont {Bunau}}, \bibinfo
  {author} {\bibfnamefont {M.}~\bibnamefont {Buongiorno~Nardelli}}, \bibinfo
  {author} {\bibfnamefont {M.}~\bibnamefont {Calandra}}, \bibinfo {author}
  {\bibfnamefont {R.}~\bibnamefont {Car}}, \bibinfo {author} {\bibfnamefont
  {C.}~\bibnamefont {Cavazzoni}}, \bibinfo {author} {\bibfnamefont
  {D.}~\bibnamefont {Ceresoli}}, \bibinfo {author} {\bibfnamefont
  {M.}~\bibnamefont {Cococcioni}}, \bibinfo {author} {\bibfnamefont
  {N.}~\bibnamefont {Colonna}}, \bibinfo {author} {\bibfnamefont
  {I.}~\bibnamefont {Carnimeo}}, \bibinfo {author} {\bibfnamefont
  {A.}~\bibnamefont {Dal~Corso}}, \bibinfo {author} {\bibfnamefont
  {S.}~\bibnamefont {de~Gironcoli}}, \bibinfo {author} {\bibfnamefont
  {P.}~\bibnamefont {Delugas}}, \bibinfo {author} {\bibfnamefont {R.~A.}\
  \bibnamefont {DiStasio}}, \bibinfo {author} {\bibfnamefont {A.}~\bibnamefont
  {Ferretti}}, \bibinfo {author} {\bibfnamefont {A.}~\bibnamefont {Floris}},
  \bibinfo {author} {\bibfnamefont {G.}~\bibnamefont {Fratesi}}, \bibinfo
  {author} {\bibfnamefont {G.}~\bibnamefont {Fugallo}}, \bibinfo {author}
  {\bibfnamefont {R.}~\bibnamefont {Gebauer}}, \bibinfo {author} {\bibfnamefont
  {U.}~\bibnamefont {Gerstmann}}, \bibinfo {author} {\bibfnamefont
  {F.}~\bibnamefont {Giustino}}, \bibinfo {author} {\bibfnamefont
  {T.}~\bibnamefont {Gorni}}, \bibinfo {author} {\bibfnamefont
  {J.}~\bibnamefont {Jia}}, \bibinfo {author} {\bibfnamefont {M.}~\bibnamefont
  {Kawamura}}, \bibinfo {author} {\bibfnamefont {H.-Y.}\ \bibnamefont {Ko}},
  \bibinfo {author} {\bibfnamefont {A.}~\bibnamefont {Kokalj}}, \bibinfo
  {author} {\bibfnamefont {E.}~\bibnamefont {Küçükbenli}}, \bibinfo {author}
  {\bibfnamefont {M.}~\bibnamefont {Lazzeri}}, \bibinfo {author} {\bibfnamefont
  {M.}~\bibnamefont {Marsili}}, \bibinfo {author} {\bibfnamefont
  {N.}~\bibnamefont {Marzari}}, \bibinfo {author} {\bibfnamefont
  {F.}~\bibnamefont {Mauri}}, \bibinfo {author} {\bibfnamefont {N.~L.}\
  \bibnamefont {Nguyen}}, \bibinfo {author} {\bibfnamefont {H.-V.}\
  \bibnamefont {Nguyen}}, \bibinfo {author} {\bibfnamefont {A.}~\bibnamefont
  {Otero-de-la Roza}}, \bibinfo {author} {\bibfnamefont {L.}~\bibnamefont
  {Paulatto}}, \bibinfo {author} {\bibfnamefont {S.}~\bibnamefont {Poncé}},
  \bibinfo {author} {\bibfnamefont {D.}~\bibnamefont {Rocca}}, \bibinfo
  {author} {\bibfnamefont {R.}~\bibnamefont {Sabatini}}, \bibinfo {author}
  {\bibfnamefont {B.}~\bibnamefont {Santra}}, \bibinfo {author} {\bibfnamefont
  {M.}~\bibnamefont {Schlipf}}, \bibinfo {author} {\bibfnamefont {A.~P.}\
  \bibnamefont {Seitsonen}}, \bibinfo {author} {\bibfnamefont {A.}~\bibnamefont
  {Smogunov}}, \bibinfo {author} {\bibfnamefont {I.}~\bibnamefont {Timrov}},
  \bibinfo {author} {\bibfnamefont {T.}~\bibnamefont {Thonhauser}}, \bibinfo
  {author} {\bibfnamefont {P.}~\bibnamefont {Umari}}, \bibinfo {author}
  {\bibfnamefont {N.}~\bibnamefont {Vast}}, \bibinfo {author} {\bibfnamefont
  {X.}~\bibnamefont {Wu}},\ and\ \bibinfo {author} {\bibfnamefont
  {S.}~\bibnamefont {Baroni}},\ }\href
  {https://doi.org/10.1088/1361-648X/aa8f79} {\bibfield  {journal} {\bibinfo
  {journal} {Journal of Physics: Condensed Matter}\ }\textbf {\bibinfo {volume}
  {29}},\ \bibinfo {pages} {465901} (\bibinfo {year} {2017})}\BibitemShut
  {NoStop}%
\bibitem [{\citenamefont {Giannozzi}\ \emph {et~al.}(2020)\citenamefont
  {Giannozzi}, \citenamefont {Baseggio}, \citenamefont {Bonfà}, \citenamefont
  {Brunato}, \citenamefont {Car}, \citenamefont {Carnimeo}, \citenamefont
  {Cavazzoni}, \citenamefont {de~Gironcoli}, \citenamefont {Delugas},
  \citenamefont {Ferrari~Ruffino}, \citenamefont {Ferretti}, \citenamefont
  {Marzari}, \citenamefont {Timrov}, \citenamefont {Urru},\ and\ \citenamefont
  {Baroni}}]{10.1063/5.0005082}%
  \BibitemOpen
  \bibfield  {author} {\bibinfo {author} {\bibfnamefont {P.}~\bibnamefont
  {Giannozzi}}, \bibinfo {author} {\bibfnamefont {O.}~\bibnamefont {Baseggio}},
  \bibinfo {author} {\bibfnamefont {P.}~\bibnamefont {Bonfà}}, \bibinfo
  {author} {\bibfnamefont {D.}~\bibnamefont {Brunato}}, \bibinfo {author}
  {\bibfnamefont {R.}~\bibnamefont {Car}}, \bibinfo {author} {\bibfnamefont
  {I.}~\bibnamefont {Carnimeo}}, \bibinfo {author} {\bibfnamefont
  {C.}~\bibnamefont {Cavazzoni}}, \bibinfo {author} {\bibfnamefont
  {S.}~\bibnamefont {de~Gironcoli}}, \bibinfo {author} {\bibfnamefont
  {P.}~\bibnamefont {Delugas}}, \bibinfo {author} {\bibfnamefont
  {F.}~\bibnamefont {Ferrari~Ruffino}}, \bibinfo {author} {\bibfnamefont
  {A.}~\bibnamefont {Ferretti}}, \bibinfo {author} {\bibfnamefont
  {N.}~\bibnamefont {Marzari}}, \bibinfo {author} {\bibfnamefont
  {I.}~\bibnamefont {Timrov}}, \bibinfo {author} {\bibfnamefont
  {A.}~\bibnamefont {Urru}},\ and\ \bibinfo {author} {\bibfnamefont
  {S.}~\bibnamefont {Baroni}},\ }\href {https://doi.org/10.1063/5.0005082}
  {\bibfield  {journal} {\bibinfo  {journal} {The Journal of Chemical Physics}\
  }\textbf {\bibinfo {volume} {152}},\ \bibinfo {pages} {154105} (\bibinfo
  {year} {2020})}\BibitemShut {NoStop}%
\bibitem [{\citenamefont {Urdshals}\ and\ \citenamefont
  {Matas}(2023)}]{urdshals_2023_7836577}%
  \BibitemOpen
  \bibfield  {author} {\bibinfo {author} {\bibfnamefont {E.}~\bibnamefont
  {Urdshals}}\ and\ \bibinfo {author} {\bibfnamefont {M.}~\bibnamefont
  {Matas}},\ }\href {https://doi.org/10.5281/zenodo.7836577} {\bibinfo {title}
  {{QEdark-EFT-Graphene}}} (\bibinfo {year} {2023})\BibitemShut {NoStop}%
\bibitem [{\citenamefont {T{\"{o}}bbens}\ \emph {et~al.}(2001)\citenamefont
  {T{\"{o}}bbens}, \citenamefont {St{\"{u}}{\ss }er}, \citenamefont {Knorr},
  \citenamefont {Mayer},\ and\ \citenamefont {Lampert}}]{tobbens2001}%
  \BibitemOpen
  \bibfield  {author} {\bibinfo {author} {\bibfnamefont {D.}~\bibnamefont
  {T{\"{o}}bbens}}, \bibinfo {author} {\bibfnamefont {N.}~\bibnamefont
  {St{\"{u}}{\ss }er}}, \bibinfo {author} {\bibfnamefont {K.}~\bibnamefont
  {Knorr}}, \bibinfo {author} {\bibfnamefont {H.}~\bibnamefont {Mayer}},\ and\
  \bibinfo {author} {\bibfnamefont {G.}~\bibnamefont {Lampert}},\ }in\ \href
  {https://doi.org/10.4028/www.scientific.net/MSF.378-381.288} {\emph {\bibinfo
  {booktitle} {European Powder Diffraction EPDIC 7}}},\ \bibinfo {series}
  {Materials Science Forum}, Vol.\ \bibinfo {volume} {378}\ (\bibinfo
  {publisher} {Trans Tech Publications Ltd},\ \bibinfo {year} {2001})\ pp.\
  \bibinfo {pages} {288--293}\BibitemShut {NoStop}%
\bibitem [{\citenamefont {Hohenberg}\ and\ \citenamefont
  {Kohn}(1964)}]{PhysRev.136.B864}%
  \BibitemOpen
  \bibfield  {author} {\bibinfo {author} {\bibfnamefont {P.}~\bibnamefont
  {Hohenberg}}\ and\ \bibinfo {author} {\bibfnamefont {W.}~\bibnamefont
  {Kohn}},\ }\href {https://doi.org/10.1103/PhysRev.136.B864} {\bibfield
  {journal} {\bibinfo  {journal} {Physical Review}\ }\textbf {\bibinfo {volume}
  {136}},\ \bibinfo {pages} {B864} (\bibinfo {year} {1964})}\BibitemShut
  {NoStop}%
\bibitem [{\citenamefont {{Kohn}}\ and\ \citenamefont
  {{Sham}}(1965)}]{1965PhRv..140.1133K}%
  \BibitemOpen
  \bibfield  {author} {\bibinfo {author} {\bibfnamefont {W.}~\bibnamefont
  {{Kohn}}}\ and\ \bibinfo {author} {\bibfnamefont {L.~J.}\ \bibnamefont
  {{Sham}}},\ }\href {https://doi.org/10.1103/PhysRev.140.A1133} {\bibfield
  {journal} {\bibinfo  {journal} {Physical Review}\ }\textbf {\bibinfo {volume}
  {140}},\ \bibinfo {pages} {1133} (\bibinfo {year} {1965})}\BibitemShut
  {NoStop}%
\bibitem [{\citenamefont {Perdew}\ and\ \citenamefont
  {Zunger}(1981)}]{PhysRevB.23.5048}%
  \BibitemOpen
  \bibfield  {author} {\bibinfo {author} {\bibfnamefont {J.~P.}\ \bibnamefont
  {Perdew}}\ and\ \bibinfo {author} {\bibfnamefont {A.}~\bibnamefont
  {Zunger}},\ }\href {https://doi.org/10.1103/PhysRevB.23.5048} {\bibfield
  {journal} {\bibinfo  {journal} {Phys. Rev. B}\ }\textbf {\bibinfo {volume}
  {23}},\ \bibinfo {pages} {5048} (\bibinfo {year} {1981})}\BibitemShut
  {NoStop}%
\bibitem [{\citenamefont {Hamann}(2013)}]{PhysRevB.88.085117}%
  \BibitemOpen
  \bibfield  {author} {\bibinfo {author} {\bibfnamefont {D.~R.}\ \bibnamefont
  {Hamann}},\ }\href {https://doi.org/10.1103/PhysRevB.88.085117} {\bibfield
  {journal} {\bibinfo  {journal} {Phys. Rev. B}\ }\textbf {\bibinfo {volume}
  {88}},\ \bibinfo {pages} {085117} (\bibinfo {year} {2013})}\BibitemShut
  {NoStop}%
\bibitem [{\citenamefont {Matas}\ \emph {et~al.}(2026)\citenamefont {Matas},
  \citenamefont {Gallo~Rosso}, \citenamefont {Cammarata}, \citenamefont {Hoch},
  \citenamefont {Blanco}, \citenamefont {Conrad}, \citenamefont {Essig},
  \citenamefont {Linden},\ and\ \citenamefont {Winslow}}]{data}%
  \BibitemOpen
  \bibfield  {author} {\bibinfo {author} {\bibfnamefont {M.}~\bibnamefont
  {Matas}}, \bibinfo {author} {\bibfnamefont {A.}~\bibnamefont {Gallo~Rosso}},
  \bibinfo {author} {\bibfnamefont {A.}~\bibnamefont {Cammarata}}, \bibinfo
  {author} {\bibfnamefont {N.}~\bibnamefont {Hoch}}, \bibinfo {author}
  {\bibfnamefont {C.}~\bibnamefont {Blanco}}, \bibinfo {author} {\bibfnamefont
  {J.}~\bibnamefont {Conrad}}, \bibinfo {author} {\bibfnamefont
  {R.}~\bibnamefont {Essig}}, \bibinfo {author} {\bibfnamefont
  {T.}~\bibnamefont {Linden}},\ and\ \bibinfo {author} {\bibfnamefont
  {L.}~\bibnamefont {Winslow}},\ }\href
  {https://doi.org/10.5281/zenodo.20487678} {10.5281/zenodo.20487678} (\bibinfo
  {year} {2026})\BibitemShut {NoStop}%
\bibitem [{\citenamefont {Trickle}\ \emph {et~al.}(2020)\citenamefont
  {Trickle}, \citenamefont {Zhang}, \citenamefont {Zurek}, \citenamefont
  {Inzani},\ and\ \citenamefont {Griffin}}]{Trickle:2019nya}%
  \BibitemOpen
  \bibfield  {author} {\bibinfo {author} {\bibfnamefont {T.}~\bibnamefont
  {Trickle}}, \bibinfo {author} {\bibfnamefont {Z.}~\bibnamefont {Zhang}},
  \bibinfo {author} {\bibfnamefont {K.~M.}\ \bibnamefont {Zurek}}, \bibinfo
  {author} {\bibfnamefont {K.}~\bibnamefont {Inzani}},\ and\ \bibinfo {author}
  {\bibfnamefont {S.~M.}\ \bibnamefont {Griffin}},\ }\href
  {https://doi.org/10.1007/JHEP03(2020)036} {\bibfield  {journal} {\bibinfo
  {journal} {JHEP}\ }\textbf {\bibinfo {volume} {03}},\ \bibinfo {pages}
  {036}},\ \Eprint {https://arxiv.org/abs/1910.08092} {arXiv:1910.08092
  [hep-ph]} \BibitemShut {NoStop}%
\bibitem [{\citenamefont {Dreyer}\ \emph {et~al.}(2024)\citenamefont {Dreyer},
  \citenamefont {Essig}, \citenamefont {Fernandez-Serra}, \citenamefont
  {Singal},\ and\ \citenamefont {Zhen}}]{Dreyer:2023ovn}%
  \BibitemOpen
  \bibfield  {author} {\bibinfo {author} {\bibfnamefont {C.~E.}\ \bibnamefont
  {Dreyer}}, \bibinfo {author} {\bibfnamefont {R.}~\bibnamefont {Essig}},
  \bibinfo {author} {\bibfnamefont {M.}~\bibnamefont {Fernandez-Serra}},
  \bibinfo {author} {\bibfnamefont {A.}~\bibnamefont {Singal}},\ and\ \bibinfo
  {author} {\bibfnamefont {C.}~\bibnamefont {Zhen}},\ }\href
  {https://doi.org/10.1103/PhysRevD.109.115008} {\bibfield  {journal} {\bibinfo
   {journal} {Phys. Rev. D}\ }\textbf {\bibinfo {volume} {109}},\ \bibinfo
  {pages} {115008} (\bibinfo {year} {2024})},\ \Eprint
  {https://arxiv.org/abs/2306.14944} {arXiv:2306.14944 [hep-ph]} \BibitemShut
  {NoStop}%
\bibitem [{\citenamefont {Cappellini}\ \emph {et~al.}(1993)\citenamefont
  {Cappellini}, \citenamefont {Del~Sole}, \citenamefont {Reining},\ and\
  \citenamefont {Bechstedt}}]{PhysRevB.47.9892}%
  \BibitemOpen
  \bibfield  {author} {\bibinfo {author} {\bibfnamefont {G.}~\bibnamefont
  {Cappellini}}, \bibinfo {author} {\bibfnamefont {R.}~\bibnamefont
  {Del~Sole}}, \bibinfo {author} {\bibfnamefont {L.}~\bibnamefont {Reining}},\
  and\ \bibinfo {author} {\bibfnamefont {F.}~\bibnamefont {Bechstedt}},\ }\href
  {https://doi.org/10.1103/PhysRevB.47.9892} {\bibfield  {journal} {\bibinfo
  {journal} {Phys. Rev. B}\ }\textbf {\bibinfo {volume} {47}},\ \bibinfo
  {pages} {9892} (\bibinfo {year} {1993})}\BibitemShut {NoStop}%
\bibitem [{\citenamefont {Knapen}\ \emph {et~al.}(2022)\citenamefont {Knapen},
  \citenamefont {Kozaczuk},\ and\ \citenamefont {Lin}}]{Knapen:2021bwg}%
  \BibitemOpen
  \bibfield  {author} {\bibinfo {author} {\bibfnamefont {S.}~\bibnamefont
  {Knapen}}, \bibinfo {author} {\bibfnamefont {J.}~\bibnamefont {Kozaczuk}},\
  and\ \bibinfo {author} {\bibfnamefont {T.}~\bibnamefont {Lin}},\ }\href
  {https://doi.org/10.1103/PhysRevD.105.015014} {\bibfield  {journal} {\bibinfo
   {journal} {Phys. Rev. D}\ }\textbf {\bibinfo {volume} {105}},\ \bibinfo
  {pages} {015014} (\bibinfo {year} {2022})},\ \Eprint
  {https://arxiv.org/abs/2104.12786} {arXiv:2104.12786 [hep-ph]} \BibitemShut
  {NoStop}%
\bibitem [{\citenamefont {Dreyer}\ \emph {et~al.}(2026)\citenamefont {Dreyer},
  \citenamefont {Essig}, \citenamefont {Fernandez-Serra}, \citenamefont
  {Hott},\ and\ \citenamefont {Singal}}]{Dreyer:2026bmz}%
  \BibitemOpen
  \bibfield  {author} {\bibinfo {author} {\bibfnamefont {C.}~\bibnamefont
  {Dreyer}}, \bibinfo {author} {\bibfnamefont {R.}~\bibnamefont {Essig}},
  \bibinfo {author} {\bibfnamefont {M.}~\bibnamefont {Fernandez-Serra}},
  \bibinfo {author} {\bibfnamefont {M.}~\bibnamefont {Hott}},\ and\ \bibinfo
  {author} {\bibfnamefont {A.}~\bibnamefont {Singal}},\ }\href@noop {}
  {\bibfield  {journal} {\bibinfo  {journal} {arXiv e-prints}\ } (\bibinfo
  {year} {2026})},\ \Eprint {https://arxiv.org/abs/2603.12326}
  {arXiv:2603.12326 [hep-ph]} \BibitemShut {NoStop}%
\bibitem [{\citenamefont {Griffin}\ \emph {et~al.}(2021)\citenamefont
  {Griffin}, \citenamefont {Inzani}, \citenamefont {Trickle}, \citenamefont
  {Zhang},\ and\ \citenamefont {Zurek}}]{Griffin:2021znd}%
  \BibitemOpen
  \bibfield  {author} {\bibinfo {author} {\bibfnamefont {S.~M.}\ \bibnamefont
  {Griffin}}, \bibinfo {author} {\bibfnamefont {K.}~\bibnamefont {Inzani}},
  \bibinfo {author} {\bibfnamefont {T.}~\bibnamefont {Trickle}}, \bibinfo
  {author} {\bibfnamefont {Z.}~\bibnamefont {Zhang}},\ and\ \bibinfo {author}
  {\bibfnamefont {K.~M.}\ \bibnamefont {Zurek}},\ }\href
  {https://doi.org/10.1103/PhysRevD.104.095015} {\bibfield  {journal} {\bibinfo
   {journal} {Phys. Rev. D}\ }\textbf {\bibinfo {volume} {104}},\ \bibinfo
  {pages} {095015} (\bibinfo {year} {2021})},\ \Eprint
  {https://arxiv.org/abs/2105.05253} {arXiv:2105.05253 [hep-ph]} \BibitemShut
  {NoStop}%
\bibitem [{\citenamefont {Angle}\ \emph {et~al.}(2011)\citenamefont {Angle}
  \emph {et~al.}}]{XENON10:2011prx}%
  \BibitemOpen
  \bibfield  {author} {\bibinfo {author} {\bibfnamefont {J.}~\bibnamefont
  {Angle}} \emph {et~al.} (\bibinfo {collaboration} {XENON10}),\ }\href
  {https://doi.org/10.1103/PhysRevLett.107.051301} {\bibfield  {journal}
  {\bibinfo  {journal} {Phys. Rev. Lett.}\ }\textbf {\bibinfo {volume} {107}},\
  \bibinfo {pages} {051301} (\bibinfo {year} {2011})},\ \bibinfo {note}
  {[Erratum: Phys.Rev.Lett. 110, 249901 (2013)]},\ \Eprint
  {https://arxiv.org/abs/1104.3088} {arXiv:1104.3088 [astro-ph.CO]}
  \BibitemShut {NoStop}%
\bibitem [{\citenamefont {Aprile}\ \emph {et~al.}(2019)\citenamefont {Aprile}
  \emph {et~al.}}]{XENON:2019gfn}%
  \BibitemOpen
  \bibfield  {author} {\bibinfo {author} {\bibfnamefont {E.}~\bibnamefont
  {Aprile}} \emph {et~al.} (\bibinfo {collaboration} {XENON Collaboration}),\
  }\href {https://doi.org/10.1103/PhysRevLett.123.251801} {\bibfield  {journal}
  {\bibinfo  {journal} {Phys. Rev. Lett.}\ }\textbf {\bibinfo {volume} {123}},\
  \bibinfo {pages} {251801} (\bibinfo {year} {2019})},\ \Eprint
  {https://arxiv.org/abs/1907.11485} {arXiv:1907.11485 [hep-ex]} \BibitemShut
  {NoStop}%
\bibitem [{\citenamefont {Aprile}\ \emph {et~al.}(2023)\citenamefont {Aprile}
  \emph {et~al.}}]{XENON:2023cxc}%
  \BibitemOpen
  \bibfield  {author} {\bibinfo {author} {\bibfnamefont {E.}~\bibnamefont
  {Aprile}} \emph {et~al.} (\bibinfo {collaboration} {XENON}),\ }\href
  {https://doi.org/10.1103/PhysRevLett.131.041003} {\bibfield  {journal}
  {\bibinfo  {journal} {Phys. Rev. Lett.}\ }\textbf {\bibinfo {volume} {131}},\
  \bibinfo {pages} {041003} (\bibinfo {year} {2023})},\ \Eprint
  {https://arxiv.org/abs/2303.14729} {arXiv:2303.14729 [hep-ex]} \BibitemShut
  {NoStop}%
\bibitem [{\citenamefont {Yu}\ \emph {et~al.}(2022)\citenamefont {Yu},
  \citenamefont {Chen}, \citenamefont {Han}, \citenamefont {Fan},\ and\
  \citenamefont {Pei}}]{yu2022liquid}%
  \BibitemOpen
  \bibfield  {author} {\bibinfo {author} {\bibfnamefont {H.}~\bibnamefont
  {Yu}}, \bibinfo {author} {\bibfnamefont {T.}~\bibnamefont {Chen}}, \bibinfo
  {author} {\bibfnamefont {Z.}~\bibnamefont {Han}}, \bibinfo {author}
  {\bibfnamefont {J.}~\bibnamefont {Fan}},\ and\ \bibinfo {author}
  {\bibfnamefont {Q.}~\bibnamefont {Pei}},\ }\href@noop {} {\bibfield
  {journal} {\bibinfo  {journal} {ACS Applied Nano Materials}\ }\textbf
  {\bibinfo {volume} {5}},\ \bibinfo {pages} {14572} (\bibinfo {year}
  {2022})}\BibitemShut {NoStop}%
\bibitem [{\citenamefont {Villalpando}\ \emph {et~al.}(2024)\citenamefont
  {Villalpando} \emph {et~al.}}]{Villalpando:2023ate}%
  \BibitemOpen
  \bibfield  {author} {\bibinfo {author} {\bibfnamefont {E.~M.}\ \bibnamefont
  {Villalpando}} \emph {et~al.},\ }\href
  {https://doi.org/10.1088/1538-3873/ad2865} {\bibfield  {journal} {\bibinfo
  {journal} {Publ. Astron. Soc. Pac.}\ }\textbf {\bibinfo {volume} {136}},\
  \bibinfo {pages} {045001} (\bibinfo {year} {2024})},\ \Eprint
  {https://arxiv.org/abs/2311.00813} {arXiv:2311.00813 [astro-ph.IM]}
  \BibitemShut {NoStop}%
\bibitem [{\citenamefont {Romani}\ \emph {et~al.}(2024)\citenamefont {Romani},
  \citenamefont {Chang}, \citenamefont {Mahapatra}, \citenamefont {Platt},
  \citenamefont {Reed}, \citenamefont {Rydstrom}, \citenamefont {Sadoulet},
  \citenamefont {Serfass},\ and\ \citenamefont {Pyle}}]{Romani:2024rfh}%
  \BibitemOpen
  \bibfield  {author} {\bibinfo {author} {\bibfnamefont {R.~K.}\ \bibnamefont
  {Romani}}, \bibinfo {author} {\bibfnamefont {Y.-Y.}\ \bibnamefont {Chang}},
  \bibinfo {author} {\bibfnamefont {R.}~\bibnamefont {Mahapatra}}, \bibinfo
  {author} {\bibfnamefont {M.}~\bibnamefont {Platt}}, \bibinfo {author}
  {\bibfnamefont {M.}~\bibnamefont {Reed}}, \bibinfo {author} {\bibfnamefont
  {I.}~\bibnamefont {Rydstrom}}, \bibinfo {author} {\bibfnamefont
  {B.}~\bibnamefont {Sadoulet}}, \bibinfo {author} {\bibfnamefont
  {B.}~\bibnamefont {Serfass}},\ and\ \bibinfo {author} {\bibfnamefont
  {M.}~\bibnamefont {Pyle}},\ }\href {https://doi.org/10.1063/5.0234265}
  {\bibfield  {journal} {\bibinfo  {journal} {Appl. Phys. Lett.}\ }\textbf
  {\bibinfo {volume} {125}},\ \bibinfo {pages} {232601} (\bibinfo {year}
  {2024})},\ \Eprint {https://arxiv.org/abs/2408.11158} {arXiv:2408.11158
  [physics.ins-det]} \BibitemShut {NoStop}%
\bibitem [{\citenamefont {De~Lucia}\ \emph {et~al.}(2024)\citenamefont
  {De~Lucia}, \citenamefont {Bo}, \citenamefont {Di~Giorgi}, \citenamefont
  {Lari}, \citenamefont {Puglia},\ and\ \citenamefont
  {Paolucci}}]{DeLucia:2024sxp}%
  \BibitemOpen
  \bibfield  {author} {\bibinfo {author} {\bibfnamefont {M.}~\bibnamefont
  {De~Lucia}}, \bibinfo {author} {\bibfnamefont {P.~D.}\ \bibnamefont {Bo}},
  \bibinfo {author} {\bibfnamefont {E.}~\bibnamefont {Di~Giorgi}}, \bibinfo
  {author} {\bibfnamefont {T.}~\bibnamefont {Lari}}, \bibinfo {author}
  {\bibfnamefont {C.}~\bibnamefont {Puglia}},\ and\ \bibinfo {author}
  {\bibfnamefont {F.}~\bibnamefont {Paolucci}},\ }\href
  {https://doi.org/10.3390/instruments8040047} {\bibfield  {journal} {\bibinfo
  {journal} {Instruments}\ }\textbf {\bibinfo {volume} {8}},\ \bibinfo {pages}
  {47} (\bibinfo {year} {2024})},\ \Eprint {https://arxiv.org/abs/2411.01968}
  {arXiv:2411.01968 [physics.ins-det]} \BibitemShut {NoStop}%
\bibitem [{\citenamefont {Gol’tsman}\ \emph {et~al.}(2001)\citenamefont
  {Gol’tsman}, \citenamefont {Okunev}, \citenamefont {Chulkova},
  \citenamefont {Lipatov}, \citenamefont {Semenov}, \citenamefont {Smirnov},
  \citenamefont {Voronov}, \citenamefont {Dzardanov}, \citenamefont
  {Williams},\ and\ \citenamefont {Sobolewski}}]{10.1063/1.1388868}%
  \BibitemOpen
  \bibfield  {author} {\bibinfo {author} {\bibfnamefont {G.~N.}\ \bibnamefont
  {Gol’tsman}}, \bibinfo {author} {\bibfnamefont {O.}~\bibnamefont {Okunev}},
  \bibinfo {author} {\bibfnamefont {G.}~\bibnamefont {Chulkova}}, \bibinfo
  {author} {\bibfnamefont {A.}~\bibnamefont {Lipatov}}, \bibinfo {author}
  {\bibfnamefont {A.}~\bibnamefont {Semenov}}, \bibinfo {author} {\bibfnamefont
  {K.}~\bibnamefont {Smirnov}}, \bibinfo {author} {\bibfnamefont
  {B.}~\bibnamefont {Voronov}}, \bibinfo {author} {\bibfnamefont
  {A.}~\bibnamefont {Dzardanov}}, \bibinfo {author} {\bibfnamefont
  {C.}~\bibnamefont {Williams}},\ and\ \bibinfo {author} {\bibfnamefont
  {R.}~\bibnamefont {Sobolewski}},\ }\href {https://doi.org/10.1063/1.1388868}
  {\bibfield  {journal} {\bibinfo  {journal} {Applied Physics Letters}\
  }\textbf {\bibinfo {volume} {79}},\ \bibinfo {pages} {705} (\bibinfo {year}
  {2001})}\BibitemShut {NoStop}%
\bibitem [{\citenamefont {Mazin}\ \emph {et~al.}(2012)\citenamefont {Mazin},
  \citenamefont {Bumble}, \citenamefont {Meeker}, \citenamefont {O’Brien},
  \citenamefont {McHugh},\ and\ \citenamefont
  {Langman}}]{mazin2012superconducting}%
  \BibitemOpen
  \bibfield  {author} {\bibinfo {author} {\bibfnamefont {B.~A.}\ \bibnamefont
  {Mazin}}, \bibinfo {author} {\bibfnamefont {B.}~\bibnamefont {Bumble}},
  \bibinfo {author} {\bibfnamefont {S.~R.}\ \bibnamefont {Meeker}}, \bibinfo
  {author} {\bibfnamefont {K.}~\bibnamefont {O’Brien}}, \bibinfo {author}
  {\bibfnamefont {S.}~\bibnamefont {McHugh}},\ and\ \bibinfo {author}
  {\bibfnamefont {E.}~\bibnamefont {Langman}},\ }\href@noop {} {\bibfield
  {journal} {\bibinfo  {journal} {Optics express}\ }\textbf {\bibinfo {volume}
  {20}},\ \bibinfo {pages} {1503} (\bibinfo {year} {2012})}\BibitemShut
  {NoStop}%
\bibitem [{\citenamefont {Ramanathan}\ and\ \citenamefont
  {Kurinsky}(2020)}]{Ramanathan:2020fwm}%
  \BibitemOpen
  \bibfield  {author} {\bibinfo {author} {\bibfnamefont {K.}~\bibnamefont
  {Ramanathan}}\ and\ \bibinfo {author} {\bibfnamefont {N.}~\bibnamefont
  {Kurinsky}},\ }\href {https://doi.org/10.1103/PhysRevD.102.063026} {\bibfield
   {journal} {\bibinfo  {journal} {Phys. Rev. D}\ }\textbf {\bibinfo {volume}
  {102}},\ \bibinfo {pages} {063026} (\bibinfo {year} {2020})},\ \Eprint
  {https://arxiv.org/abs/2004.10709} {arXiv:2004.10709 [astro-ph.IM]}
  \BibitemShut {NoStop}%
\bibitem [{\citenamefont {Giffin}\ \emph {et~al.}(2026)\citenamefont {Giffin},
  \citenamefont {Lillard}, \citenamefont {Munbodh},\ and\ \citenamefont
  {Yu}}]{Giffin:2025hdx}%
  \BibitemOpen
  \bibfield  {author} {\bibinfo {author} {\bibfnamefont {P.}~\bibnamefont
  {Giffin}}, \bibinfo {author} {\bibfnamefont {B.}~\bibnamefont {Lillard}},
  \bibinfo {author} {\bibfnamefont {P.}~\bibnamefont {Munbodh}},\ and\ \bibinfo
  {author} {\bibfnamefont {T.-T.}\ \bibnamefont {Yu}},\ }\href
  {https://doi.org/10.1103/6kyn-kqhc} {\bibfield  {journal} {\bibinfo
  {journal} {Phys. Rev. D}\ }\textbf {\bibinfo {volume} {113}},\ \bibinfo
  {pages} {115018} (\bibinfo {year} {2026})},\ \Eprint
  {https://arxiv.org/abs/2511.10764} {arXiv:2511.10764 [hep-ph]} \BibitemShut
  {NoStop}%
\end{thebibliography}%
\bibliographystyle{apsrev4-2}

\end{document}